\documentclass[
reprint,
onecolumn,
superscriptaddress,
longbibliography,
amsmath,
amssymb,
aps,
physrev
]{revtex4-2}
\usepackage[dvipsnames,table]{xcolor}
\usepackage{graphicx}
\usepackage{bm}
\makeatletter\let\SIsave@makecaption\@makecaption\makeatother
\usepackage{subcaption}
\usepackage{hyperref}
\usepackage{xr-hyper}
\usepackage{gensymb}
\usepackage{siunitx}
\usepackage{verbatim}
\DeclareSIUnit\atom{atom}
\DeclareSIUnit\angstrom{\text{\AA}}
\usepackage{soul}
\usepackage{cleveref}
\hypersetup{
    colorlinks=true,
    citecolor=PineGreen,
    urlcolor=RoyalBlue,
    linkcolor=blue,
    pdftitle={TMDC-HEA}
}

\usepackage{threeparttable}
\usepackage{dcolumn}
\usepackage{booktabs}
\usepackage{multirow}
\usepackage{placeins}
\begin{document}
\makeatletter
\let\auto@bib@innerbib\@empty
\let\auto@bib\@empty
\makeatother
\makeatletter
\let\SIsave@maketitle\maketitle
\let\SIsave@and\and
\let\SIsave@affiliation\affiliation
\let\SIsave@author\author
\let\SIsave@AFjoin\@AF@join
\let\SIsave@email\email
\let\SIsave@thanks\thanks
\let\SIsave@abstract\abstract
\let\SIsave@endabstract\endabstract
\makeatother

\title{Prediction and characterisation of cation topology of pentanary high-entropy alloy (MoWNbTaV)S$_2$}
\author{Shubham Agarwal}
\email{agarwshu@fel.cvut.cz}
\affiliation{Department of Control Engineering, Faculty of Electrical Engineering,
Czech Technical University in Prague, Technicka 2, 16627 Prague 6,
Czech Republic}

\author{Tahir Wahab}
\affiliation{Department of Control Engineering, Faculty of Electrical Engineering,
Czech Technical University in Prague, Technicka 2, 16627 Prague 6,
Czech Republic}

\author{Jesper Byggmästar}
\affiliation{Department of Physics, University of Helsinki, P.O. Box 43, FI-00014, Finland}

\author{Antonio Cammarata}
\email{cammaant@fel.cvut.cz}
\affiliation{Department of Control Engineering, Faculty of Electrical Engineering,
Czech Technical University in Prague, Technicka 2, 16627 Prague 6,
Czech Republic}

\author{Tomas Polcar}
\affiliation{School of Engineering,
University of Southampton,
SO17 1BJ, Southampton,
United Kingdom}

\date{\today}
\begin{abstract}

High-entropy transition-metal dichalcogenides offer properties tuned by composition, yet two basic questions remain open: which arrangement of the metals forms under synthesis conditions, and how it can be recognised in experiment. Density functional theory (DFT) cannot sample the vast space of possible arrangements, and the ground-state energies usually employed to screen them do not take into account the temperatures at which these materials are made. This work solves both issues for the pentanary (MoWNbTaV)S\textsubscript{2} monolayer with a Gaussian approximation potential trained on DFT data by iterative learning. Gibbs free energy calculations reveal that vibrational entropy reshapes the stability ranking, while five selected configurations retain their trigonal-prismatic framework during nanosecond molecular dynamics up to \qty{1300}{\kelvin}. Raman and infrared spectra combine parent-monolayer signatures with disorder-activated modes, providing fingerprints for experimental identification.
The predicted lattice parameter follows Vegard's law and agrees within \qty{0.8}{\percent} with X{-}ray diffraction on a sputtered alloy film, which identifies the solid solution but not the arrangement of its metals: distinguishing between arrangements requires the vibrational fingerprints. The potential and its development workflow enable further atomistic simulations and offer a transferable approach to screening related two-dimensional high-entropy materials.
\end{abstract}
\maketitle
\section{INTRODUCTION}
\label{sec:Intro}

The past two decades have witnessed a fundamental shift in material design, moving from the long-held “one principal element” philosophy towards the exploration of vast multidimensional compositional spaces.
A prime exemplar of this new paradigm is the class of high{-}entropy alloys (HEAs), first introduced in 2004 \cite{Yeh2004,Cantor2004}.
By combining five or more principal elements in nearly equimolar ratios, HEAs leverage configurational entropy to overcome the enthalpy of mixing, thereby stabilizing single{‑}phase solid solutions, rather than intermetallic compounds or phase-separated mixtures.
This entropic stabilization effect has unlocked an extraordinary range of mechanically robust, thermally stable, and corrosion-resistant materials.
Importantly, the HEA concept has transcended its original metallic alloy domain, finding a successful extension to oxides \cite{Rost2015}, carbides \cite{Castle2018, Zaid2020a}, borides \cite{Gild2016}, nitrides \cite{Zaid2021a, Zaid2021b}, and most recently, to two{-}dimensional (2D) layered materials \cite{Nemani2023, Divilov2024, Ying2021}.

Among 2D materials, transition-metal dichalcogenides (TMDCs), \textit{MX\textsubscript{2}} with \textit{M} a group IV, V, or VI transition metal and \textit{X} = S, Se, or Te, are especially versatile.
Prototypical members such as MoS\textsubscript{2} and WS\textsubscript{2} exhibit remarkable optoelectronic properties \cite{Chiu2014}, including direct bandgaps in the monolayer limit ($\sim$1.8 eV for MoS\textsubscript{2} \cite{Lee2012}), strong excitonic effects \cite{Ramasubramaniam2012}, valley polarization \cite{Zeng2012}, and high carrier mobilities \cite{Zhang2012}.
These features drove their use in nanoelectronics \cite{Butler2013,Cazalilla2014, Mak2014, Li2014, Piacentini2023, Daus2021, Pu2012} and in electrocatalysis for hydrogen evolution \cite{Henckel2018, Wu2021, Wang2022, Akhound2025} and CO\textsubscript{2} reduction \cite{Cavin2021, Pedersen2023}. 
Yet the small chemical palette of binary TMDCs limits their properties. 
Mixing several transition metals on a shared cation sublattice opens a far larger design space, in which local bonding, and therefore electronic and catalytic behavior, can be tuned by composition and by atomic arrangement.

Despite their promise, understanding the stability of high-entropy TMDCs remains challenging because chemical disorder generates numerous possible atomic arrangements with different local environments.
Even at identical overall compositions, variations in atomic coordination, bond strengths, and lattice distortions can significantly influence structural energetics and vibrational behavior.
The relative thermodynamical stability of individual atomic configurations depends on their Gibbs free-energy contributions, whereas configurational entropy plays a major role at elevated temperatures.
Consequently, configurations favored at low temperatures may not necessarily remain thermodynamically preferred at higher temperatures.
Entropy-assisted stabilization has been predicted for multicomponent disulfides \cite{Deshpande2022}, and compositionally complex TMDCs have been synthesized \cite{Wang2022,Cavin2021,Chin2025,SWang2025}.
Mapping the configurational landscape and its temperature dependence, however, requires extensive sampling, reliable lattice dynamics, and finite-temperature simulations, a combination that is prohibitively expensive with density functional theory (DFT) alone.

Machine-learned interatomic potentials (MLIPs) address this bottleneck by learning the potential-energy surface from first-principles data, delivering near-DFT accuracy at a small fraction of the cost \cite{Behler2007, Shapeev2016, Mishin2021}. 
Among the various MLIP frameworks \cite{Behler2007, Shapeev2016, Mishin2021}, the Gaussian Approximation Potential (GAP) \cite{Bartok2010, Bartok2013, Bartok2015} is a particularly powerful approach.
GAP provides a rigorous Bayesian framework for uncertainty quantification, employs a sparse Gaussian process to balance accuracy and computational cost, and has been successfully applied to diverse systems ranging from elemental metals \cite{Byggmastar2019, Byggmastar2020a} and alloys \cite{Nikoulis2021} to high{-}entropy alloys \cite{Byggmastar2021, Byggmastar2022a, Fellman2026}, liquids \cite{Byggmastar2022b}, and MXenes \cite{Byggmastar2026}.
However, to the best of our knowledge, no study reported a GAP force field specifically developed for multi{-}cationic 2D TMDC HEAs, particularly for the (MoWNbTaV)S\textsubscript{2} system. 
Existing MLIP studies on 2D HEAs \cite{Zhang2020, Liu2021, Zhang2020b, Dai2020, Pei2020} either focused on fine-tuning universal potentials (e.g., CHGNet \cite{Deng2023}, MACE \cite{Batatia2022}, MatterSim \cite{Yang2024}) or employed other MLIP architectures, such as neural network potentials \cite{MartinBarrios2024, Kocer2021}.
Although these efforts are valuable, universal potentials often lack the specialized accuracy required for subtle energy differences (e.g., mixing energies and defect formation energies) in a specific HEA composition, and fine{-}tuning can inadvertently compromise transferability.

In this work, we train a dedicated GAP force field for (MoWNbTaV)S\textsubscript{2}, and use it to explore its thermodynamic stability together with structural and vibrational properties.
The potential is trained on a diverse DFT database refined by iterative learning and validated against DFT.
We then use it to (i) sample the configurational landscape, (ii) compute temperature-dependent Gibbs free energies to identify the arrangements favored in the \qtyrange{500}{1000}{\celsius}  window typical of experimental synthesis, and (iii) test the thermal stability of these arrangements with MD.
Finally, work-function calculations and simulated infrared, Raman, and X-ray diffraction (XRD) spectra link the stable configurations to measurable electronic, vibrational, and structural signatures.
%
\section{METHODS}\label{sec:Methods}
\subsection{Simulation Methods}
\label{sec:SM}
We perform Density Functional Theory calculations using the projector augmented wave (PAW) method \cite{Blochel1994, Kresse1999} as implemented in the Vienna Ab initio Simulation Package (\textsc{vasp}, version 6.5.0) \cite{Kresse1993, Kresse1994, Kresse1996a, Kresse1996b}.
The exchange-correlation functional is treated within the Perdew-Burke-Ernzerhof (PBE) formulation of the generalized gradient approximation (GGA)\cite{Perdew1996}.
For the transition metals, we use the following valence configurations: Ta is treated with the PAW pseudopotential that explicitly includes the 5p semicore states (`Ta\_pv'); for Mo, W, Nb, and V, we employ the `\_sv' pseudopotentials to include the outermost \textit{s} and \textit{p} semicore states (e.g., \textit{4s} and \textit{4p} for Mo and Nb).
Sulfur is described using the standard PAW potential, and a plane-wave kinetic energy cutoff of \qty{500}{\electronvolt} is applied.
To drive the setup, we set the convergence criteria for the electronic self-consistent field (SCF) loop to \qty{e-8}{\electronvolt}, continuing ionic relaxations until the Hellmann-Feynman forces on each atom fall below \qty{e-3}{\electronvolt\per\angstrom}.
A Gaussian smearing of \qty{0.01}{\electronvolt} accelerates SCF convergence.
All calculations are performed on monolayer TMDC systems in a cell with a height of \qty{25.0}{\angstrom} perpendicular to the layer, which leaves a vacuum of about \qty{21.87}{\angstrom} between periodic images of the layer and avoids spurious interactions between them.
To model the pentanary alloy, we use a large unit cell (\autoref{fig:crystal_str}) for which it is sufficient to use only the $\Gamma$-point to sample the Brillouin zone.
%

We carry out molecular dynamics (MD) simulations using the Large-scale Atomic/ Molecular Massively Parallel Simulator (LAMMPS) code \cite{Thompson2022} interfaced with the QUIP/GAP library.
Using a timestep of \qty{1}{\femto\second}, we control the temperature with a Nosé–Hoover thermostat over production runs that span \qty{3}{\nano\second}.
We perform these simulations at \qtylist{300;600;900;1273}{\kelvin} to probe the temperature-dependent stability and configurational disorder of the pentanary high-entropy alloy. 
Finally, we perform phonon calculations using Phonopy \cite{Togo1,Togo2} interfaced with VASP and LAMMPS via the finite-displacement method.
%
\subsection{Atomic Structure}\label{sec:AS}
\begin{figure}[t]
    \centering
    \includegraphics[width=0.6\columnwidth]{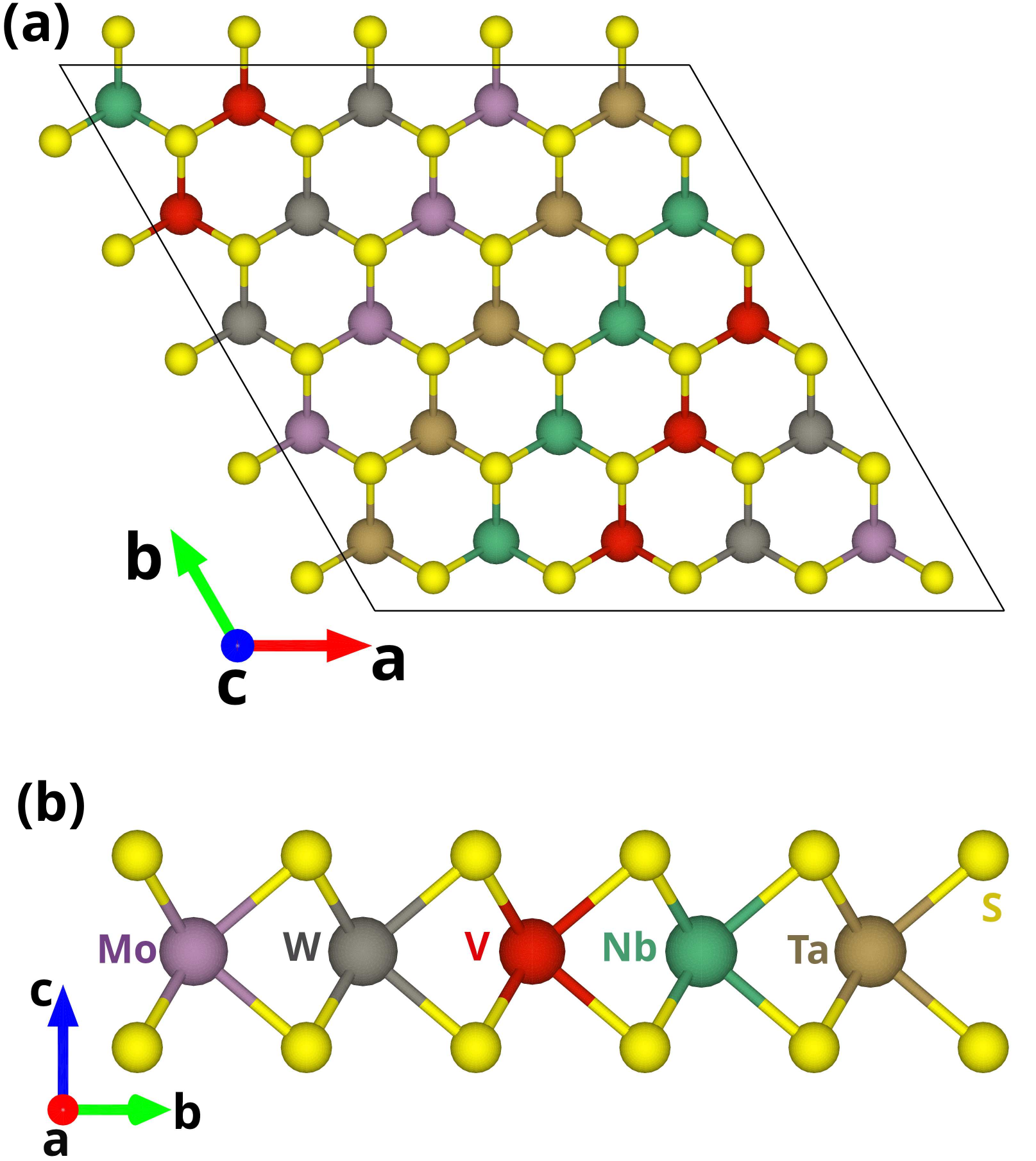}
     \captionsetup{justification=raggedright,singlelinecheck=false}
    \caption{Atomic structure of the (MoWNbTaV)\textsubscript{0.2}S\textsubscript{2} high-entropy monolayer.
  (a) Top view of the $5\times5\times1$ supercell (black parallelogram; $\mathbf{a}$ and
  $\mathbf{b}$ are the in-plane lattice vectors and $\mathbf{c}$ is normal to the layer),
  containing 25 metal atoms, five each of Mo, Nb, Ta, V and W, and 50 S atoms
  (75 atoms in total).
  (b) Side view along $\mathbf{a}$ showing the S--M--S trilayer, in which every metal atom is
  surrounded by six S atoms in a trigonal-prismatic arrangement. 
  Mo, W, V, Nb, Ta and S atoms are shown in purple, grey, red, green, brown and yellow,
  respectively.}
    \label{fig:crystal_str}
\end{figure}

For the pentanary alloy (MoWNbTaV)S\textsubscript{2}, a $5\times5\times1$ supercell is built from the 3{-}atom primitive cell (one metal + two sulfur atoms) as shown in \autoref{fig:crystal_str}, then including a total of 75 atoms (25 metal sites and 50 sulfur atoms).
To represent the equimolar composition, each of the five transition metals is placed on exactly five of the 25 metal sites (i.e., 20~at.\% each).
Given the large number of possible configurations, we generate independent random configurations using a custom random substitution algorithm.
Additionally, to more accurately capture the chemical disorder characteristic of high{-}entropy alloys while maintaining computational feasibility, we generate special quasi{-}random structures (SQSs) \cite{Zunger1990} using the alloy theoretic automated toolkit (ATAT) \cite{vandeWalle2002}.
The \textsc{vesta} \cite{vesta} and \textsc{ovito} \cite{ovito} software are used for visualization of structures and dynamical trajectories.
%
%
\subsection{Machine{-}Learning Potential}\label{sec:MLP}
We use the Gaussian Approximation Potential \cite{Bartok2010, Bartok2013, Bartok2015} framework combined with the QUantum mechanics Interatomic Potential \textsc{(QUIP)} code \cite{QUIP} with a suitable combination of two{-}body and three{-}body descriptors.
For a more detailed description about the GAP/QUIP methodology, we refer the reader to read \cite{Byggmastar2020a, Byggmastar2021} and the references therein. 
Within this approach, the total energy of an \textit{N}{-}atom configuration is expressed as a sum of two{-}body (2b) and three{-}body (3b) contributions:
%
\begin{equation}
E_{\mathrm{GAP}}^{\mathrm{tot}} =
\sum_{i<j}^{N} E^{\mathrm{rep}}_{ij}(r_{ij})
+ \delta_{2b}^{2} \sum_{i<j}^{N}  \sum_{p=1}^{M_{2b}} \alpha_{ij,p} \, K_{2b}(r_{ij}, r_p) 
+ \delta_{3b}^{2} \sum_{i,\, j<k}^{N}  \sum_{t=1}^{M_{3b}} \alpha_{ijk,t} \, K_{3b}(q_{ijk}, q_t)
\label{eq:e_gap}
\end{equation}
where the subscripts 2b and 3b distinguish the two{-}body and three{-}body descriptors.
We summarized the hyperparameters associated with each descriptor (i.e., 2b and 3b) term in \autoref{tab:gap_params}, together with a brief explanations.
%
\begin{table}[htbp]
\caption{\raggedright Hyperparameters used during training of the potential. 
$R_{\text{cut}}$: cutoff radius; $\delta$: energy-scale; 
$N_{\text{sparse}}$: number of sparse points 
(representative atomic environments taken from the training configurations). 
A more detailed explanation of the parameters is provided in Ref. \citenum{Bartok2015}.}
\label{tab:gap_params}
\begin{ruledtabular}
\begin{tabular}{lcc}
Parameter & Two-body (2b) & Three-body (3b) \\
\hline
$R_{\text{cut}}$ & \qty{5.0}{\angstrom} & \qty{4.5}{\angstrom} \\
$\delta$ & \qty{1.0}{\electronvolt} & \qty{0.1}{\electronvolt} \\
$N_{\text{sparse}}$ & 20 & 400 \\
Sparse Method & Uniform & Uniform \\
\end{tabular}
\end{ruledtabular}
\end{table}

The cut-off radii set the size of the atomic environment that the potential can resolve. To determine
appropriate cut-off distances, we examine several values between \qty{3.0}{\angstrom} and \qty{8.0}{\angstrom}.
Increasing the two-body cut-off beyond \qty{5.0}{\angstrom} does not improve the fit: the energy RMSE remains
\qtyrange[range-units=single]{1.17}{1.22}{\milli\electronvolt\per atom} (\qty{1.18}{\milli\electronvolt\per atom}
at \qty{5.0}{\angstrom}) and the force RMSE \qtyrange[range-units=single]{0.099}{0.100}{\electronvolt\per\angstrom},
while the number of neighbours per atom increases from 19 at \qty{5.0}{\angstrom} to 33 at \qty{6.0}{\angstrom}.
A three-body cut-off of \qty{5.0}{\angstrom} lowers the errors only slightly, to
\qty{1.04}{\milli\electronvolt\per atom} and \qty{0.096}{\electronvolt\per\angstrom}, whereas larger values
increase the force error, in particular on S (\qtyrange[range-units=single]{0.115}{0.123}{\electronvolt\per\angstrom}
against \qty{0.097}{\electronvolt\per\angstrom}); at the same time the number of triplets per atom grows from 141
at \qty{4.5}{\angstrom} to 165 at \qty{5.0}{\angstrom} and 520 at \qty{6.0}{\angstrom}.
The cut-offs of \qty{5.0}{\angstrom} for the two-body and \qty{4.5}{\angstrom} for the three-body interactions
therefore provide a balance between accuracy and computational cost.
The chosen values also have a clear physical meaning. The two-body cut-off of \qty{5.0}{\angstrom} contains the
first two M--S shells, the first M--M shell and the S--S pairs across the layer, and lies just above a gap in the
interatomic distances (\qtyrange[range-units=single]{4.60}{4.97}{\angstrom}) that separates these shells from the
next ones. The three-body cut-off of \qty{4.5}{\angstrom} contains the complete coordination of each metal, its six
S and six metal neighbours, and therefore all the bond angles of the trigonal prisms and of the metal network.
The number of sparse points (\textit{N\textsubscript{sparse}}) are selected based on systematic convergence tests.
%
\section{Results and discussion}
\label{sec:Results}
\subsection{Training datapool and ML potential}
\label{sec:Training}
A reliable machine-learned force field (MLFF) requires a training set that spans all bonding environments encountered in the target simulations.
This is particularly important for multi-cation TMDC alloys, where chemical disorder, competing phases, and local compositional fluctuations generate a wide range of atomic environments. 
%
%
Our training database for (MoWNbTaV)S\textsubscript{2} is built in three stages. 
First, we generate structures at the equiatomic composition by systematically permuting the five metal species on the cation sublattice, together with the corresponding pure elemental end‑members. 
Second, we create special quasi‑random structures (SQSs) to emulate the chemical disorder of a random solid solution.
Third, a subset of configurations are lattice‑distorted by isotropic compression and expansion. 
We use this initial database to train a first version of the ML potential.
The training set is then extended through iterative learning.
Using the initial potential, we first relax the structures via energy and force minimization, and subsequently propagate them in NVT molecular dynamics (MD) simulations over a temperature range of \qtyrange{300}{1000}{\kelvin}. 
During these runs, we observe that the preliminary potential occasionally drives atoms into unphysically close contacts, forming dense clusters. 
Incorporating these problematic frames from MD runs into our database, after computing their DFT energies and forces, enables successful retraining of the potential. 
Two iterative cycles suffice to eliminate unphysical clustering and yield a potential that provides stable structures across the full temperature range, bringing the final training set to 2,164 configurations with 162,300 total atoms.

\begin{figure}[t]
  \centering
  \includegraphics[width=\textwidth]{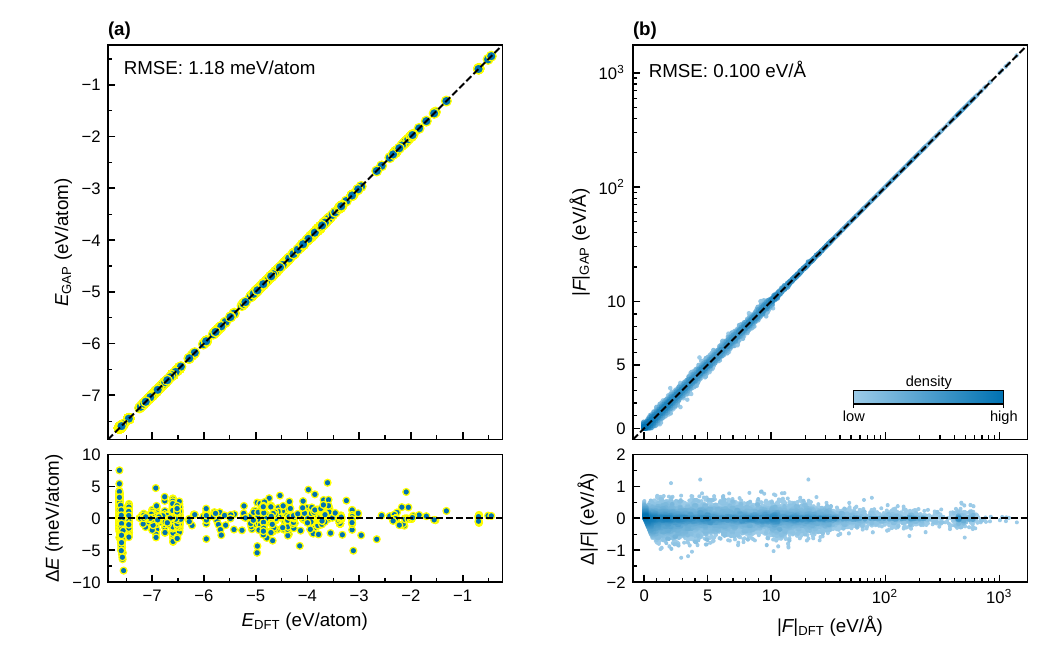}
  \captionsetup{justification=raggedright,singlelinecheck=false}
  \caption{Comparison between DFT and the GAP (a) energies and (b) forces for a training set of (MoWNbTaV)\textsubscript{0.2}S\textsubscript{2} alloy.
  (a) Energy per atom; the lower panel shows the residual
  $\Delta E = E_{\mathrm{GAP}} - E_{\mathrm{DFT}}$.
  (b) Magnitude of the atomic forces, $|F|$, for all atoms (Mo, Nb, Ta, V, W and S); both
  axes are linear up to \qty{10}{\electronvolt\per\angstrom} and logarithmic above, and points are coloured by local data density on a logarithmic scale. The lower panel shows the residual
  $\Delta|F| = |F|_{\mathrm{GAP}} - |F|_{\mathrm{DFT}}$.
  Dashed lines mark perfect agreement.}
  \label{fig:rmse_e_f}
\end{figure}

During training, different weights are assigned to the structure types through the regularization parameters 
$\sigma_{\mathrm{energy}} = \qty{1}{\milli\electronvolt\per\atom}$ and $\sigma_{\mathrm{force}} = \qty{0.01}{\electronvolt\per\angstrom}$. 
The resulting root-mean-square error (RMSE) of the energy on the training set, \qty{1.18}{\milli\electronvolt\per\atom}, is close to $\sigma_{\mathrm{energy}}$, whereas the force RMSE, \qty{0.10}{\electronvolt\per\angstrom} for the force magnitude (\qty{0.09}{\electronvolt\per\angstrom} for the individual Cartesian components), is larger than $\sigma_{\mathrm{force}}$; the small $\sigma_{\mathrm{force}}$ gives the forces a large weight in the fit. 
\autoref{fig:rmse_e_f} shows parity plots comparing DFT reference values with GAP predictions for (a) energies and (b) forces. 
The errors are small and uniform over the whole energy range of the training set, from the equiatomic configurations to the distorted structures. 
Species-wise force RMSEs for Mo, W, Nb, Ta, V and S are provided in the Supporting Information, section S1 ``Accuracy of the GAP forces on the training set''.
Since low training-set errors do not guarantee that the potential is transferable to properties not included in the fit, we test it separately, on the optimised structures of all configurations in \autoref{sec:validation} and at finite temperature in \autoref{sec:MD}.
%
%
\subsection{Validation of ML Potential}
\label{sec:validation}
\begin{figure}[t]
  \centering
  \includegraphics[width=0.8\columnwidth]{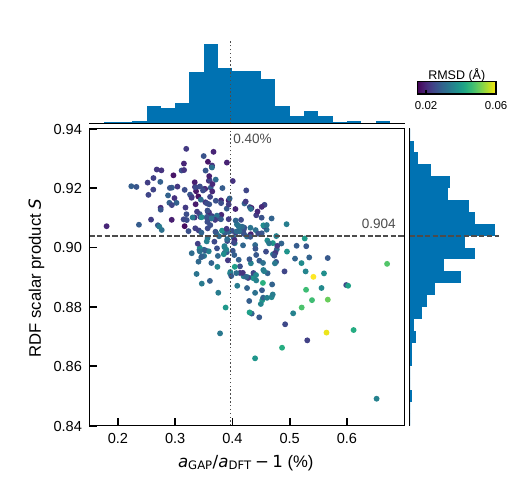}
  \captionsetup{justification=raggedright,singlelinecheck=false}
  \caption{RDF scalar product $S$ between the DFT{-} and GAP{-}optimised structures of all 283 configurations, as a function of the relative difference of their lattice parameters; the histograms show the distributions of the two
  quantities and the dashed and dotted lines their means. The colour gives the root{-}mean-square deviation
  (RMSD) of the atomic positions after placing the GAP structure in the DFT cell.}
  \label{fig:dft_vs_gap}
\end{figure}

Low training errors do not guarantee that a potential reproduces the DFT structures. 
We therefore optimise all 283 configurations with the GAP force field, relaxing both the atomic positions and the in-plane lattice parameters, and compare each of them with its DFT-optimised counterpart.
As a global measure of structural similarity, we compute the scalar product $S$ of the radial distribution functions (RDFs) of the two structures with the MAISE package\cite{Hajinazar2021}; $S=1$ for identical structures and decreases as their interatomic distances differ.
For the 283 configurations, $S$ ranges from \numrange{0.849}{0.933}, with a mean of \num{0.904}
(\autoref{fig:dft_vs_gap}).
These values are below \num{1} mainly because the GAP lattice is slightly
expanded: the GAP lattice parameter, \qtyrange{3.226}{3.239}{\angstrom}, is larger than the DFT one,
\qtyrange{3.214}{3.221}{\angstrom}, by $\Delta a = a_\mathrm{GAP}/a_\mathrm{DFT}-1 = 0.18$--$0.67\%$
(on average \qty{0.40}{\percent}).
MAISE resolves interatomic distances with a Gaussian smearing of only \qty{0.008}{\angstrom}, so that even
this small expansion lowers $S$: configurations with a larger expansion tend to have a smaller $S$, and expanding each DFT structure by its own $\Delta a$, without any other change, already gives $S = 0.906$ on average, the same as the mean value obtained with the GAP structures.
Once the lattice mismatch is removed by placing each GAP structure in the corresponding DFT cell, the
atomic positions differ by a root-mean-square deviation (RMSD) of only \qtyrange{0.017}{0.060}{\angstrom}
(on average \qty{0.029}{\angstrom}), and the mean M{--}S bond lengths by \qtyrange{0.001}{0.006}{\angstrom}.
The GAP thus reproduces the DFT-optimised geometry of every configuration, apart from a small lattice
expansion of \qtyrange{0.2}{0.7}{\percent}.
%
%
\subsection{Configurations favoured at experimental temperatures}
\label{sec:gibbs}
\begin{figure*}[t]
  \centering
  \includegraphics[width=0.8\textwidth]{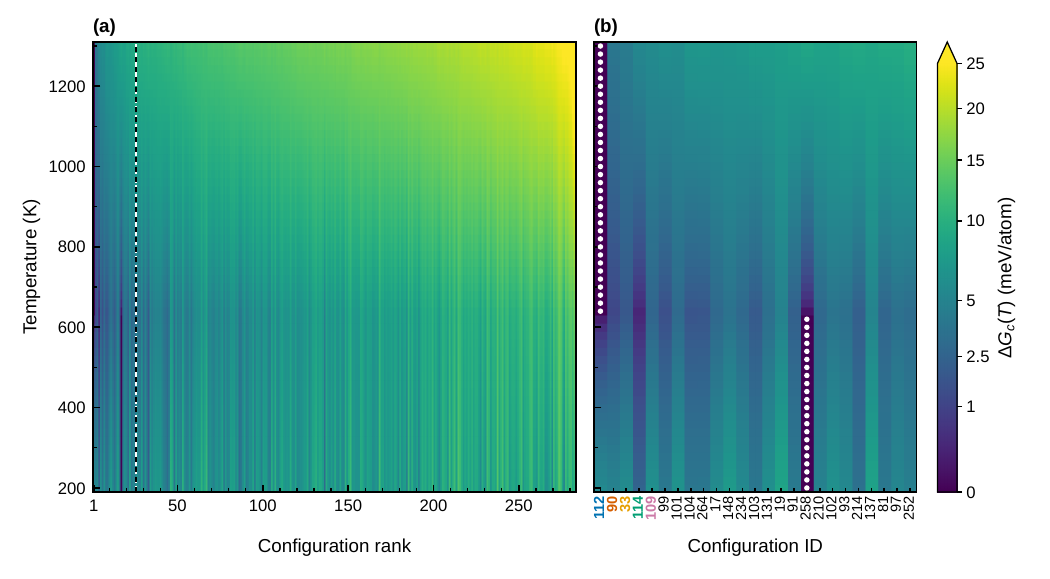}
  \captionsetup{justification=raggedright,singlelinecheck=false}
  \caption{Relative Gibbs free energy of the 283 (MoWNbTaV)\textsubscript{0.2}S\textsubscript{2} configurations from 200 to \qty{1300}{\kelvin}. (a) $\Delta G_c(T) = G_c(T) - G_{\min}(T)$ per atom, where $G_{\min}(T)$ is the lowest Gibbs free energy among all configurations at temperature $T$. Columns are ranked by their mean $\Delta G_c$ over \qtyrange{1140}{1273}{\kelvin} (\qtyrange{867}{1000}{\celsius}), the upper part of the experimental synthesis window (\qtyrange{500}{1000}{\celsius}), so rank~1 is the most stable configuration in this range. The dashed line marks the 25 top-ranked (most stable) configurations, which are enlarged in (b) and labelled by structure ID; coloured IDs are the five configurations for which we report the Gibbs free energy in \autoref{fig:gibbs_selected_config}. White dots in (b) mark the lowest-$G$ configuration at each temperature: configuration~112 above \qty{\approx635}{\kelvin} and configuration~258 below. The colour scale is square-root scaled and saturates at \qty{25}{\milli\electronvolt\per\atom}.}
  \label{fig:gibbs_heatmap}
\end{figure*}

Two-dimensional TMDC HEAs are typically grown at \qtyrange{500}{1000}{\celsius} (\qtyrange{773}{1273}{\kelvin})\cite{SWang2025,Cavin2021, Wang2022,Chin2025,Suneel2023, Suneel2024,Ying2021,Zahra2020}, so the relevant question is which arrangements of the five metals are favoured under growth conditions rather than at \qty{0}{\kelvin}. 
The answer to this question requires the calculation of Gibbs free energies, where the entropic contribution can be evaluated in terms of phonon contributions.
%
%
We therefore compute the Gibbs free energy of all 283 configurations between 200 and \qty{1300}{\kelvin}, a range that brackets the synthesis window.
%
For each configuration, the phonon frequencies $\nu_{\mathbf{q}s}$ are obtained from GAP forces,
and the vibrational contribution is evaluated in the harmonic approximation by treating every
normal mode $(\mathbf{q},s)$ as an independent quantum harmonic oscillator. 
The vibrational internal energy and entropy per cell are \cite{Togo2015}
\begin{equation}
  U_{\mathrm{vib}}(T) = \frac{1}{N_q}\sum_{\mathbf{q},s}
  \left[\frac{1}{2}h\nu_{\mathbf{q}s}
  + \frac{h\nu_{\mathbf{q}s}}{e^{h\nu_{\mathbf{q}s}/k_BT}-1}\right],
  \label{eq:Uvib}
\end{equation}
\begin{equation}
  S_{\mathrm{vib}}(T) = \frac{k_B}{N_q}\sum_{\mathbf{q},s}
  \left[\frac{x_{\mathbf{q}s}}{e^{x_{\mathbf{q}s}}-1}
  - \ln\!\left(1-e^{-x_{\mathbf{q}s}}\right)\right],
  \qquad x_{\mathbf{q}s} = \frac{h\nu_{\mathbf{q}s}}{k_BT},
  \label{eq:Svib}
\end{equation}
where the sum runs over all branches $s$ and the sampled wavevectors $\mathbf{q}$ (here only the $\Gamma$ point of the 75-atom cell, $N_q = 1$), excluding the three zero-frequency acoustic modes at $\Gamma$. 
Twenty-five configurations show a single imaginary mode at $\Gamma$ (between $-0.18$ and \qty{-1.37}{\tera\hertz}), which are then excluded, i.e. none of them is among the most stable configurations discussed below.
The first term of \autoref{eq:Uvib} is the zero-point energy; the second is the thermal excitation energy. With the vibrational Helmholtz free energy
$F_{\mathrm{vib}} = U_{\mathrm{vib}} - TS_{\mathrm{vib}}$, the Gibbs free energy of configuration $c$ is
%
\begin{equation}
  G_c(T) = U_c^{\mathrm{static}} + pV_c + F_c^{\mathrm{vib}}(T),
  \label{eq:G}
\end{equation}
where $U_c^{\mathrm{static}}$ is the GAP energy of configuration $c$ relaxed at $p = \qty{1.77}{\pascal}$ and $V_c$ is its relaxed cell volume. 
At this pressure the $pV_c$ term is below \qty{e-7}{\electronvolt} per cell, so $G_c$ in practice equals the total Helmholtz free energy of the configuration,
$F_c = U_c^{\mathrm{static}} + F_c^{\mathrm{vib}}$.
The frequencies are evaluated once, at the relaxed geometry; thermal expansion and anharmonic effects are therefore not included. 
Since all configurations have the same composition, their ideal configurational entropy of mixing is identical and cancels in the relative free energy
%
\begin{equation}
  \Delta G_c(T) = G_c(T) - G_{\min}(T), \qquad
  G_{\min}(T) = \min_{c'} G_{c'}(T),
  \label{eq:dG}
\end{equation}
which we report per atom.
\autoref{fig:gibbs_heatmap} maps $\Delta G_c(T)$ for all configurations.\\

Although they share the same composition and parent lattice, their free energies span
\qty{13.5}{\milli\electronvolt\per\atom} at \qty{200}{\kelvin}, and this spread widens to
\qty{32}{\milli\electronvolt\per\atom} at \qty{1300}{\kelvin}, while the median $\Delta G_c$ grows from 7.4 to
\qty{15.7}{\milli\electronvolt\per\atom}.
The vibrational free energy therefore does not lower the free energy of all configurations by the same amount:
arrangements of the metals with softer phonon spectra gain more vibrational entropy, so that their free energy
decreases faster with temperature, and the free energies of the different configurations spread further apart as
the temperature rises.
Only a small group of configurations remain close to the lowest free energy at each temperature, $G_{\min}(T)$:
averaged over the whole synthesis window, the 25 most stable configurations lie within
\qty{7}{\milli\electronvolt\per\atom} of $G_{\min}$, whereas the median over all 283 configurations is
\qty{12}{\milli\electronvolt\per\atom}.

\begin{figure}[t]
  \centering
  \includegraphics[width=0.5\columnwidth]{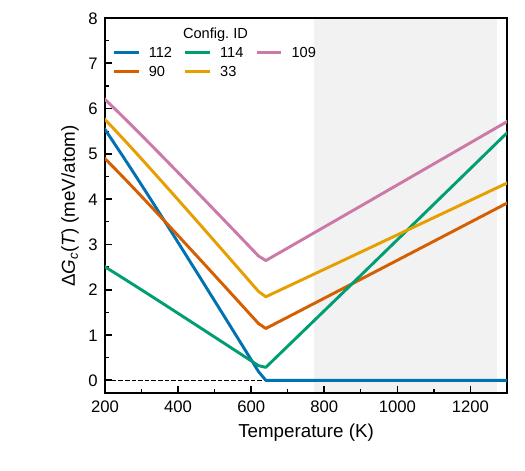}
  \captionsetup{justification=raggedright,singlelinecheck=false}
  \caption{Gibbs free energy of the five configurations with the lowest $G$ at every temperature from
\qty{1140}{\kelvin} to \qty{1273}{\kelvin} (112, 90, 33, 114 and 109), relative to the lowest-$G$
configuration of all 283, $\Delta G_c(T) = G_c(T) - G_{\min}(T)$, per atom. The grey area marks the
experimental synthesis window (\qtyrange{773}{1273}{\kelvin}; \qtyrange{500}{1000}{\celsius}).
The kink at $\approx$\qty{635}{\kelvin} marks the change of the lowest-$G$ configuration from 258
(not shown) to 112.}
  \label{fig:gibbs_selected_config}
\end{figure}

The identity of the most stable configuration changes with temperature.
Configuration~258 has the lowest Gibbs free energy ($\Delta G_c = 0$) below \qty{\approx 635}{\kelvin},
whereas configuration~112 has the lowest Gibbs free energy above \qty{\approx 635}{\kelvin} and up to
\qty{1300}{\kelvin}, i.e.\ throughout the synthesis window (white dots in \autoref{fig:gibbs_heatmap}b).
The static energy is therefore a poor guide to stability under growth conditions.
Ranking the configurations by their static energy and by their Gibbs free energy makes this explicit:
configuration~258 is the first in static energy but only the 20th in Gibbs free energy at \qty{1300}{\kelvin},
whereas configuration~112 is only the 192nd of 283 in static energy but the first in Gibbs free energy
throughout the synthesis window.
Of the 25 configurations with the lowest static energy, only 4 are among the 25 with the lowest Gibbs free
energy averaged over the synthesis window.
\autoref{fig:gibbs_selected_config} follows the five configurations with the lowest Gibbs free energy at every
temperature from \qtyrange{1140}{1273}{\kelvin}, the upper part of the synthesis window: 112, 90, 33, 114 and 109.
The kink of all curves at \qty{\approx635}{\kelvin} reflects the change of the reference $G_{\min}$ from configuration~258 to~112. 
Above it, $\Delta G_c$ grows almost linearly with temperature, as expected in the high-temperature limit of \autoref{eq:Uvib} and \autoref{eq:Svib}, in which the free-energy difference between two configurations becomes $\Delta U^{\mathrm{static}} - T\Delta S_{\mathrm{vib}}$ with a temperature-independent $\Delta S_{\mathrm{vib}}$.
Configuration~112 has a higher static energy than each of its competitors, by 0.4 (109), 0.6 (33), 1.6 (90), 4.9 (114) and \qty{8.5}{\milli\electronvolt\per\atom} (258), but a larger vibrational entropy: linear fits over \qtyrange{780}{1260}{\kelvin} give $\Delta S_{\mathrm{vib}} = 0.054$ (109), 0.044 (33), 0.049 (90), 0.091 (114) and 0.153~$k_B$/atom (258) in its favour (Tables S2 and S3 in the Supporting Information). 
The intercepts of these fits reproduce the static-energy differences to within \qty{0.15}{\milli\electronvolt\per\atom}, confirming that the high-temperature limit holds in this range.
Thus, configuration~112 is stabilised by vibrational entropy and is the most stable of the sampled configurations throughout the synthesis window. 
This entropic advantage is, however, small: the lead of configuration~112 over the runner-up grows only from \qty{1.4}{\milli\electronvolt\per\atom} at \qty{780}{\kelvin} to
\qty{3.7}{\milli\electronvolt\per\atom} at \qty{1260}{\kelvin}, which at the low end of the window is comparable to the GAP energy RMSE of \qty{1.18}{\milli\electronvolt\per\atom}. 
The order of the four following configurations also changes within the window (see Supporting Information, section S2: Free energy analysis of the configurations, for a detailed analysis).
Moreover, the 283 configurations sample only a small part of the $\sim6\times10^{14}$ possible arrangements of the metals in the $5\times5$ cell, and the ideal configurational entropy, 
$-k_BT\ln 5$ per cation (\qty{\approx-139}{\milli\electronvolt} per cation, i.e.\ \qty{-46}{\milli\electronvolt\per\atom}, at \qty{1000}{\kelvin}), is the same for all of them and absent from $\Delta G_c$. 
The ranking therefore identifies which local arrangements are favoured within the solid solution, rather than the stability of the solid solution itself. 
We carry these five most stable configurations, 112, 90, 33, 114 and 109, forward to molecular{-}dynamics simulations and analysis below.
%
%
%
\subsection{Molecular dynamics of the selected configurations}
\label{sec:MD}

\begin{figure}[t]
  \centering
  \includegraphics[width=0.75\textwidth]{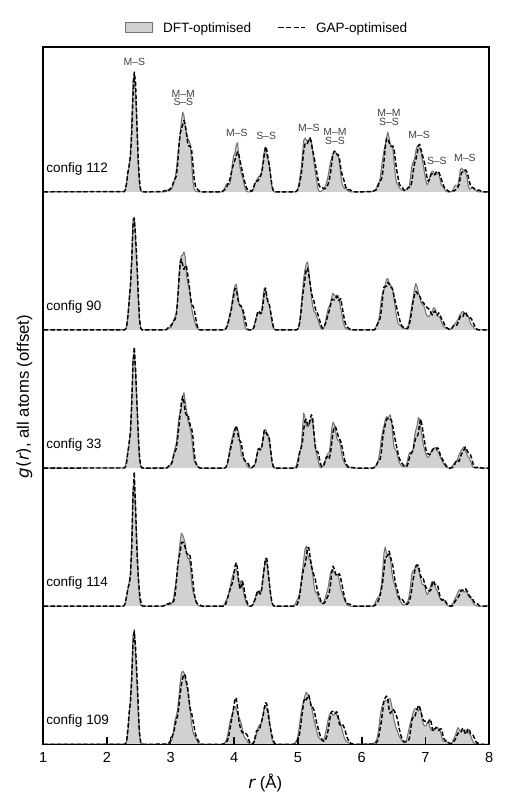}
  \captionsetup{justification=raggedright,singlelinecheck=false}
  \caption{Radial distribution function of all atom pairs, $g(r)$, of the DFT-optimised (grey) and
  GAP-optimised (dashed) structures of the five configurations at (curves offset vertically; Gaussian
  broadening of \qty{0.02}{\angstrom}). The labels give the atom pairs that form each peak. The $g(r)$ are
  normalised with the areal density of the layer.}
  \label{fig:rdf}
\end{figure}

The free energies of \autoref{sec:gibbs} rest on the harmonic phonons of the GAP at \qty{0}{\kelvin}.
We now test whether the GAP remains reliable for the five selected configurations at the temperatures of
synthesis, where the atoms move far from their equilibrium positions.
During MD, the system explores configurations that are not in the training set, and a single poorly described
region of the potential-energy surface is enough to drive atoms into unphysical contacts; this is the
failure we met with the first version of the potential (\autoref{sec:Training}).
At \qty{0}{\kelvin}, the GAP{-} and DFT{-}optimised structures of the five configurations agree as closely as
those of the whole set (\autoref{sec:validation}): their RDFs coincide peak by peak up to \qty{8}{\angstrom}
(\autoref{fig:rdf}) and $S = 0.900$--$0.931$.
The two methods differ in two separate respects, which we quantify separately: the GAP lattice parameter is
\qtyrange{0.35}{0.53}{\percent} larger than the DFT one, and the positions of the atoms within the cell, compared
after expressing both structures in the same cell, differ by an RMSD of only \qtyrange{0.019}{0.031}{\angstrom}.

%
\begin{figure}[h]
  \centering
  \captionsetup{justification=raggedright,singlelinecheck=false}
  \includegraphics[width=\textwidth]{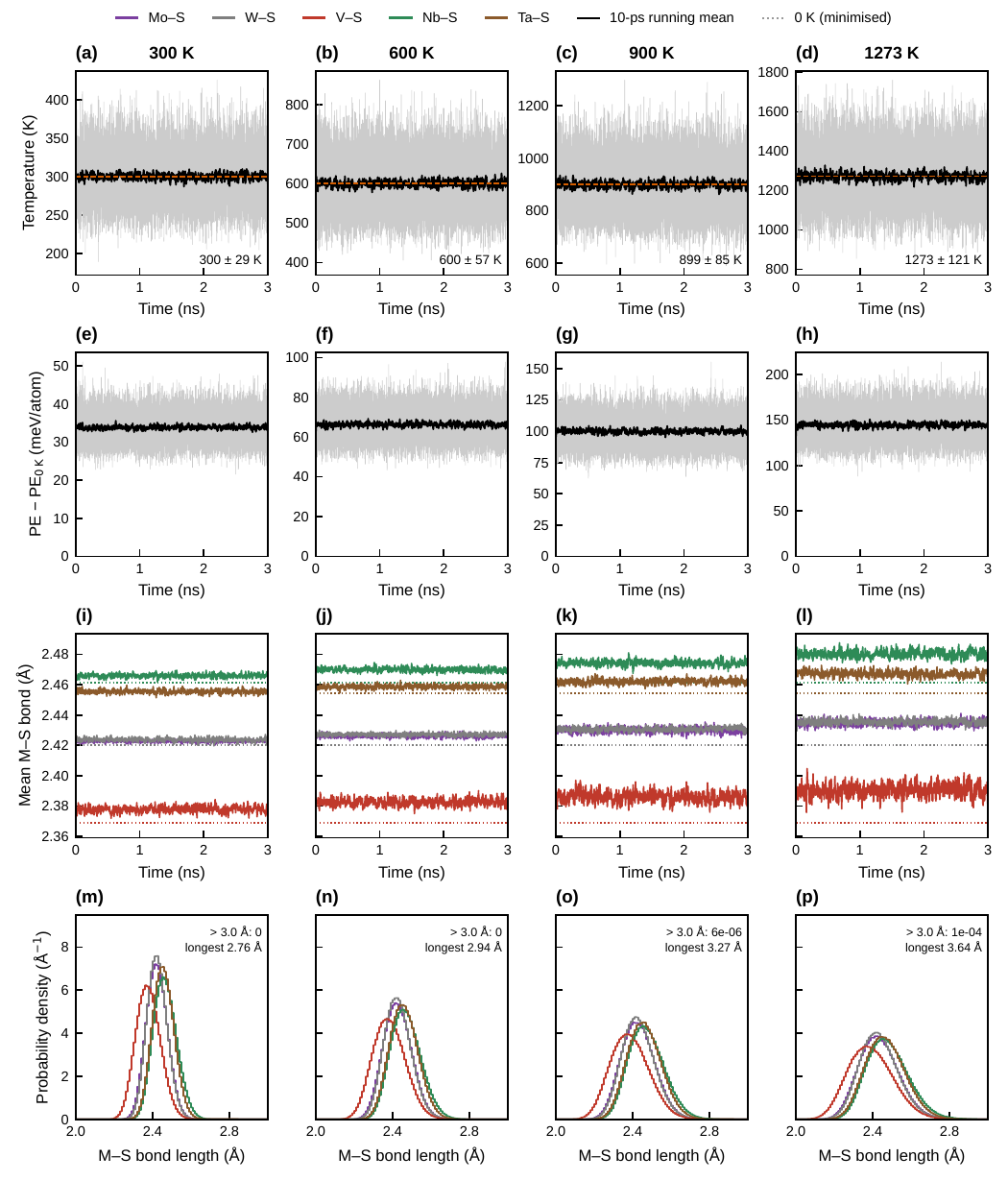}
  \caption{NVT molecular dynamics of configuration 112 with the GAP potential at 300, 600, 900 and
  \qty{1273}{\kelvin} (columns; \qty{3}{\nano\second}, time step \qty{1}{\femto\second}).
  (a{-}d) Instantaneous temperature (grey) with its \qty{10}{\pico\second} running mean (black); the dashed line is the set temperature.
  (e{-}h) Potential energy relative to the 0~K minimised structure.
  (i{-}l) Mean bond length of each M{--}S pair (\qty{10}{\pico\second} running mean); the dotted lines are the \qty{0}{\kelvin} values.
  (m{-}p) Distribution of all M{--}S bond lengths over the run; the text gives the fraction of bonds longer than
  \qty{3.0}{\angstrom} and the longest bond observed.}
  \label{fig:md112_overview}
\end{figure}

\begin{figure}[h]
  \centering
  \includegraphics[width=0.9\textwidth]{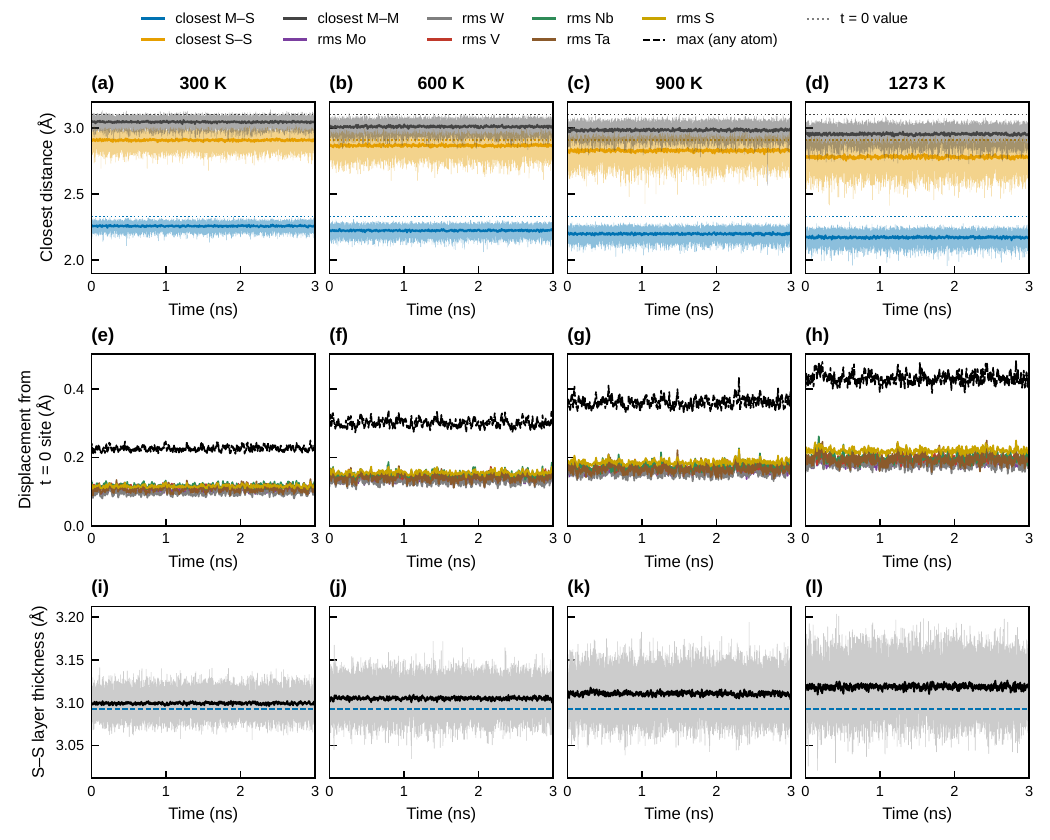}
  \captionsetup{justification=raggedright,singlelinecheck=false}
  \caption{Structural integrity of configuration 112 during the runs of \autoref{fig:md112_overview}.
  (a{-}d) Closest M{--}S, S{--}S and M{--}M distances in each frame (light colour lines) and their \qty{10}{\pico\second} running means
  (dark colour lines); the dotted lines are the values at $t=0$.
  (e{-}h) Root-mean-square displacement of each element from its site at $t=0$ (centre-of-mass motion removed)
  and the largest displacement of any atom (dashed).
  (i{-}l) Layer thickness, i.e.\ the distance between the top and bottom S planes; the dashed line is the value
  at $t=0$.}
  \label{fig:md112_checks}
\end{figure}

%
\begin{table}[htbp]
  \caption{\raggedright DFT- and GAP-optimised structures of the five configurations: lattice parameter $a$, its relative difference $\Delta a$, mean M--S bond length $\bar d_{\text{M--S}}$, root-mean-square deviation (RMSD) of the atomic positions after placing the GAP structure in the DFT cell, and RDF scalar product $S$ of the two optimised structures (Gaussian smearing \qty{0.008}{\angstrom}, cut-off \qty{6.0}{\angstrom}. $S = 1$ for identical structures.}
  \label{tab:static}
  \begin{ruledtabular}
  \begin{tabular}{lccccccc}
    Config. & \multicolumn{2}{c}{$a$ (\AA)} & $\Delta a$ & \multicolumn{2}{c}{$\bar d_{\text{M--S}}$ (\AA)} & RMSD & $S$ \\
     & DFT & GAP & (\%) & DFT & GAP & (\AA) & \\
    \hline
    112 & 3.219 & 3.233 & $+$0.44 & 2.421 & 2.425 & 0.031 & 0.906 \\
    90 & 3.219 & 3.230 & $+$0.35 & 2.421 & 2.424 & 0.027 & 0.931 \\
    33 & 3.218 & 3.231 & $+$0.38 & 2.422 & 2.425 & 0.020 & 0.916 \\
    114 & 3.217 & 3.231 & $+$0.43 & 2.422 & 2.424 & 0.019 & 0.913 \\
    109 & 3.216 & 3.233 & $+$0.53 & 2.422 & 2.425 & 0.027 & 0.900 \\
  \end{tabular}
  \end{ruledtabular}
\end{table}
Starting from the GAP{-}optimised structures, we run \qty{3}{\nano\second} of NVT dynamics for each configuration at
\qtylist{300;600;900;1273}{\kelvin}. 
%
%
\autoref{fig:md112_overview} and \autoref{fig:md112_checks} show the results for
configuration 112; the other four configurations behave in the same way (Figures S4-S11 and Tables S7-S9
of the Supporting Information).
%
The temperature remains stable over the \qty{3}{ns} of each run [\autoref{fig:md112_overview}(a-d)].
The potential energy fluctuates around a constant mean [\autoref{fig:md112_overview}(e-h)]: its linear drift is at most \qty{0.04} {\milli\electronvolt\per\atom} per ns, two orders of magnitude smaller than its thermal fluctuations (\qtyrange{3.3}{15.2}{\milli\electronvolt\per\atom}). 
The runs therefore reach a stationary state, with no slow structural change hidden under the fluctuations.

\autoref{fig:md112_overview}(i-l) follows the mean length of the six M{--}S bonds of each metal. 
The bonds keep the order of the \qty{0}{\kelvin} structure at all temperatures, 
V{--}S $<$ Mo{--}S $\approx$ W{--}S $<$ Ta{--}S $<$ Nb{--}S,
and lengthen only slightly on heating, 
by \qtyrange{0.013}{0.021}{\angstrom} between \num{0} and \qty{1273}{\kelvin}
(less than \qty{1}{\percent}). 
The \qty{0}{\kelvin} bond lengths of the GAP{-}minimised structure agree with those of the DFT{-}relaxed
structure to within \qty{0.004}{\angstrom}. 
The thermal spread of a single bond grows from
\qtyrange{0.052}{0.063}{\angstrom} at \qty{300}{\kelvin} to \qtyrange{0.102}{0.115}{\angstrom} at
\qty{1273}{\kelvin}, a factor of \num{1.8}{--}\num{2.0}, close to the factor $\sqrt{1273/300}=2.06$ of a harmonic bond. 
The bond{-}length distributions [\autoref{fig:md112_overview}(m-p)] stay single-peaked and centred on the \qty{0}{\kelvin} values. 
Even at \qty{1273}{\kelvin} only one bond length in \num{e4} exceeds \qty{3.0}{\angstrom}; such
stretches, up to \qty{3.64}{\angstrom} in \qty{3}{\nano\second}, are isolated events that last no more than
\qty{0.2}{\pico\second} before the bond returns to its usual length.

We test the integrity of the structure directly by monitoring the metal{--}sulfur bonds, the shortest interatomic distances and the thickness of the layer along each trajectory (\autoref{fig:md112_checks}).
Throughout every run, each metal maintains its first coordination shell it has at $t=0$, and no atom moves off its crystallographic site.
The closest distances stay well above any unphysical contact, over all four temperatures,
M{--}S $\geq$ \qty{1.96}{\angstrom}, S{--}S $\geq$ \qty{2.41}{\angstrom} and M{--}M $\geq$ \qty{2.56}{\angstrom}
[\autoref{fig:md112_checks}(a-d)], so the clustering seen with the first version of the potential does not
occur. 
The atoms vibrate about their lattice sites without diffusing
[\autoref{fig:md112_checks}(e-h)]: the root-mean-square displacement of the metals increases from
\qty{0.11}{\angstrom} at \qty{300}{\kelvin} to \qty{0.19}{\angstrom} at \qty{1273}{\kelvin}, about \qty{6}{\percent} of the metal{--}metal distance, and no atom ever moves more than \qty{0.94}{\angstrom} from its initial site.
The layer thickness [\autoref{fig:md112_checks}(i-l)] fluctuates around a mean that rises by only
\qty{0.03}{\angstrom}, from \qty{3.093}{\angstrom} at \qty{0}{\kelvin} to \qty{3.119}{\angstrom} at \qty{1273}{\kelvin}.

\begin{figure}[h]
  \centering
  \includegraphics[width=0.7\columnwidth]{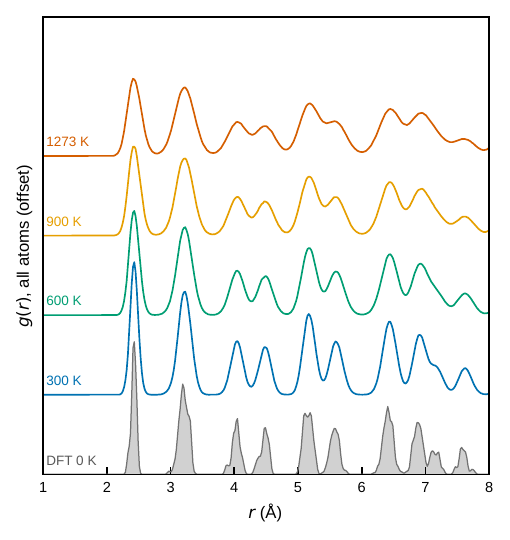}
  \captionsetup{justification=raggedright,singlelinecheck=false}
  \caption{Radial distribution function of all atom pairs of configuration 112, averaged over each
  \qty{3}{ns} NVT run with the GAP at \qtylist{300;600;900;1273}{\kelvin} (curves offset vertically). The grey
  curve at the bottom is the DFT-optimised structure (Gaussian broadening of \qty{0.02}{\angstrom}, scaled
  to the height of the \qty{300}{K} curve); the atom pairs of each peak are given in \autoref{fig:rdf}.}
  \label{fig:md112_rdfT}
\end{figure}

The RDFs averaged over the MD runs confirm that the structure is retained (\autoref{fig:md112_rdfT}; for
the other configurations, see Figure-S13 and Figure-S14, and Table-S7
of the Supporting Information).
On heating from \num{300} to \qty{1273}{\kelvin}, the peaks
broaden (the width of the first M{--}S peak grows from \qty{0.16}{\angstrom} to \qty{0.26}{\angstrom}) but do
not shift significantly, no new peak appears, and integrating the M{--}S RDF over the
first shell gives six S neighbours per metal at every temperature. 
The structure is thus retained up to \qty{1273}{\kelvin}, with no sign of amorphisation, clustering or segregation of the metals on the time scale of the simulation.
%
All five configurations give the same picture:
averaged over all 150 M{--}S bonds, the bond length is
\qtylist{2.429;2.433;2.437;2.442}{\angstrom} at \qtylist{300;600;900;1273}{\kelvin} in every configuration, to
within \qty{0.001}{\angstrom}, and the bonds of each metal lengthen and broaden with temperature in the
same way in all five configurations (Figure-S12 and Table-S8 of the Supporting Information).

Taken together, these runs show that the GAP force fields describes the selected configurations reliably from room temperature to \qty{1273}{\kelvin}, beyond the temperature range of its training data.

%
\subsection{Fingerprints of the solid solution: comparison with experiment}
\label{sec:fingerprints}

The 5 selected systems are characterised by distinct geometric arrangements of the transition metal cations;
such distinct atom geometries produce distinct Raman and infrared signatures which can be compared with corresponding experimental data to characterise synthesised HEA samples.
To this aim, we consider the Raman and infrared measurements of our produced (MoWNbTaV)S\textsubscript{2} films and compare them with the simulated spectra of the 5 selected atomic models.
The films were deposited by reactive high{-}power impulse magnetron sputtering (HiPIMS)\cite{cada2011} in a custom{-}built ultra{-}high{-}vacuum cylindrical chamber from an equiatomic V{-}Nb{-}Mo{-}Ta{-}W target in an Ar/H\textsubscript{2}S atmosphere on Si substrate; here we use the film deposited with an H\textsubscript{2}S flow of \num{20}~sccm.
Raman spectrum were recorded with a Renishaw Raman Microscope RM 1000 (Renishaw, UK) with a wavelength of \qty{514.5}{\nano\metre}.
The XRD pattern was recorded with an Empyrean diffractometer (Malvern Panalytical, UK) in grazing{-}incidence
geometry (GIXRD), with Cu~K$\alpha$ radiation ($\lambda = \qty{0.154}{\nano\metre}$) at an incidence angle of \qty{2.5}{\degree}, for $2\theta$ from
\qty{10}{\degree} to \qty{90}{\degree} in steps of \qty{0.05}{\degree}.
The calculations describe a single, defect{-}free monolayer, whereas the film consists of stacked, nanocrystalline
layers.
Rather than comparing the two directly, we therefore fit the measured data with models built from the
calculated structures, which separates the intrinsic properties of the solid solution from those of the film.
%

\subsubsection{Raman and infrared signatures}
\label{sec:vib}
Raman and infrared (IR) spectra are used as standard vibrational fingerprints of TMDC monolayers
and their alloys. 
The two techniques are complementary: in a 1H{-}MS\textsubscript{2} monolayer, the in{-}plane E$'$
mode is both Raman and IR active, the out{-}of{-}plane A$_1'$ mode is Raman active only, and the
out{-}of{-}plane A$_2''$ mode is IR active only. 
To provide a reference for experiments, we simulate both spectra for the five pristine MS\textsubscript{2} monolayers (M = Mo, W, V, Nb, Ta) and for the most stable configuration discussed in \autoref{sec:gibbs}, configuration 112, and asked to what extent the HEA spectra can be understood in terms of the signals of its constituents.
The other four configurations are discussed in the Supporting Information, Section S4 ``Raman and IR spectra of the selected configurations''.

\begin{figure*}[t]
  \centering
  \includegraphics[width=\textwidth]{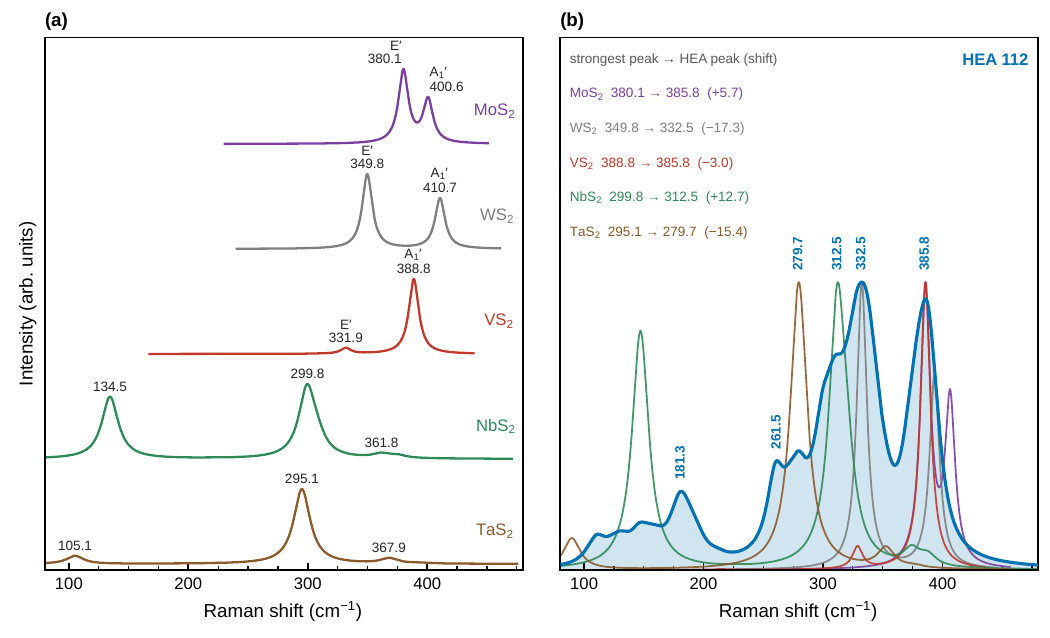}
  \captionsetup{justification=raggedright,singlelinecheck=false}
  \caption{Simulated Raman spectra. (a) Pristine MS$_2$ monolayers (M = Mo, W, V, Nb, Ta), offset
  vertically; numbers give the peak positions in \qty[per-mode=reciprocal]{1}{\per\centi\metre} and, for 
  MoS\textsubscript{2}, WS\textsubscript{2} and VS\textsubscript{2}, the symmetry of the mode. (b) HEA configuration~112 (filled) with the spectra of the pristine monolayers, each rigidly shifted so that its strongest peak coincides with the nearest peak of the HEA spectrum; the shifts are listed in the panel, and the numbers above the HEA spectrum give its peak positions in \qty[per-mode=reciprocal]{1}{\per\centi\metre}. All spectra are normalised to their own maximum.}
  \label{fig:raman}
\end{figure*}

\begin{figure}[t]
  \centering
  \includegraphics[width=\textwidth]{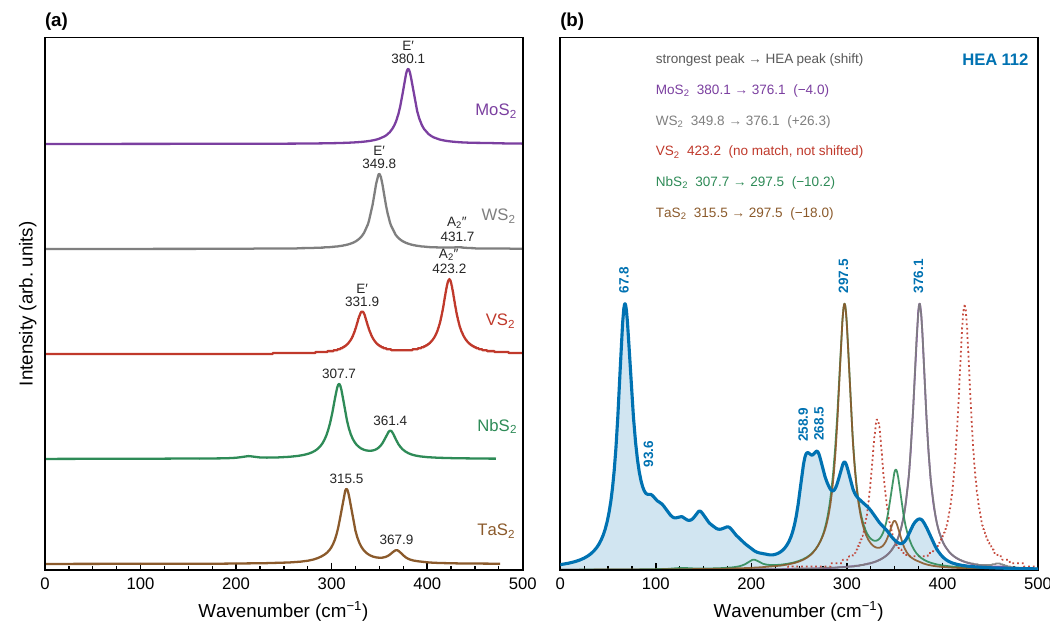}
  \captionsetup{justification=raggedright,singlelinecheck=false}
  \caption{Simulated IR spectra. 
  (a) Pristine MS$_2$ monolayers (M = Mo, W, V, Nb, Ta), offset
  vertically; numbers give the peak positions in \qty[per-mode=reciprocal]{1}{\per\centi\metre} and, for MoS\textsubscript{2}, WS\textsubscript{2} and VS\textsubscript{2}, the symmetry of the mode. 
  (b) HEA configuration~112 (filled) with the spectra of the pristine
  monolayers, each rigidly shifted so that its strongest peak coincides with the nearest peak of the
  HEA spectrum; the shifts are listed in the panel, and the numbers above the HEA spectrum give its
  peak positions in \qty[per-mode=reciprocal]{1}{\per\centi\metre}. The VS\textsubscript{2} spectrum (dotted) has no HEA peak within
  \qty{40}{\per\centi\metre} of its strongest peak and is not shifted. All spectra are normalised to
  their own maximum.}
  \label{fig:ir}
\end{figure}

We compute the spectra following the method of Skelton \textit{et al.} \cite{Skelton2017} with the \textsc{Phonopy-Spectroscopy} tool \cite{phonopy-spectroscopy}.
The $\Gamma$-point phonon modes of each cell are obtained with the aid of the \textsc{Phonopy} package from DFT forces.
The IR intensity of each mode is calculated also using the Born effective charges projected onto the mode eigenvector, and the Raman intensity from the change of the high-frequency dielectric tensor for two structures displaced along the Raman-active mode eigenvector, followed by an orientational average;
the Born charges and dielectric tensors are calculated within the density-functional perturbation theory framework.
%
%
Each mode is represented by a Lorentzian function with an artificial width used for visualisation only, so band widths are not discussed.
Since the cells contain 3 (MoS\textsubscript{2}, WS\textsubscript{2}, VS\textsubscript{2}), 6 (NbS\textsubscript{2}), 12 (TaS\textsubscript{2}) or 75 atoms (HEA), each spectrum is normalised to its own maximum.
%
%
For the pristine monolayers (\autoref{fig:raman}a, \autoref{fig:ir}a), the comparison of the two
spectra identifies the modes directly, because the E$'$ peak appears at the same frequency in both.
MoS\textsubscript{2} and WS\textsubscript{2} show E$'$ at
\qty[per-mode=reciprocal]{380}{\per\centi\metre} and \qty[per-mode=reciprocal]{350}{\per\centi\metre}
in both spectra, and A$_1'$ at
\qty[per-mode=reciprocal]{401}{\per\centi\metre} and \qty[per-mode=reciprocal]{411}{\per\centi\metre}
in Raman only; their A$_2''$ modes, at
\qty[per-mode=reciprocal]{\approx 466}{\per\centi\metre} and \qty[per-mode=reciprocal]{\approx 432}{\per\centi\metre},
are very weak in IR.
The Raman frequencies are within \qtyrange[per-mode=reciprocal]{2}{7}{\per\centi\metre}
of the measured monolayer values\cite{Lee2010,Berkdemir2013}. 
For VS\textsubscript{2}, the peak at \qty[per-mode=reciprocal]{332}{\per\centi\metre},
strong in IR and weak in Raman, is E$'$, so that the strong Raman peak at
\qty[per-mode=reciprocal]{389}{\per\centi\metre} is A$_1'$ and the strong IR peak at
\qty[per-mode=reciprocal]{423}{\per\centi\metre} is A$_2''$. In monolayers of H-phase
VS\textsubscript{2} grown by chemical vapour deposition, the A$_1'$ and E$'$ peaks are measured at
about 404 and \qty[per-mode=reciprocal]{382}{\per\centi\metre}\cite{Su2020}. The calculated A$_1'$
frequency is \qty[per-mode=reciprocal]{15}{\per\centi\metre} lower, whereas the calculated E$'$
frequency is \qty[per-mode=reciprocal]{50}{\per\centi\metre} lower; H{-}phase VS\textsubscript{2} is
predicted to be ferromagnetic, whereas our calculations are non{-}spin{-}polarised.
NbS\textsubscript{2} and TaS\textsubscript{2} are not stable in the trigonal-prismatic structure of
the other three monolayers and relax to lower symmetry, with point groups C$_{2v}$ (NbS\textsubscript{2},
6-atom cell) and C$_1$ (TaS\textsubscript{2}, 12-atom cell). Their modes therefore carry different
labels, and more of them are Raman and IR active. They show their strongest peaks at
\qty[per-mode=reciprocal]{300}{\per\centi\metre} and \qty[per-mode=reciprocal]{295}{\per\centi\metre}
in Raman and at
\qty[per-mode=reciprocal]{308}{\per\centi\metre} and \qty[per-mode=reciprocal]{316}{\per\centi\metre}
in IR, and a common peak at
\qtyrange[per-mode=reciprocal]{361}{368}{\per\centi\metre} in both. The strongest Raman peaks lie
close to the E$^1_{2g}$ modes measured at \qty[per-mode=reciprocal]{304}{\per\centi\metre} in
2H-NbS\textsubscript{2}\cite{McMullan1983} and at \qty[per-mode=reciprocal]{285}{\per\centi\metre}
in bulk and about \qty[per-mode=reciprocal]{290}{\per\centi\metre} in few-layer
2H-TaS\textsubscript{2}\cite{ValenciaIbanez2022}, whose frequency increases as the number of
layers decreases.

%

In the HEA, the chemical disorder lowers the symmetry of the cell, so that many more modes become
Raman and IR active and merge into bands (\autoref{fig:raman}b, \autoref{fig:ir}b). 
The Raman spectrum of configuration 112 shows a high-frequency band at 
\qty[per-mode=reciprocal]{386}{\per\centi\metre}, a main band at
\qtyrange[per-mode=reciprocal]{260}{355}{\per\centi\metre} with its maximum at \qty[per-mode=reciprocal]{332}{\per\centi\metre}, and weaker features at \qtyrange[per-mode=reciprocal]{100}{220}{\per\centi\metre}. 
The IR spectrum has its main band at
\qtyrange[per-mode=reciprocal]{259}{298}{\per\centi\metre}, a weaker band at \qty[per-mode=reciprocal]{376}{\per\centi\metre} and a strong low-frequency peak at \qty[per-mode=reciprocal]{68}{\per\centi\metre}. 
As in the pristine monolayers, the two spectra emphasise different modes: the high-frequency Raman band lies \qty[per-mode=reciprocal]{10}{\per\centi\metre} above the corresponding IR band, as the Raman-only A$_1'$ mode lies above E$'$ in MoS\textsubscript{2} and VS\textsubscript{2}.
To resolve these bands in terms of the pristine signals, we rigidly shifted each pristine spectrum
so that its strongest peak coincides with the nearest HEA peak (\autoref{fig:raman}b, \autoref{fig:ir}b). 
In Raman, all five signals are found within \qty[per-mode=reciprocal]{18}{\per\centi\metre} of their pristine positions: the
MoS\textsubscript{2} (+5.7) and VS\textsubscript{2} (\qty[per-mode=reciprocal]{-3.0}{\per\centi\metre}) peaks fall on the high-frequency band, the WS\textsubscript{2} peak (\qty[per-mode=reciprocal]{-17.3}{\per\centi\metre}) on the maximum of the main band, and the NbS\textsubscript{2} (+12.7) and TaS\textsubscript{2} (\qty[per-mode=reciprocal]{-15.4}{\per\centi\metre}) peaks on its lower part. 
In IR, the MoS\textsubscript{2} E$'$ peak is almost unshifted (\qty[per-mode=reciprocal]{-4.0}{\per\centi\metre}) and the NbS\textsubscript{2} and TaS\textsubscript{2} peaks fall on the main band (\num{-10.2} and \qty[per-mode=reciprocal]{-18.0}{\per\centi\metre}), whereas the WS\textsubscript{2} peak requires a larger shift (\qty[per-mode=reciprocal]{+26.3}{\per\centi\metre}) and the VS\textsubscript{2} A$_2''$ peak has no counterpart within
\qty[per-mode=reciprocal]{40}{\per\centi\metre}. 
The same picture holds for the other four configurations (Figs. S16 and S18, Tables S10 and S11 of Supporting Information): in Raman, the shifts are at most \qty[per-mode=reciprocal]{22}{\per\centi\metre}, and
the MoS\textsubscript{2}-like signal is found in all of them at \qtyrange[per-mode=reciprocal]{373}{386}{\per\centi\metre} in Raman and
\qtyrange[per-mode=reciprocal]{371}{382}{\per\centi\metre} in IR, with shifts of only $-7$ to $+$\qty[per-mode=reciprocal]{6}{\per\centi\metre} and
$-9$ to $+$\qty[per-mode=reciprocal]{2}{\per\centi\metre}, respectively.
The HEA spectra thus contain the signatures of the five constituents, displaced by the different
local environment of each metal in the alloy, together with features that have no counterpart in
the pristine monolayers: the Raman features below \qty[per-mode=reciprocal]{220}{\per\centi\metre} and the strong IR
peaks below \qty[per-mode=reciprocal]{100}{\per\centi\metre}, which appear in four of the five configurations
(Fig. S17 of Supporting Information) and arise from the disorder-activated modes of the alloy.

Across the five configurations, the band positions are similar in both spectra, but the relative
heights of the bands differ (Figs. S15 and S17 of the Supporting Information): the strongest Raman feature lies between
\num{284} and \qty[per-mode=reciprocal]{344}{\per\centi\metre}
for four configurations but at 
\qty[per-mode=reciprocal]{209}{\per\centi\metre}
for configuration~114 (with an equally strong band at \qty[per-mode=reciprocal]{300}{\per\centi\metre}), and the strongest IR feature at 
\qty[per-mode=reciprocal]{68}{\per\centi\metre}
for configuration~112 but between 
\num{259} and \qty[per-mode=reciprocal]{285}{\per\centi\metre}
for the others. 
The peak positions therefore identify the HEA and relate its bands to the constituents, whereas the shape of the spectra carries the signature of the specific arrangement of the metals.
%
%
%
\subsubsection{X-ray Diffraction (XRD)}
\label{sec:xrd}
Powder X{-}ray diffraction (XRD) patterns are simulated with VESTA from the DFT{-}optimised
structures, for Cu~K$\alpha$ radiation (K$\alpha_1$ and K$\alpha_2$, \qtylist{1.54059;1.54443}{\angstrom},
intensity ratio 2:1) and $2\theta$ from \qty{1}{\degree} to \qty{120}{\degree} in steps of \qty{0.01}{\degree}.
Because the simulation cell repeats the monolayer every \qty{25}{\angstrom} along $z$, only the in{-}plane
($l=0$) reflections $(100)$, $(110)$ and $(200)$ belong to the monolayer, while other reflections arise from
the replication of the cell along the non-periodic direction (Figure S19 of SI).
%
Lattice parameters are obtained from the positions of the in{-}plane reflections with Bragg's law,
$d_{hk0} = \lambda_{\mathrm{K}\alpha_1}/(2\sin\theta)$, and $a = 2d_{hk0}\sqrt{(h^2+hk+k^2)/3}$.
%

We fit the GIXRD pattern of the film (\autoref{sec:fingerprints}) with a model built from configuration~112.
Its DFT{-}optimised monolayer, with the atomic positions kept fixed, is stacked into crystallites of a few
layers, in which each layer either sits directly on the layer below or is shifted by one third of a lattice
vector, at random (a stacking{-}fault model \cite{warren1941,Hendricks1942}).
The stack is a geometric model of the film: the layers are not relaxed together, neither with DFT nor with the
GAP, and the interlayer distance and stacking are fitted to the measured pattern; relaxing the stack could
shift the peaks slightly, which we do not expected to affect the results in a significant way.
The scattered intensity is computed along the reciprocal{-}lattice rods of the stacked layers from the
structure factor of the monolayer, including the Debye{-}Waller factor of each element obtained from the MD at
\qty{300}{\kelvin} (\autoref{sec:MD}), and averaged over the orientations of the crystallites.
The in{-}plane lattice parameter, the interlayer distance, the number of layers, the in{-}plane crystallite
size, the probability of a stacking fault, the preferred orientation, an additional displacement parameter
for static disorder and the background (the smooth intensity underneath the peaks) are refined by least
squares; the model and the corrections for the GIXRD geometry are described in Section S5 ``X-ray Diffraction'' of the Supporting Information.
%

\begin{figure}[t]
  \centering
  \captionsetup{justification=raggedright,singlelinecheck=false}
  \includegraphics[width=\columnwidth]{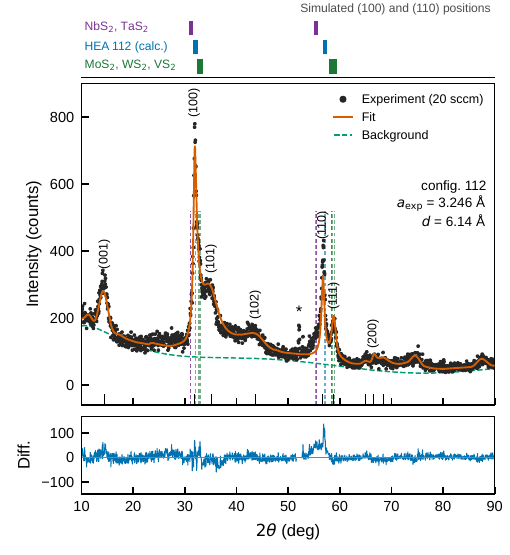}
  \caption{Experimental GIXRD pattern (Cu K$\alpha$, incidence angle \qty{2.5}{\degree}) of the
  (MoWNbTaV)S\textsubscript{2} film deposited with 20 sccm H\textsubscript{2}S (dots) and its fit with the
  stacking-fault model built from configuration~112 (line); the dashed line is the background and the lower
  panel the difference. Ticks below the pattern mark the reflections of the layered film with
  $a_\mathrm{exp} = \qty{3.246}{\angstrom}$ and the interlayer distance \qty{6.14}{\angstrom}, indexed in the
  one-layer cell (reflections with $h-k \neq 3n$ are broadened into a band by stacking faults); * marks a sharp
  reflection that does not belong to the film and was excluded from the fit. Top, and coloured dashed lines in
  the pattern: positions of the $(100)$ and $(110)$ reflections in the simulated patterns of the pristine
  monolayers and of configuration~112 (Fig. S19, and  Table S12 of SI).}
  \label{fig:xrdmain}
\end{figure}

In the simulated patterns, the pristine monolayers fall into two groups (top of \autoref{fig:xrdmain}):
the $(100)$ reflection lies at
\qty{31.14}{\degree}{--}\qty{31.16}{\degree} in NbS\textsubscript{2} and TaS\textsubscript{2} ($a = \qty{3.31}{\angstrom}$), and at
\qty{32.71}{\degree}{--}\qty{33.02}{\degree} in MoS\textsubscript{2}, WS\textsubscript{2} and VS\textsubscript{2} ($a = $ \qtyrange{3.13}{3.16}{\angstrom}).
The HEA shows a single set of in{-}plane reflections between the two groups, with $(100)$ at
\qty{32.05}{\degree}, $(110)$ at \qty{57.16}{\degree} and $(200)$ at \qty{67.02}{\degree},
rather than the reflections of its five parents.
Its in{-}plane lattice parameter, $a = \qty{3.220}{\angstrom}$, is only \qty{0.2}{\percent} larger than the
average of the five parent lattice parameters, \qty{3.214}{\angstrom}; the alloy thus follows Vegard's law closely.
The other four configurations give the same pattern (Figure S19 and S20):
over all five configurations the
$(100)$ and $(110)$ reflections vary by only
\qty{0.04}{\degree} and \qty{0.11}{\degree},
and the lattice parameter by
\qty{0.006}{\angstrom}, from \qtyrange{3.217}{3.223}{\angstrom} (Table:S12).
Like the work function (\autoref{sec:wf}), the lattice parameter is therefore set by the composition and not by the arrangement of the metals.
%

The experimental pattern confirms this picture (\autoref{fig:xrdmain}).
It shows a single set of in{-}plane reflections, with a sharp $(100)$ reflection at \qty{31.91}{\degree} and
$(110)$ at \qty{56.74}{\degree}, between those of the two groups of parent compounds, and no separate
reflections of the parent phases, as expected for a solid solution.
The fit with configuration~112 reproduces the whole pattern ($R_\mathrm{wp} = 0.124$) with an in{-}plane
lattice parameter $a_\mathrm{exp} = \qty{3.246}{\angstrom}$, \qty{0.8}{\percent} larger than the DFT value and
\qty{0.4}{\percent} larger than the GAP value.
A difference of this size is within the usual accuracy of DFT lattice parameters and may further reflect
the composition and defects of the film and the thermal expansion between \qty{0}{\kelvin} and room temperature.
The film is nanocrystalline: its crystallites contain on average about seven layers
(\qty{4.5}{\nano\metre}), separated by \qty{6.14}{\angstrom}, close to the interlayer distance of bulk
MoS\textsubscript{2} and WS\textsubscript{2}, and about half of the layers are shifted with respect to the
layer below, with no regular 2H stacking.
These stacking faults turn the $(10l)$ reflections into a continuous band, which gives the $(100)$ reflection
its long tail on the high{-}angle side and leaves only broad maxima near the $(101)$ and $(102)$ positions,
while the $(00l)$ and $(11l)$ reflections stay sharp.
The thermal displacements from the MD (\qtyrange{0.24}{0.34}{\angstrom\squared} for $B = 8\pi^2U$ at
\qty{300}{\kelvin}) account for only a small part of the decrease of the intensities with angle; the fit
requires an additional $B \approx \qty{1.5}{\angstrom\squared}$, showing that static disorder in the film
dominates.
The other four configurations fit the pattern equally well, with the same $a_\mathrm{exp}$ and
$R_\mathrm{wp} = $ 0.123{--}0.125 (Figure:S23 and Table:S13).
XRD thus identifies the (MoWNbTaV)S\textsubscript{2} solid solution and confirms its predicted lattice
parameter, whereas distinguishing between arrangements of the metals requires the vibrational fingerprints of \autoref{sec:vib}.
%
%
\subsection{Work function}
\label{sec:wf}
\begin{figure}[t]
  \centering
  \includegraphics[width=\columnwidth]{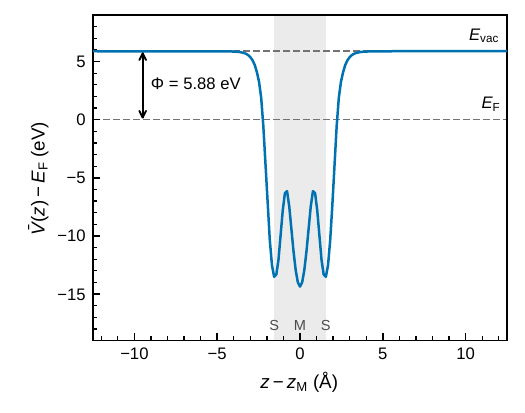}
  \captionsetup{justification=raggedright,singlelinecheck=false}
  \caption{Planar-averaged electrostatic potential $\bar{V}(z)$ of HEA configuration~112 along the
  direction normal to the layer, relative to the Fermi energy $E_\mathrm{F}$; $z_\mathrm{M}$ is the
  position of the metal plane and the shaded area marks the layer between the two sulfur planes.
  The work function $\Phi$ is the difference between the vacuum level $E_\mathrm{vac}$, the flat
  value of $\bar{V}$ far from the layer, and $E_\mathrm{F}$.}
  \label{fig:wf_planar}
\end{figure}

The work function $\Phi$ is defined as the minimum energy needed to move an electron from the Fermi
level to the vacuum level.
It sets the band alignment at contacts in heterostructures and is therefore a key quantity for
electronic and catalytic applications of TMDC monolayers; for this reason, we evaluate the work
function for the selected systems.
We calculate it as $\Phi = E_\mathrm{vac} - E_\mathrm{F}$, where $E_\mathrm{F}$ is the Fermi energy
and $E_\mathrm{vac}$ the vacuum level; the latter corresponds to the constant value reached by the
planar-averaged electrostatic potential $\bar{V}(z)$ in the vacuum region of the unit cell, as shown
for configuration~112 in \autoref{fig:wf_planar}.
$\bar{V}(z)$ is flat over more than \qty{7}{\angstrom} on both sides of the layer, and the vacuum
levels on the two sides agree to within \qty{0.01}{\milli\electronvolt} in all systems
(Fig. S24 of the Supporting Information), so that the layers carry no net dipole and no dipole
correction is needed.
The calculated values of the work function are listed in \autoref{tab:wf}.
Among the pristine monolayers, $\Phi$ ranges from \qty{5.756}{\electronvolt} (WS\textsubscript{2})
to \qty{6.044}{\electronvolt} (NbS\textsubscript{2}).
The HEA configuration~112 has $\Phi = \qty{5.882}{\electronvolt}$, within the range of its
constituents and slightly below the values of the metallic NbS\textsubscript{2}, VS\textsubscript{2}
and TaS\textsubscript{2}.
The five configurations of \autoref{sec:gibbs} give $\Phi$ between \num{5.854} and
\qty{5.882}{\electronvolt} (Table S14 of Supporting Information), a spread of only \qty{28}{\milli\electronvolt}.
Their vacuum levels agree within \qty{6}{\milli\electronvolt}, so the small differences come from
the Fermi level; their potential profiles are indistinguishable on the scale of
\autoref{fig:wf_planar} (Fig. S24 of Supporting Information).
The work function is thus essentially independent of the arrangement of the metals, and
$\Phi \approx \qty{5.87}{\electronvolt}$ can be taken as a property of the
(MoWNbTaV)S\textsubscript{2} monolayer. 
%
%
\begin{table}[t]
  \caption{\raggedright Work function $\Phi$ of the pristine MS$_2$ monolayers and of the
  (MoWNbTaV)S\textsubscript{2} HEA.}
  \label{tab:wf}
  \begin{ruledtabular}
  \begin{tabular}{lc}
    System & $\Phi$ (\unit{\electronvolt}) \\
    \hline
    MoS$_2$ & 5.939 \\
    WS$_2$  & 5.756 \\
    VS$_2$  & 5.949 \\
    NbS$_2$ & 6.044 \\
    TaS$_2$ & 5.898 \\
    \hline
    HEA, configuration 112 & 5.882 \\
    HEA, five configurations$^{a}$ & 5.854--5.882 \\
  \end{tabular}
  \end{ruledtabular}
  \raggedright\footnotesize $^{a}$Configurations 112, 90, 33, 114 and 109 (Table S9).
\end{table}

\section{Conclusion}
\label{sec:conclusion}

We have determined which arrangements of the metals the pentanary (MoWNbTaV)S\textsubscript{2}
monolayer is likely to form under synthesis conditions, and how they can be recognised in experiments.
To this end, we trained a Gaussian approximation potential on DFT data by iterative learning.
In nanosecond molecular dynamics up to \qty{1273}{\kelvin}, the five selected configurations retain
their trigonal-prismatic framework: every metal keeps its sixfold sulfur coordination, and heating
only broadens the interatomic distances.
Gibbs free energy calculations show that vibrational entropy plays a fundamental role in the
formation of specific cation arrangements, indicating that internal energies alone are a poor guide
to the possible cation distributions under growth conditions.
Of the 25 configurations with the lowest internal energy, only four are among the 25 most stable at
synthesis temperatures, so that a screening based on internal energies would miss most of the
configurations that matter.
For the five configurations with the lowest Gibbs free energy, we predicted the experimental
signatures by which they can be identified.
Their Raman and infrared spectra retain the modes of all five parent monolayers, each shifted by the
local chemical environment of its metal, together with other modes absent in every parent pristine
monolayer; therefore, a single spectrum both identifies the alloy and reports on its constituents.
The in-plane lattice parameter, \qty{3.22}{\angstrom}, follows Vegard's law to within
\qty{0.2}{\percent}, and the work function, \qty{5.87}{\electronvolt}, varies by only
\qty{28}{\milli\electronvolt} between configurations; both are set by the composition rather than by
the arrangement of the transition metals.
X-ray diffraction on a sputtered (MoWNbTaV)S\textsubscript{2} film confirms the predicted lattice
parameter within \qty{0.8}{\percent} and shows a single set of in-plane reflections, as expected for
a solid solution.
A stacking-fault model built from the calculated monolayer reproduces the whole pattern and shows
that the film consists of nanocrystalline stacks of about seven disordered layers, whose intensities
are reduced mainly by static disorder rather than by thermal vibrations.
X-ray diffraction thus identifies the solid solution and confirms its predicted lattice parameter,
but it cannot distinguish between arrangements of the metals, which requires the vibrational
fingerprints.
These results show that the configurations of high-entropy two-dimensional materials must be
predicted from free energies at the temperatures at which they are made, and that the vibrational
spectra, rather than the lattice parameter or the work function, are the probe that distinguishes
between them.
The identified computational protocol (i.e.\ a machine-learned potential validated against DFT and
combined with harmonic free energies, molecular dynamics and calculated fingerprints) can be promptly
applied to the prediction and characterisation of HEAs with chemical compositions other than the one
considered here.
The same potential enables us to study the thermal transport and mechanical properties of monolayers
with different cation ordering, which will be the subject of future work.

\begin{acknowledgments}
\label{sec:acknow}
This work was co-funded by the European Union under the project ``Robotics and advanced industrial production'' (reg. no. CZ.02.01.01/00/22\_008/0004590).
This work was also done with the support of the Czech Science Foundation (project No. 25-17542S),
and by the Ministry of Education, Youth and Sports of the Czech Republic through the e-INFRA CZ (ID:90254).
The access to the computational infrastructure of the OP VVV funded project CZ.02.1.01/0.0/0.0/16\_019/0000765 ``Research Center for Informatics'' is also acknowledged.
JB was supported by funding from the Research council of Finland through the OCRAMLIP project, grant number \num{354234}.
\end{acknowledgments}
%

%
%
\clearpage
\setcounter{figure}{0}
\setcounter{table}{0}
\setcounter{equation}{0}
\setcounter{section}{0}
\renewcommand{\thefigure}{S\arabic{figure}}
\renewcommand{\thetable}{S\arabic{table}}
\renewcommand{\theequation}{S\arabic{equation}}
\renewcommand{\thesection}{S\arabic{section}}
\renewcommand{\thesubsection}{\thesection.\arabic{subsection}}
\renewcommand{\theHfigure}{S\arabic{figure}}
\renewcommand{\theHtable}{S\arabic{table}}
\renewcommand{\theHequation}{S\arabic{equation}}
\renewcommand{\theHsection}{S\arabic{section}}
\renewcommand{\theHsubsection}{S\arabic{section}.\arabic{subsection}}
\hypersetup{citecolor=teal,urlcolor=blue,linkcolor=blue}
\sisetup{range-phrase={\ifmmode\ \else\space\fi\text{to}\ifmmode\ \else\space\fi}}
\makeatletter
\expandafter\renewcommand\expandafter\section\expandafter
  {\expandafter\@fb@secFB\section}
\newcommand\@fb@secFB{\FloatBarrier
  \gdef\@fb@afterHHook{\@fb@topbarrier \gdef\@fb@afterHHook{}}}
\g@addto@macro\@afterheading{\@fb@afterHHook}
\gdef\@fb@afterHHook{}
\let\@makecaption\SIsave@makecaption
\let\maketitle\SIsave@maketitle
\let\and\SIsave@and
\let\affiliation\SIsave@affiliation
\let\author\SIsave@author
\let\@AF@join\SIsave@AFjoin
\let\email\SIsave@email
\let\thanks\SIsave@thanks
\let\abstract\SIsave@abstract
\let\endabstract\SIsave@endabstract
\setcounter{affil}{0}
\setcounter{collab}{0}
\def\@date{}
\patchcmd{\titleblock@produce}{\label{FirstPage}}{\label{FirstPageSI}}{}{\errmessage{FirstPage patch failed}}
\patchcmd{\frontmatter@footnotemark}{{frontmatter.#1}}{{SIfrontmatter.#1}}{}{\errmessage{fnmark patch failed}}
\patchcmd{\frontmatter@makefntext}{frontmatter.\expandafter\the}{SIfrontmatter.\expandafter\the}{}{\errmessage{fntext patch failed}}
\makeatother
\title{Prediction and characterisation of cation topology of pentanary high-entropy alloy (MoWNbTaV)S$_2$ \\---\\Supporting Information}
\author{Shubham Agarwal}
\email{agarwshu@fel.cvut.cz}
\affiliation{Department of Control Engineering, Faculty of Electrical Engineering,
Czech Technical University in Prague, Technicka 2, 16627 Prague 6,
Czech Republic}

\author{Tahir Wahab}
\affiliation{Department of Control Engineering, Faculty of Electrical Engineering,
Czech Technical University in Prague, Technicka 2, 16627 Prague 6,
Czech Republic}

\author{Jesper Byggmästar}
\affiliation{Department of Physics, University of Helsinki, P.O. Box 43, FI-00014, Finland}

\author{Antonio Cammarata}
\email{cammaant@fel.cvut.cz}
\affiliation{Department of Control Engineering, Faculty of Electrical Engineering,
Czech Technical University in Prague, Technicka 2, 16627 Prague 6,
Czech Republic}

\author{Tomas Polcar}
\affiliation{School of Engineering,
University of Southampton,
SO17 1BJ, Southampton,
United Kingdom}
%
%
\maketitle

\section{Accuracy of the GAP Forces on the training set}
\label{sec:gap_forces}
\begin{figure}[p]
  \centering
  \includegraphics[width=\textwidth]{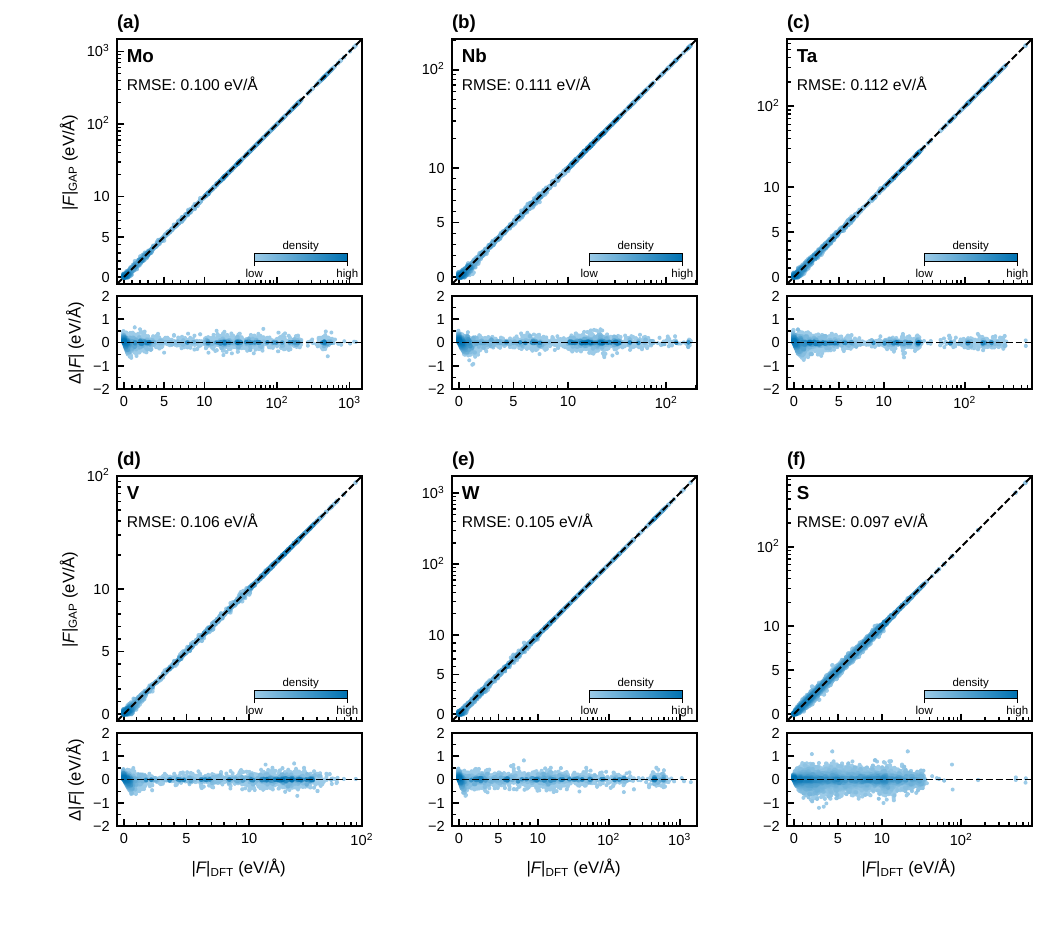}
  \caption{Species-wise comparison between DFT and the GAP for the magnitude of the atomic
  forces, $|F|$, for (a) Mo, (b) Nb, (c) Ta, (d) V, (e) W and (f) S in the training set. Each panel shows the parity plot
  (top) and the residual $\Delta|F| = |F|_{\mathrm{GAP}} - |F|_{\mathrm{DFT}}$ (bottom); the RMSE of
  $|F|$ is given in each panel. Both axes of the parity plots, and the horizontal axes of the
  residual plots, are linear up to \qty{10}{\electronvolt\per\angstrom} and logarithmic above.
  Points are coloured by local data density on a logarithmic scale, and dashed lines mark perfect
  agreement.}
  \label{fig:SI_force_mag}
\end{figure}

\begin{figure}[p]
  \centering
  \includegraphics[width=\textwidth]{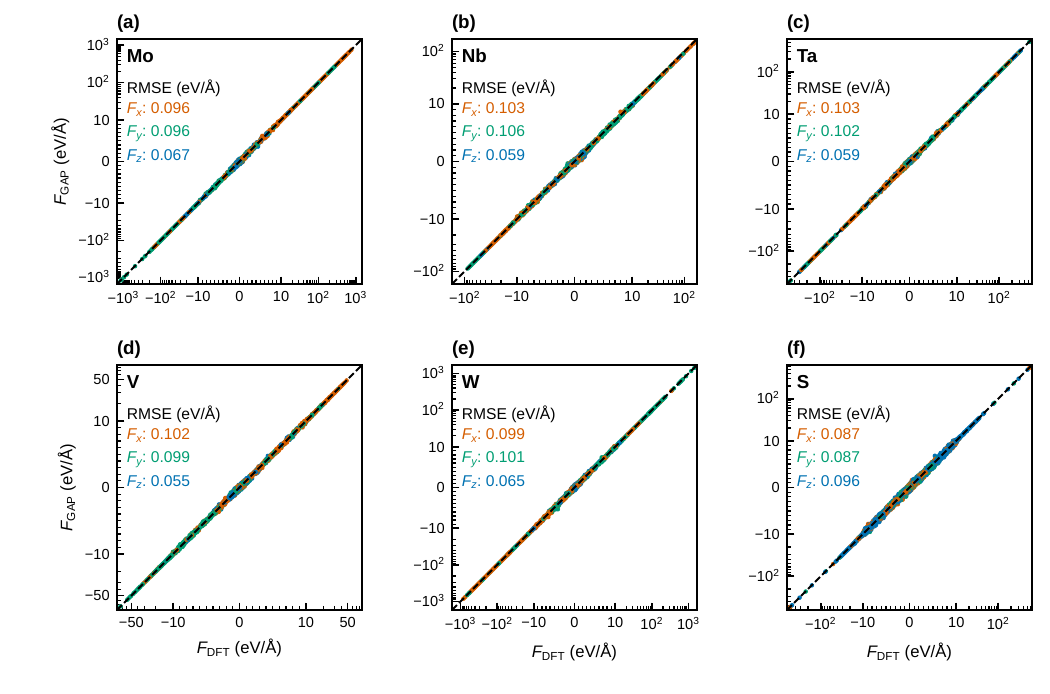}
  \caption{Species-wise comparison between DFT and the GAP for the Cartesian force components
  $F_x$ (orange), $F_y$ (green) and $F_z$ (blue) for (a) Mo, (b) Nb, (c) Ta, (d) V, (e) W and (f) S 
  in the training set, where $z$ is normal to the
  layer. The RMSE of each component is given in each panel. Both axes are linear between
  \qty{-10}{\electronvolt\per\angstrom} and \qty{10}{\electronvolt\per\angstrom} and logarithmic
  outside this range; dashed lines mark perfect agreement.}
  \label{fig:SI_force_comp}
\end{figure}

The training set contains 2164 configurations of 75 atoms (162\,300 atoms in total). 
For the forces, three RMSE definitions are in common use, and we report all of them in \autoref{tab:SI_forces}: the RMSE of the force magnitude,
$\mathrm{RMSE}_{|F|} = \langle (|\mathbf{F}_{\mathrm{GAP}}| - |\mathbf{F}_{\mathrm{DFT}}|)^2 \rangle^{1/2}$,
shown in the parity plots; the RMSE of the individual Cartesian components, pooled over $x$, $y$
and $z$, which is the value usually quoted for machine-learned potentials; and the vector RMSE,
$\langle |\mathbf{F}_{\mathrm{GAP}} - \mathbf{F}_{\mathrm{DFT}}|^2 \rangle^{1/2}$, which also
includes errors in the force direction. 
Over all atoms these are 0.100, 0.090 and
\qty{0.156}{\electronvolt\per\angstrom}, 
respectively, and the mean error of the force magnitude is
\qty{-0.002}{\electronvolt\per\angstrom}, 
i.e.\ the potential shows no systematic bias.
\autoref{fig:SI_force_mag} and \autoref{fig:SI_force_comp} show the force errors by species.
The five metals have very similar magnitude errors, \numrange{0.100}{0.112} \unit{\electronvolt\per\angstrom},
and S, which accounts for two thirds of the atoms, has the lowest
(\qty{0.097}{\electronvolt\per\angstrom}). 
%
\begin{table}[h]
  \caption{Force errors of the GAP on the training set, per species and over all atoms (in
  \unit{\electronvolt\per\angstrom}). $N$: number of atoms; $|F|$: force magnitude; comp.: RMSE of the
  Cartesian components pooled over $x$, $y$ and $z$; vec.: vector RMSE,
  $\langle|\mathbf{F}_{\mathrm{GAP}}-\mathbf{F}_{\mathrm{DFT}}|^2\rangle^{1/2}$.}
  \label{tab:SI_forces}
  \setlength{\tabcolsep}{4pt}
  \begin{ruledtabular}
  \begin{tabular}{lrcccccccc}
    & & \multicolumn{3}{c}{Force magnitude $|F|$} & \multicolumn{5}{c}{RMSE} \\
    Species & $N$ & RMSE & MAE & max error & $F_x$ & $F_y$ & $F_z$ & comp. & vec. \\
    \hline
    Mo  & 10\,820  & 0.100 & 0.068 & 0.649 & 0.096 & 0.096 & 0.067 & 0.088 & 0.152 \\
    Nb  & 10\,820  & 0.111 & 0.076 & 0.935 & 0.103 & 0.106 & 0.059 & 0.092 & 0.159 \\
    Ta  & 10\,820  & 0.112 & 0.077 & 0.715 & 0.103 & 0.102 & 0.059 & 0.091 & 0.157 \\
    V   & 10\,820  & 0.106 & 0.073 & 0.671 & 0.102 & 0.099 & 0.055 & 0.088 & 0.152 \\
    W   & 10\,820  & 0.105 & 0.072 & 0.795 & 0.099 & 0.101 & 0.065 & 0.090 & 0.156 \\
    S   & 108\,200 & 0.097 & 0.062 & 1.208 & 0.087 & 0.087 & 0.096 & 0.090 & 0.156 \\
    \hline
    All & 162\,300 & 0.100 & 0.065 & 1.208 & 0.092 & 0.092 & 0.086 & 0.090 & 0.156 \\
  \end{tabular}
  \end{ruledtabular}
\end{table}
%
%
%
\section{Free-energy analysis of the configurations}
\label{SI:gibbs}
\subsection{Absolute Gibbs free energy}
\begin{figure}[h]
  \centering
  \includegraphics[width=0.65\textwidth]{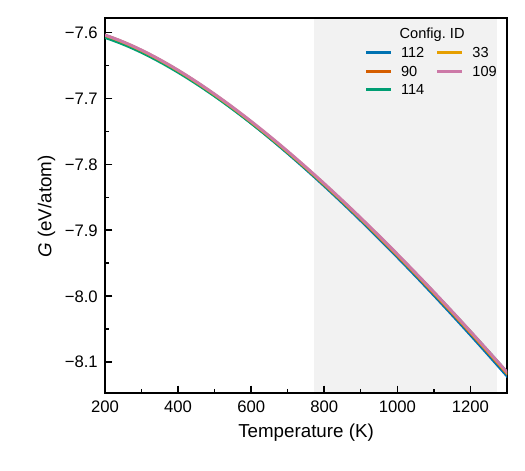}
  \caption{Gibbs free energy per atom, $G(T)$, of the five configurations shown in Fig. 4 of the
  main text (112, 90, 33, 114 and 109) from \qtyrange{200}{1300}{\kelvin}. $G$ decreases by
  $\approx$\qty{0.52}{\electronvolt\per\atom} over this range, whereas the differences between the
  configurations are at most \qty{5.7}{\milli\electronvolt\per\atom}, so the curves cannot be
  distinguished on this scale. The grey area marks the experimental synthesis window
  (\qtyrange{773}{1273}{\kelvin}; \qtyrange{500}{1000}{\degreeCelsius}).}
  \label{fig:SI_G}
\end{figure}

\autoref{fig:SI_G} shows the absolute Gibbs free energy of the five configurations of Fig. 4 of the main text. 
$G$ decreases by $\approx$\qty{0.52}{\electronvolt\per\atom} 
between 200 and \qty{1300}{\kelvin}, whereas the differences between the configurations are at most
\qty{5.7}{\milli\electronvolt\per\atom}; the relative free energy $\Delta G_c$ is therefore used in
the main text.
%
\subsection{Vibrational origin of the stability of configuration 112}
In the classical limit, the vibrational entropy difference between two configurations is set by
the ratio of their geometric-mean phonon frequencies, $\bar\nu = \exp\langle\ln\nu\rangle$, where
the average runs over all non-zero modes and there are three modes per atom:
%
\begin{equation}
  \Delta S_{\mathrm{vib}} = S_{\mathrm{vib},A} - S_{\mathrm{vib},B}
  = k_B \sum_{i} \ln\frac{\nu_{B,i}}{\nu_{A,i}} \;\; \text{(per cell)}
  \;\approx\; 3 k_B \ln\frac{\bar\nu_B}{\bar\nu_A} \;\; \text{(per atom)}.
  \label{eq:SI_dS}
\end{equation}
\autoref{tab:SI_static} and \autoref{tab:SI_vib} compare, for the four configurations that follow
configuration 112 and for configuration 258, the static-energy penalty of configuration 112, the
entropy difference obtained from the slope of $G_c - G_{112}$ over \qtyrange{780}{1260}{\kelvin}, the same
quantity computed directly from the phonon frequencies with Eq. \eqref{eq:SI_dS}, and the difference
in $\bar\nu$.
Configuration 112 has the lowest $\bar\nu$ of all 283 configurations
(\qty{6.16}{\tera\hertz}, against up to \qty{6.81}{\tera\hertz}). 
For the 256 configurations without imaginary or additional near-zero modes, the entropy differences obtained from the
frequencies agree with the fitted slopes to within 0.002 $k_B$/atom, and the intercepts of
the fits reproduce the static-energy differences to within
\qty{0.15}{\milli\electronvolt\per\atom}.
For configuration 258, the crossover temperature estimated from these quantities,
$\Delta U^{\mathrm{static}}/\Delta S_{\mathrm{vib}} \approx \qty{640}{\kelvin}$, agrees with the
crossover at $\approx$\qty{635}{\kelvin} found from the full free energies (Fig. 3 of the main text).
More generally, at \qty{1000}{\kelvin} the vibrational free energy $F^{\mathrm{vib}}$ varies by
\qty{26}{\milli\electronvolt\per\atom} across the 283 configurations, about twice the range of
$U^{\mathrm{static}}$ (\qty{14}{\milli\electronvolt\per\atom}), and the two are not aligned:
energetically less favourable configurations tend to be vibrationally softer. The vibrational
contribution therefore reorders the configurations as temperature rises.
%
\subsection{Tabulated quantities and how they were obtained}
\label{SI:tables}
\autoref{tab:SI_static}--\autoref{tab:SI_distribution} collect the quantities used in the main text
for the five configurations studied there (112, 90, 33, 114 and 109) and for configuration 258,
which has the lowest static energy and the lowest $G$ below $\approx$\qty{635}{\kelvin}. 
The five configurations are the five lowest-$G$ configurations at every temperature from
\qtyrange{1140}{1273}{\kelvin} (\qtyrange{867}{1000}{\degreeCelsius}), 
the upper part of the synthesis window 
(\qtyrange{773}{1273}{\kelvin}), 
where the set of the five lowest no longer changes. Averaged over the whole synthesis window, configuration 109 is sixth
($\langle\Delta G_c\rangle_{\mathrm{w}} = \qty{4.408}{\milli\electronvolt\per\atom}$), 
just behind configuration 99 
(\qty{4.156}{\milli\electronvolt\per\atom}); from \qty{1140}{\kelvin} 
upwards it is fifth (\autoref{tab:SI_dG}), whereas configuration 99 is sixth or seventh. 
All energies per atom are obtained by dividing the cell values by the 75 atoms of the cell. The quantities were obtained as follows.
\begin{enumerate}
  \item $U_c^{\mathrm{static}}$ and $V_c$: energy and cell volume of configuration $c$ after the
  GAP minimisation at $p = \qty{1.77}{\pascal}$; $V_c$ includes the vacuum region of the cell.
  $\Delta U^{\mathrm{static}} = U^{\mathrm{static}}_{112} - U^{\mathrm{static}}_{c}$, so that a positive
  value means that configuration 112 has the higher static energy.
  \item $G_c(T)$: Eqs. (2){--}(4) of the main text, evaluated from the $\Gamma$-point phonon
  frequencies of the 75-atom cell on a \qty{20}{\kelvin} grid from \qtyrange{200}{1300}{\kelvin};
  $\Delta G_c(T)$ follows from Eq. (5) of the main text, and
  $\Delta G^{112}_c(T) = G_c(T) - G_{112}(T)$ uses configuration 112 as a fixed reference at all
  temperatures (\autoref{tab:SI_dG}).
  \item $S_{\mathrm{vib}}(T)$: since $U_c^{\mathrm{static}}$ and $pV_c$ do not depend on temperature
  at fixed volume, $S_{\mathrm{vib}} = -\partial G_c/\partial T$, evaluated by central differences
  ($\pm\qty{20}{\kelvin}$).
  \item $\Delta S_{\mathrm{vib}}$ (fit) and $\Delta U^{\mathrm{static}}$ (fit): a linear least-squares
  fit of $(G_c - G_{112})/75$ against $T$ over the 25 grid points from \qtyrange{780}{1260}{\kelvin}.
  In the classical limit
  $G_c - G_{112} = -\Delta U^{\mathrm{static}} + T\,\Delta S_{\mathrm{vib}}$ with
  $\Delta S_{\mathrm{vib}} = S_{\mathrm{vib},112} - S_{\mathrm{vib},c}$, so the slope gives
  $\Delta S_{\mathrm{vib}}$ and the intercept gives $-\Delta U^{\mathrm{static}}$.
  \item $\bar\nu_c$ and $\Delta S_{\mathrm{vib}}$ (freq.): $\bar\nu_c = \exp\langle\ln\nu\rangle$, the
  geometric mean over the 222 non-zero $\Gamma$-point modes of configuration $c$, and
  $\Delta S_{\mathrm{vib}} = (222/75)\,k_B \ln(\bar\nu_c/\bar\nu_{112})$ per atom, i.e.\
  Eq.\eqref{eq:SI_dS} with the exact number of non-zero modes per atom.
  \item $\langle\Delta G_c\rangle_{\mathrm{w}}$ and $\langle\Delta G_c\rangle_{\mathrm{up}}$: mean of
  $\Delta G_c$ over the grid points inside the whole synthesis window (25 points,
  \qtyrange{780}{1260}{\kelvin}) and inside its upper part (7 points, \qtyrange{1140}{1260}{\kelvin}),
  respectively.
  Sorting the 283 configurations by $\langle\Delta G_c\rangle_{\mathrm{up}}$ gives the column order of
  Fig. 3 of the main text.
  \item Ranks at fixed temperature are by $G_c(T)$ among all 283 configurations (1 = lowest); the
  ranks by $U^{\mathrm{static}}$ and by $\bar\nu$ are among all 283 configurations as well
  (1 = lowest energy and lowest frequency, i.e.\ softest).
\end{enumerate}
\begin{table}[h]
  \caption{Static quantities of the selected configurations. $\Delta U^{\mathrm{static}}
  = U^{\mathrm{static}}_{112} - U^{\mathrm{static}}_{c}$ is given both from the LAMMPS energies
  (direct) and from the intercept of the linear fit of $G_c - G_{112}$ (fit). Rank: position among
  the 283 configurations by $U^{\mathrm{static}}$ (1 = lowest). $pV_c$ is evaluated with
  $p = \qty{1.7727}{\pascal}$.}
  \label{tab:SI_static}
  \begin{ruledtabular}
  \begin{tabular}{crrccccc}
    $c$ & \multicolumn{1}{c}{$U^{\mathrm{static}}_c$} & \multicolumn{1}{c}{$U^{\mathrm{static}}_c$} &
    \multicolumn{2}{c}{$\Delta U^{\mathrm{static}}$ (meV/atom)} & Rank & $V_c$ & $pV_c$ \\
        & \multicolumn{1}{c}{(eV/cell)} & \multicolumn{1}{c}{(eV/atom)} & direct & fit &
        & (\AA$^3$) & ($10^{-8}$ eV/cell) \\
    \hline
    112 & $-572.34255$ & $-7.63123$ & ---  & ---  & 192 & 5657.93 & 6.26 \\
    90  & $-572.46194$ & $-7.63283$ & 1.592 & 1.547 & 109 & 5646.50 & 6.25 \\
    33  & $-572.39000$ & $-7.63187$ & 0.633 & 0.595 & 161 & 5649.25 & 6.25 \\
    114 & $-572.70642$ & $-7.63609$ & 4.852 & 4.749 &   8 & 5649.40 & 6.25 \\
    109 & $-572.37169$ & $-7.63162$ & 0.389 & 0.338 & 175 & 5656.60 & 6.26 \\
    258 & $-572.97906$ & $-7.63972$ & 8.487 & 8.356 &   1 & 5649.43 & 6.25 \\
  \end{tabular}
  \end{ruledtabular}
\end{table}
\begin{table}[h]
  \caption{Vibrational quantities of the selected configurations. $S_{\mathrm{vib}}$ at
  \qty{1000}{\kelvin} from $-\partial G_c/\partial T$; $\Delta S_{\mathrm{vib}} =
  S_{\mathrm{vib},112} - S_{\mathrm{vib},c}$ from the slope of $G_c - G_{112}$ over
  \qtyrange{780}{1260}{\kelvin} (fit) and from the geometric-mean frequencies (freq.); $\bar\nu_c$:
  geometric-mean $\Gamma$-point frequency; Rank: position among the 283 configurations by
  $\bar\nu$ (1 = softest).}
  \label{tab:SI_vib}
  \begin{ruledtabular}
  \begin{tabular}{ccccccc}
    $c$ & $S_{\mathrm{vib}}$(\qty{1000}{\kelvin}) & \multicolumn{2}{c}{$\Delta S_{\mathrm{vib}}$ ($k_B$/atom)} &
    $\bar\nu_c$ & $\bar\nu_c/\bar\nu_{112} - 1$ & Rank \\
        & ($k_B$/atom) & fit & freq. & (THz) & (\%) & \\
    \hline
    112 & 6.585 & ---   & ---   & 6.157 & ---  &   1 \\
    90  & 6.537 & 0.049 & 0.049 & 6.260 & 1.67 &   4 \\
    33  & 6.541 & 0.044 & 0.044 & 6.250 & 1.51 &   3 \\
    114 & 6.494 & 0.091 & 0.092 & 6.351 & 3.15 &  31 \\
    109 & 6.531 & 0.054 & 0.054 & 6.271 & 1.85 &   7 \\
    258 & 6.433 & 0.153 & 0.153 & 6.484 & 5.31 & 158 \\
  \end{tabular}
  \end{ruledtabular}
\end{table}
\begin{table}[h]
  \caption{Gibbs free energy $G_c(T)$ (eV/atom) of the selected configurations at representative
  temperatures: \qty{200}{\kelvin} (lowest temperature), 780 and \qty{1260}{\kelvin} (first and last grid points
  inside the synthesis window), \qty{1000}{\kelvin}, and \qty{1300}{\kelvin} (highest temperature).}
  \label{tab:SI_G}
  \begin{ruledtabular}
  \begin{tabular}{cccccc}
    $c$ & 200 K & 780 K & 1000 K & 1260 K & 1300 K \\
    \hline
    112 & $-7.60428$ & $-7.82249$ & $-7.94074$ & $-8.09617$ & $-8.12137$ \\
    90  & $-7.60493$ & $-7.82076$ & $-7.93809$ & $-8.09243$ & $-8.11746$ \\
    33  & $-7.60406$ & $-7.82011$ & $-7.93753$ & $-8.09197$ & $-8.11701$ \\
    114 & $-7.60732$ & $-7.82111$ & $-7.93764$ & $-8.09102$ & $-8.11591$ \\
    109 & $-7.60362$ & $-7.81919$ & $-7.93643$ & $-8.09065$ & $-8.11566$ \\
    258 & $-7.60982$ & $-7.82059$ & $-7.93596$ & $-8.08797$ & $-8.11264$ \\
  \end{tabular}
  \end{ruledtabular}
\end{table}
\begin{table}[h]
  \caption{Free energy of the selected configurations relative to configuration 112,
  $\Delta G^{112}_c(T) = G_c(T) - G_{112}(T)$ (meV/atom), at fixed temperatures; its mean over the whole
  synthesis window (\qtyrange{773}{1273}{\kelvin}), $\langle\Delta G_c\rangle_{\mathrm{w}}$, and over its
  upper part (\qtyrange{1140}{1273}{\kelvin}), $\langle\Delta G_c\rangle_{\mathrm{up}}$; and the rank among
  all 283 configurations at \qty{200}{\kelvin}, at \qty{1260}{\kelvin} and by
  $\langle\Delta G_c\rangle_{\mathrm{up}}$ (1 = most stable). A negative value means that configuration $c$
  has a lower $G$ than configuration 112. Configuration 112 has the lowest $G$ of all 283 configurations
  above $\approx$\qty{635}{\kelvin}, so throughout the synthesis window $\Delta G^{112}_c$ equals
  $\Delta G_c$ of Eq. (5) of the main text.}
  \label{tab:SI_dG}
  \begin{ruledtabular}
  \begin{tabular}{cccccccccc}
    & \multicolumn{4}{c}{$\Delta G^{112}_c(T)$ (meV/atom)} & & & \multicolumn{3}{c}{Rank} \\
    $c$ & 200 K & 780 K & 1000 K & 1260 K & $\langle\Delta G_c\rangle_{\mathrm{w}}$ &
    $\langle\Delta G_c\rangle_{\mathrm{up}}$ & 200 K & 1260 K & up \\
    \hline
    112 & $0.000$  & 0.000 & 0.000 & 0.000 & 0.000 & 0.000 & 52 &  1 &  1 \\
    90  & $-0.643$ & 1.729 & 2.651 & 3.744 & 2.736 & 3.492 & 34 &  2 &  2 \\
    33  & $0.221$  & 2.375 & 3.212 & 4.203 & 3.288 & 3.974 & 61 &  3 &  3 \\
    114 & $-3.033$ & 1.379 & 3.104 & 5.148 & 3.262 & 4.676 &  3 &  4 &  4 \\
    109 & $0.666$  & 3.292 & 4.314 & 5.525 & 4.408 & 5.246 & 81 &  5 &  5 \\
    258 & $-5.538$ & 1.899 & 4.786 & 8.207 & 5.050 & 7.417 &  1 & 19 & 17 \\
  \end{tabular}
  \end{ruledtabular}
\end{table}
\begin{table}[h]
  \caption{Distribution of $\Delta G_c$ (meV/atom) over all 283 configurations at representative
  temperatures and for the window averages $\langle\Delta G_c\rangle_{\mathrm{w}}$ and
  $\langle\Delta G_c\rangle_{\mathrm{up}}$. Lowest: the configuration with $\Delta G_c = 0$; the $n$-th value is the $\Delta G_c$ of the configuration
  ranked $n$-th; $N_{<5}$ and $N_{<10}$ are the numbers of configurations within 5 and
  \qty{10}{\milli\electronvolt\per\atom} of $G_{\min}$.}
  \label{tab:SI_distribution}
  \begin{ruledtabular}
  \begin{tabular}{lcccccccc}
    & Lowest & 2nd & 5th & 25th & Median & Highest & $N_{<5}$ & $N_{<10}$ \\
    \hline
    200 K                                    & 258 & 2.187 & 3.025 & 4.707 &  7.416 & 13.456 & 37 & 245 \\
    780 K                                    & 112 & 1.379 & 2.375 & 4.660 &  9.103 & 19.017 & 29 & 174 \\
    1000 K                                   & 112 & 2.651 & 4.016 & 6.708 & 11.978 & 24.648 & 10 &  86 \\
    1260 K                                   & 112 & 3.744 & 5.525 & 9.168 & 15.266 & 31.327 &  3 &  37 \\
    1300 K                                   & 112 & 3.912 & 5.712 & 9.478 & 15.745 & 32.356 &  3 &  30 \\
    $\langle\Delta G_c\rangle_{\mathrm{w}}$  & 112 & 2.736 & 4.156 & 6.967 & 12.248 & 25.165 &  9 &  79 \\
    $\langle\Delta G_c\rangle_{\mathrm{up}}$ & 112 & 3.492 & 5.246 & 8.689 & 14.477 & 29.785 &  4 &  46 \\
  \end{tabular}
  \end{ruledtabular}
\end{table}
The numbers quoted in the main text follow directly from these tables. 
The static-energy penalty of configuration 112 (0.4, 0.6, 1.6, 4.9 and \qty{8.5}{\milli\electronvolt\per\atom} for 109, 33, 90, 114 and 258) and its entropy advantage (0.054, 0.044, 0.049, 0.091 and 0.153 $k_B$/atom) are the
``direct'' and ``fit'' columns of \autoref{tab:SI_static} and \autoref{tab:SI_vib}. 
The spread of the free energies, 13.5 and \qty{32}{\milli\electronvolt\per\atom} at 200 and \qty{1300}{\kelvin}, and the median, 7.4 and \qty{15.7}{\milli\electronvolt\per\atom}, are the ``Highest'' and ``Median'' entries of
 \autoref{tab:SI_distribution}; averaged over the whole synthesis window, the 25 most stable
configurations lie within \qty{7.0}{\milli\electronvolt\per\atom} of $G_{\min}$ and the median over all
283 configurations is \qty{12.3}{\milli\electronvolt\per\atom} (row
$\langle\Delta G_c\rangle_{\mathrm{w}}$ of \autoref{tab:SI_distribution}).
%
\subsection{Margins between the most stable configurations}
The lead of configuration 112 over the runner-up (configuration 114 below $\approx$\qty{876}{\kelvin}
and configuration 90 above) grows from \qty{1.4}{\milli\electronvolt\per\atom} at \qty{780}{\kelvin} to
\qty{3.7}{\milli\electronvolt\per\atom} at \qty{1260}{\kelvin}, which for the 75-atom cell corresponds
to 1.5--2.6 $k_BT$. The other four configurations reorder within the window, with crossings at
$\approx$876 (90/114) and \qty{1027}{\kelvin} (114/33).
Weighting the 283 configurations with Boltzmann factors $\exp(-\Delta G_c/k_BT)$ of the cell,
configuration 112 accounts for 50\% of the population at \qty{780}{\kelvin}, \qty{67}{\percent} at
\qty{1000}{\kelvin} and \qty{74}{\percent} at \qty{1260}{\kelvin}, and the five most populated configurations
account for 78{--}87\%. 
These populations refer to the sampled set and express relative preferences
among the sampled configurations, not the equilibrium state of the alloy.
The free-energy differences that decide the ranking are a few meV/atom, comparable to the GAP
energy RMSE of \qty{1.18}{\milli\electronvolt\per\atom}. 
For two configurations with independent errors, the uncertainty of their energy difference would be
$\sqrt{2}\times1.18 \approx \qty{1.7}{\milli\electronvolt\per\atom}$; errors for chemically similar
configurations are, however, likely correlated and partly cancel in $\Delta G_c$. 
The lead of configuration 112 exceeds \qty{1.7}{\milli\electronvolt\per\atom} above $\approx$\qty{830}{\kelvin},
whereas the order of configurations 90, 114, 33 and 109 lies within the uncertainty.
%
\subsection{Configurations with imaginary or near-zero modes}
Twenty-five configurations (40, 41, 59, 67, 78, 118, 125, 132, 135, 136, 147, 157, 159, 169, 179,
185, 190, 208, 230, 231, 232, 255, 262, 263 and 276) show a single imaginary mode at $\Gamma$,
between $-0.18$ and \qty{-1.37}{\tera\hertz}, and two configurations (148 and 273) show a fourth mode
with $|\nu| < \qty{0.1}{\tera\hertz}$ in addition to the three acoustic modes. 
These modes were excluded from Eqs. (2) and (3) of the main text, so the free energies of these configurations lack the contribution of one mode, which makes them appear less stable by
$\approx$\qty{1.4}{\milli\electronvolt\per\atom} at \qty{1000}{\kelvin}. 
Adding back an approximate classical contribution for the missing mode leaves configurations 112, 90 and 33 as the three most stable in the upper part of the synthesis window and does not change the order of
configurations 114 and 109; only configuration 148, which has a fourth near-zero mode, would move ahead of them 
($\langle\Delta G_c\rangle_{\mathrm{up}} \approx \qty{4.5}{\milli\electronvolt\per\atom}$,
against 4.7 and \qty{5.3}{\milli\electronvolt\per\atom} for 114 and 109). 
Because this near-zero mode makes the free energy of configuration 148 uncertain, configuration 148 was not considered further.
\section{Molecular dynamics of the selected configurations}
\label{sec:SI_md}
\begin{figure}[h]
  \centering
  \includegraphics[width=\textwidth]{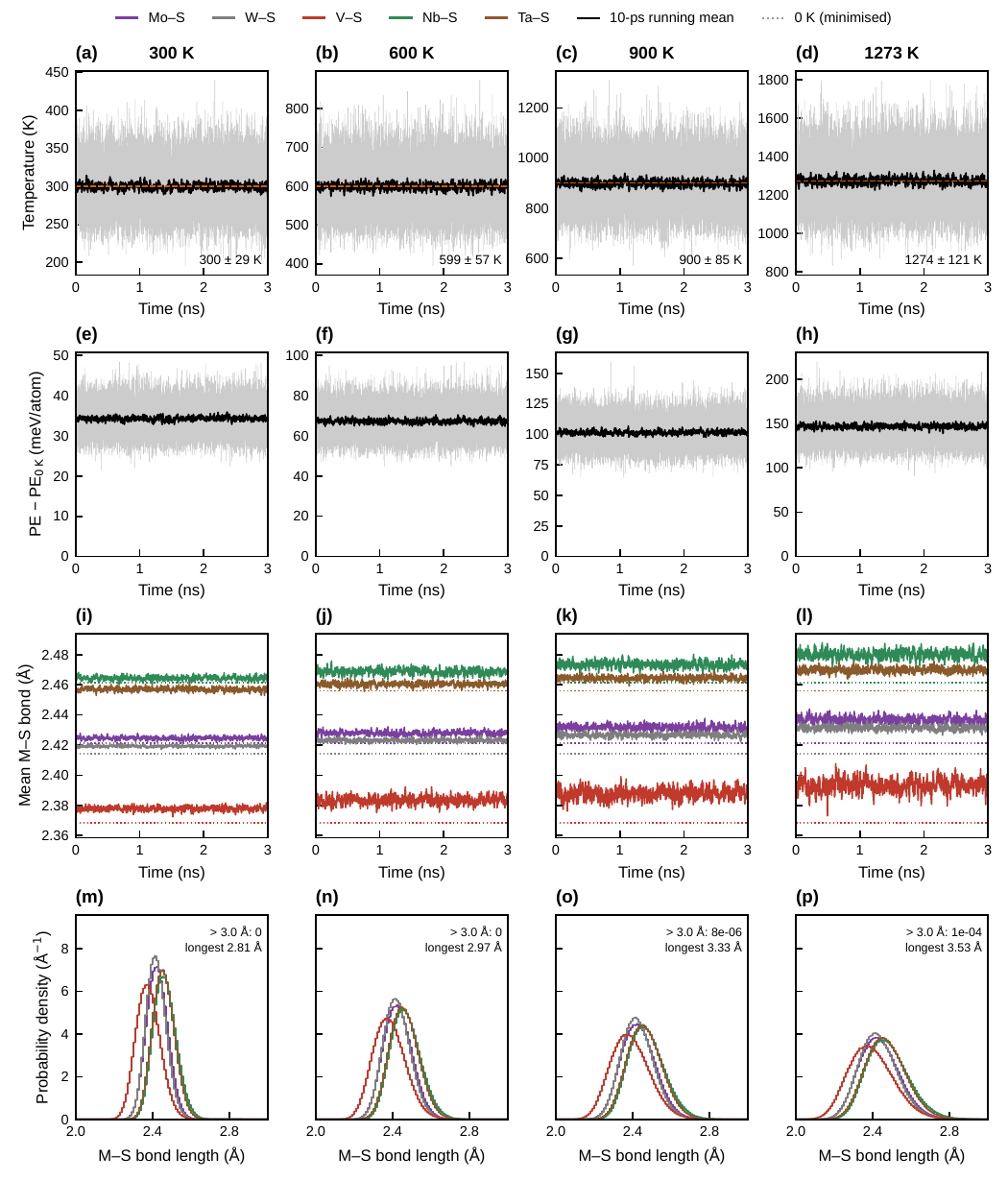}
  \caption{NVT molecular dynamics of configuration 90 with the GAP potential at 300, 600, 900 and
  \qty{1273}{K} (columns; \qty{3}{ns}, time step \qty{1}{fs}).
  (a-d) Instantaneous temperature (grey) with its \qty{10}{ps} running mean (black); the dashed line is the
  set temperature.
  (e-h) Potential energy relative to the \qty{0}{\kelvin} minimised structure.
  (i-l) Mean bond length of each M{--}S pair (\qty{10}{ps} running mean); the dotted lines are the \qty{0}{\kelvin} values.
  (m-p) Distribution of all M{--}S bond lengths over the run; the text gives the fraction of bonds longer than
  \qty{3.0}{\angstrom} and the longest bond observed.}
  \label{fig:SI_md90_overview}
\end{figure}

This section gives the MD results for configurations 90, 33, 114, and 109. 
Configuration 112 is discussed in Subsection B: Validation of ML Potential of the main text.
%
All runs use the same settings as described in the main text: 
NVT ensemble, Nos\'e--Hoover thermostat with a damping time of \qty{0.1}{ps}, time step \qty{1}{fs}, \qty{3}{ns} per
run, and the cell fixed at the GAP-relaxed lattice. 
\autoref{fig:SI_md90_overview}{--}\autoref{fig:SI_md109_checks}
have the same layout as Figure 7 and Figure 8 of the main text.
%
\autoref{fig:SI_md_bond_vs_T} compares the M{--}S bonds of the five configurations.
\autoref{fig:SI_md_rdf_T} gives the radial distribution function (RDF) of all atoms in the MD runs of
configurations 90, 33, 114 and 109, in the layout of Figure 9 of the main text.
%
and
\autoref{fig:SI_md_rdf_partial} the partial radial distribution functions of all five configurations. 
Table \ref{tab:SI_md_bonds} lists the bond lengths and Table \ref{tab:SI_md_stability} the stability indicators of all 20 runs.

The five configurations behave in the same way. 
In every run, the mean temperature is within \qty{1.2}{K} of the target, and its fluctuations agree with the canonical value $T\sqrt{2/(3N-3)}$ to within \qty{0.3}{K}. 
The drift of the potential energy is at most \qty{0.20}{meV/atom} per ns, against thermal
fluctuations of \qtyrange{3.3}{15.5}{meV/atom}. 
For each metal, the mean M{--}S bond lengths of the five configurations agree
to within \qty{0.006}{\angstrom} at every temperature; these small differences are already present at \qty{0}{\kelvin} and do not grow on heating (\autoref{fig:SI_md_bond_vs_T} and \autoref{tab:SI_md_bonds}). 
The GAP \qty{0}{\kelvin} values are \qtyrange{0.001}{0.005}{\angstrom} longer than the DFT
values, and the bonds lengthen by \qtyrange{0.013}{0.025}{\angstrom} between \num{0} and \qty{1273}{K}. 
The thermal spread of a single bond increases by a factor of \num{1.8}{--}\num{2.0} between \num{300} and \qty{1273}{K} for all metals, compared with $\sqrt{1273/300}=2.06$ for a harmonic bond. V{--}S bonds have the largest spread and W{--}S bonds the smallest.

No S atom changes side of the layer in any run, and no atom diffuses. 
The largest displacement from the initial site is below \qty{1}{\angstrom} in all runs (\autoref{tab:SI_md_stability}). 
Over all runs the closest distances are M{--}S \qty{1.94}{\angstrom}, S{--}S \qty{2.33}{\angstrom} and M{--}M \qty{2.56}{\angstrom}.
Bonds longer than \qty{3.0}{\angstrom} occur only at \qty{600}{K} and above, with a frequency of at most
$2\times10^{-4}$ at \qty{1273}{K}, and each lasts no more than \qty{0.5}{ps}.

The radial distribution functions confirm that all five configurations retain their structure up to
\qty{1273}{K}.
\autoref{tab:SI_rdf} lists the peak positions and widths of the total $g(r)$ and the M{--}S coordination number for all configurations.
In the partial RDFs (\autoref{fig:SI_md_rdf_partial}), the first M{--}S, S{--}S and M{--}M peaks lie at 
\qty{2.43}{\angstrom},
\qty{3.21}{\angstrom} and 
\qtyrange{3.21}{3.23}{\angstrom} 
at all temperatures, compared with
\qty{2.43}{\angstrom}, 
\qtyrange{3.17}{3.21}{\angstrom} and 
\qtyrange{3.17}{3.23}{\angstrom} 
in the DFT structures.
%
\begin{figure}[h]
  \centering
  \includegraphics[width=\textwidth]{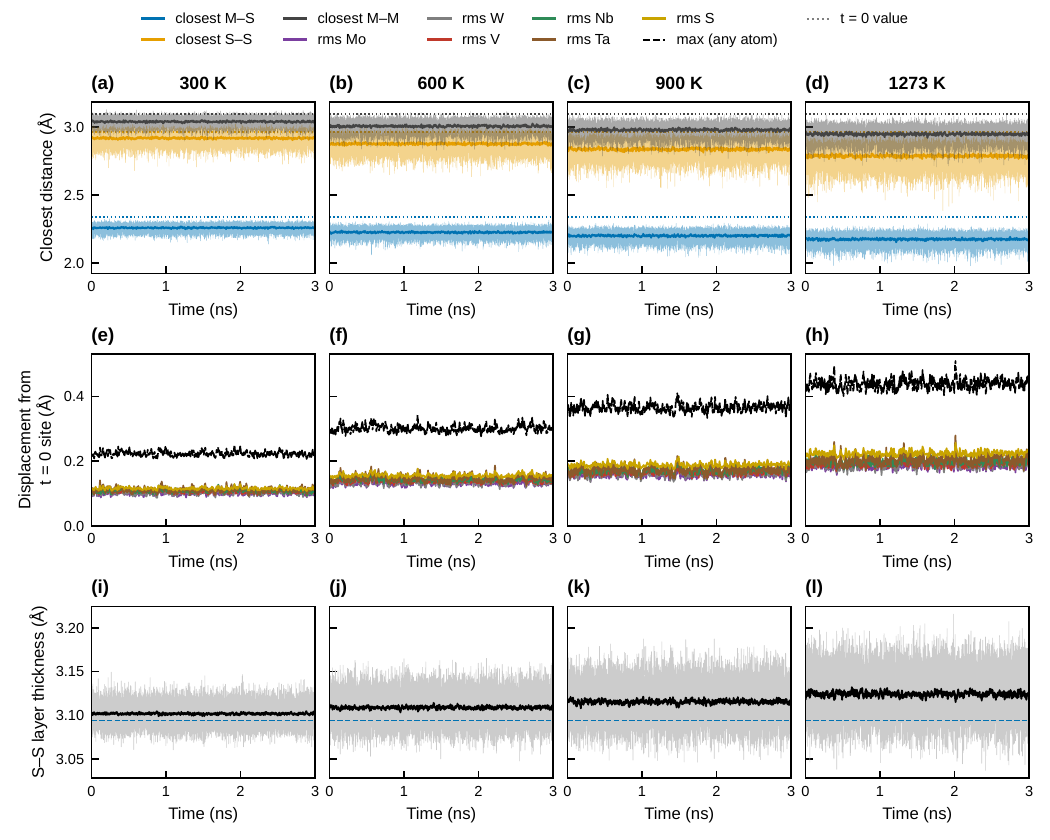}
  \caption{Structural integrity of configuration 90 during the runs of \autoref{fig:SI_md90_overview}.
  (a-d) Closest M{--}S, S{--}S and M{--}M distances in each frame (light) and their \qty{10}{ps} running means
  (dark); the dotted lines are the values at $t=0$.
  (e-h) Root-mean-square displacement of each element from its site at $t=0$ (centre-of-mass motion removed)
  and the largest displacement of any atom (dashed).
  (i-l) Layer thickness, i.e.\ the distance between the top and bottom S planes; the dashed line is the value
  at $t=0$.}
  \label{fig:SI_md90_checks}
\end{figure}
\begin{figure}[h]
  \centering
  \includegraphics[width=\textwidth]{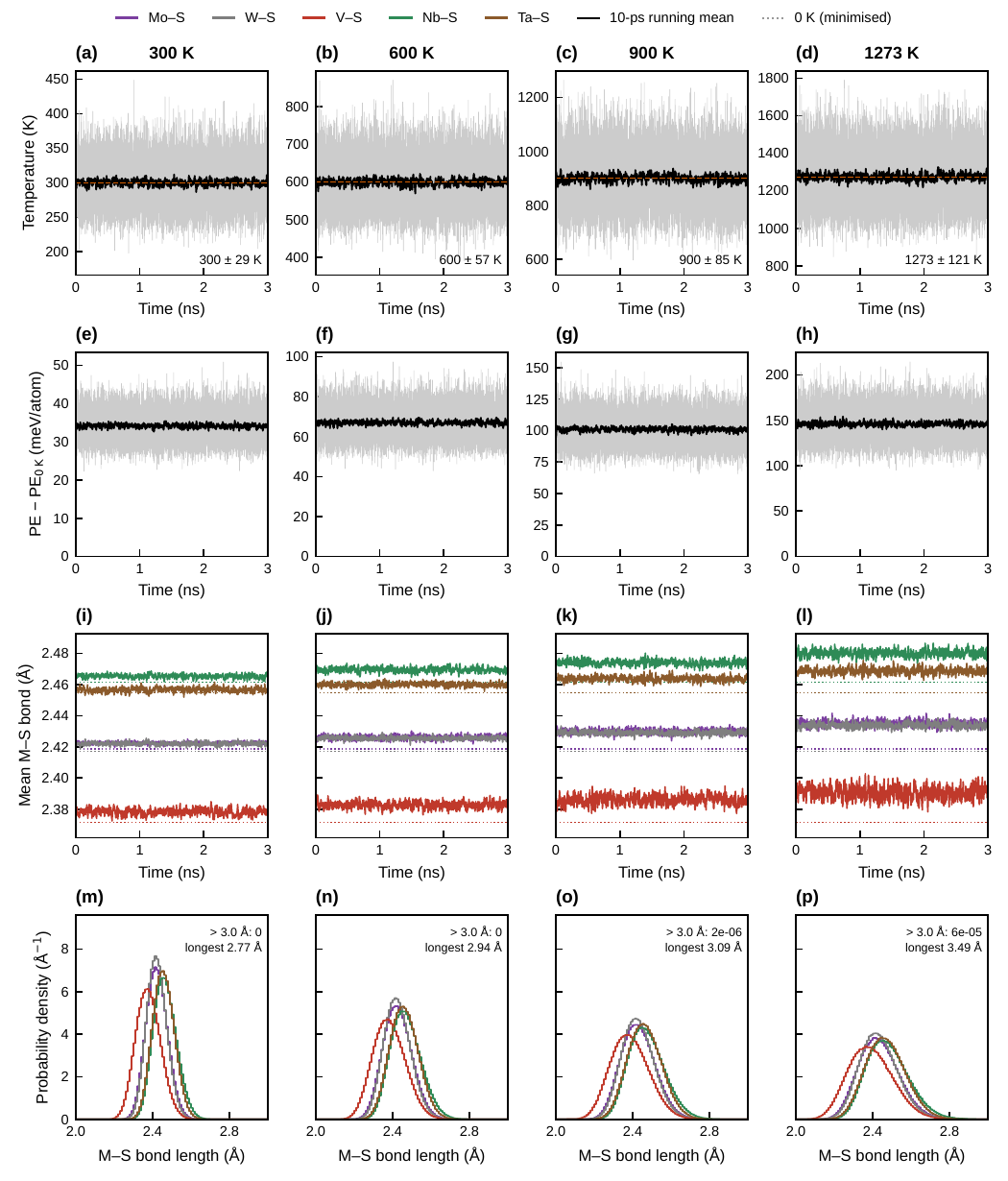}
  \caption{NVT molecular dynamics of configuration 33 with the GAP potential at 300, 600, 900 and
  \qty{1273}{K} (columns; \qty{3}{ns}, time step \qty{1}{fs}).
  (a-d) Instantaneous temperature (grey) with its \qty{10}{ps} running mean (black); the dashed line is the
  set temperature.
  (e-h) Potential energy relative to the \qty{0}{\kelvin} minimised structure.
  (i-l) Mean bond length of each M{--}S pair (\qty{10}{ps} running mean); the dotted lines are the \qty{0}{\kelvin} values.
  (m-p) Distribution of all M{--}S bond lengths over the run; the text gives the fraction of bonds longer than
  \qty{3.0}{\angstrom} and the longest bond observed.}
  \label{fig:SI_md33_overview}
\end{figure}
\begin{figure}[h]
  \centering
  \includegraphics[width=\textwidth]{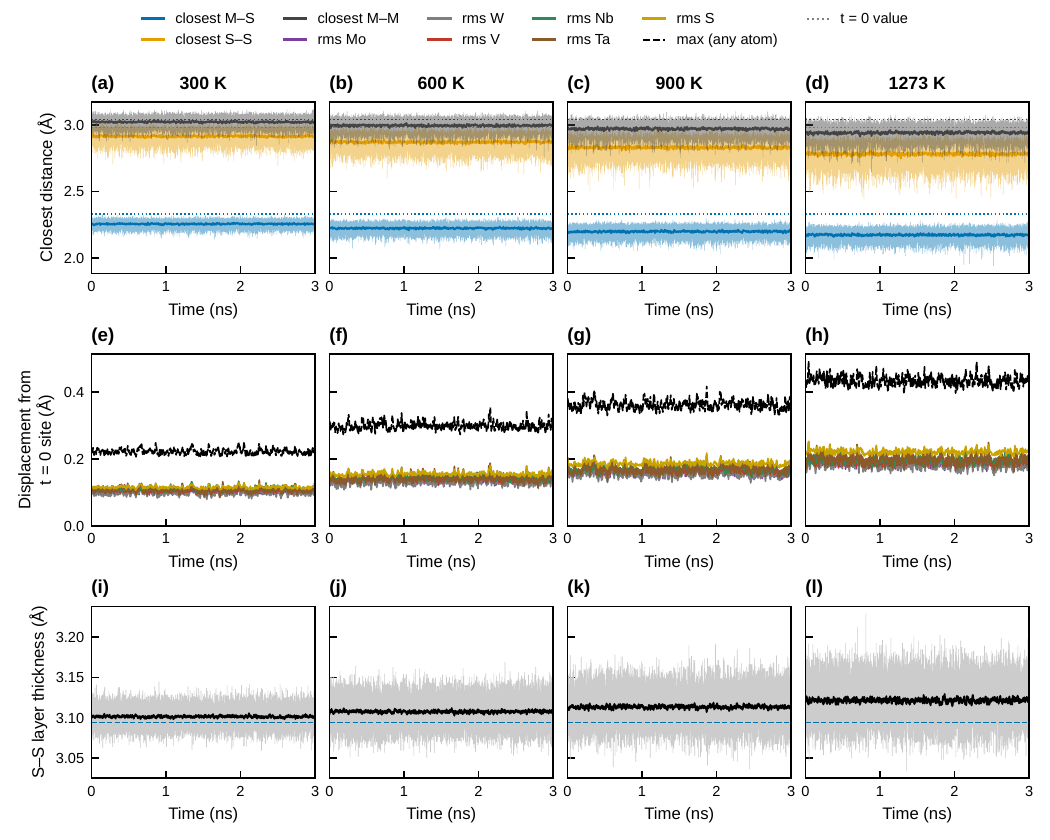}
  \caption{Structural integrity of configuration 33 during the runs of \autoref{fig:SI_md33_overview}.
  (a-d) Closest M{--}S, S{--}S and M{--}M distances in each frame (light) and their \qty{10}{ps} running means
  (dark); the dotted lines are the values at $t=0$.
  (e-h) Root-mean-square displacement of each element from its site at $t=0$ (centre-of-mass motion removed)
  and the largest displacement of any atom (dashed).
  (i-l) Layer thickness, i.e.\ the distance between the top and bottom S planes; the dashed line is the value
  at $t=0$.}
  \label{fig:SI_md33_checks}
\end{figure}
\begin{figure}[h]
  \centering
  \includegraphics[width=\textwidth]{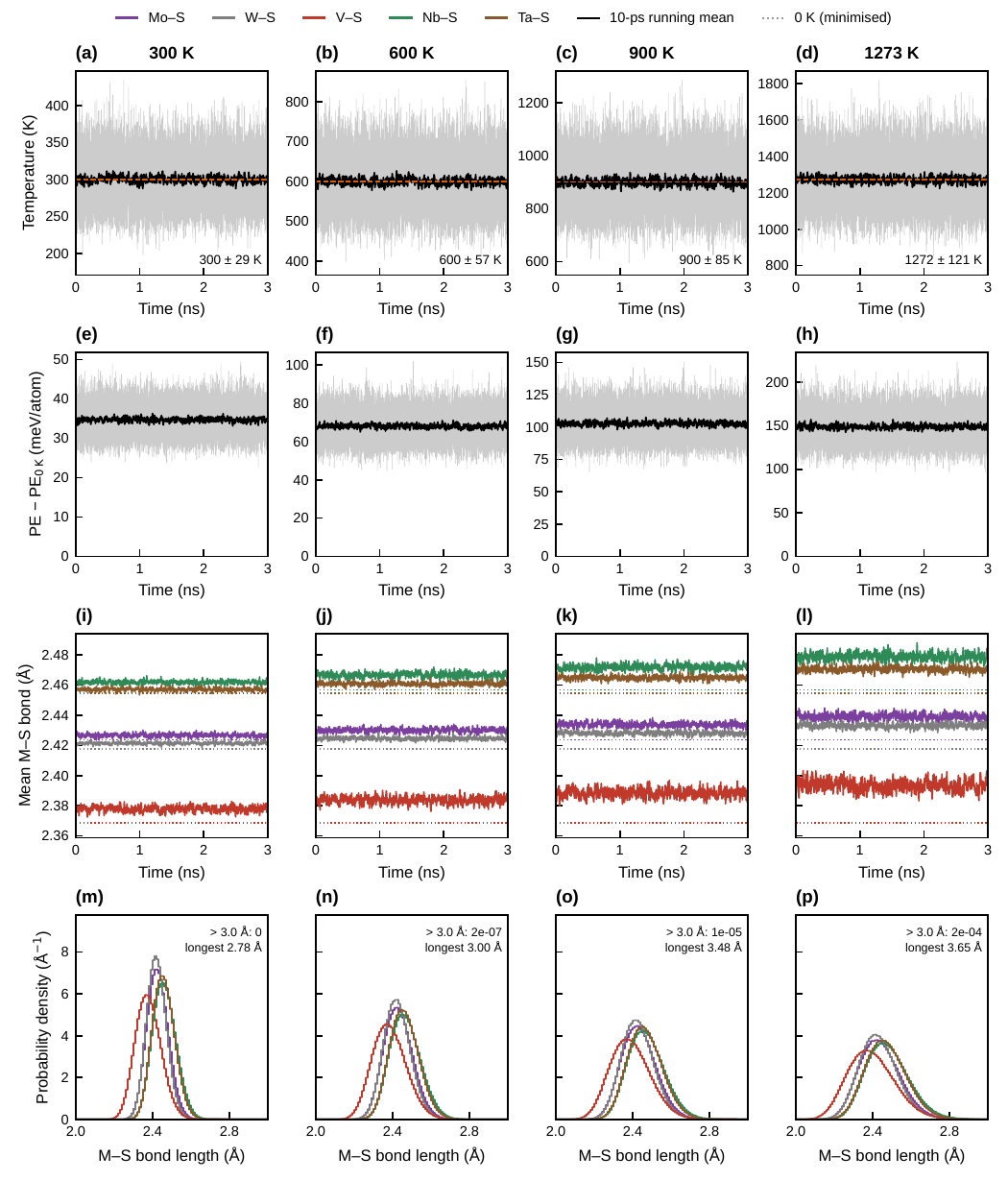}
  \caption{NVT molecular dynamics of configuration 114 with the GAP potential at 300, 600, 900 and
  \qty{1273}{K} (columns; \qty{3}{ns}, time step \qty{1}{fs}).
  (a-d) Instantaneous temperature (grey) with its \qty{10}{ps} running mean (black); the dashed line is the
  set temperature.
  (e-h) Potential energy relative to the \qty{0}{\kelvin} minimised structure.
  (i-l) Mean bond length of each M{--}S pair (\qty{10}{ps} running mean); the dotted lines are the \qty{0}{\kelvin} values.
  (m-p) Distribution of all M{--}S bond lengths over the run; the text gives the fraction of bonds longer than
  \qty{3.0}{\angstrom} and the longest bond observed.}
  \label{fig:SI_md114_overview}
\end{figure}
\begin{figure}[h]
  \centering
  \includegraphics[width=\textwidth]{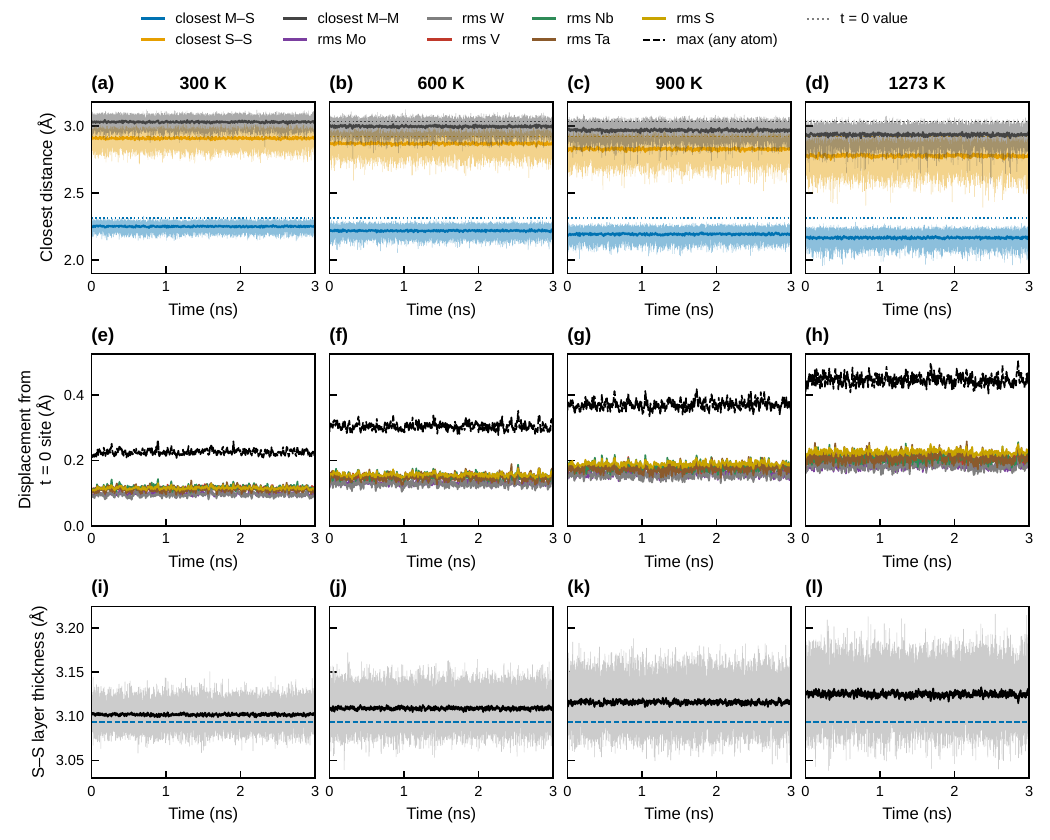}
  \caption{Structural integrity of configuration 114 during the runs of \autoref{fig:SI_md114_overview}.
  (a-d) Closest M{--}S, S{--}S and M{--}M distances in each frame (light) and their \qty{10}{ps} running means
  (dark); the dotted lines are the values at $t=0$.
  (e-h) Root-mean-square displacement of each element from its site at $t=0$ (centre-of-mass motion removed)
  and the largest displacement of any atom (dashed).
  (i-l) Layer thickness, i.e.\ the distance between the top and bottom S planes; the dashed line is the value
  at $t=0$.}
  \label{fig:SI_md114_checks}
\end{figure}
\begin{figure}[h]
  \centering
  \includegraphics[width=\textwidth]{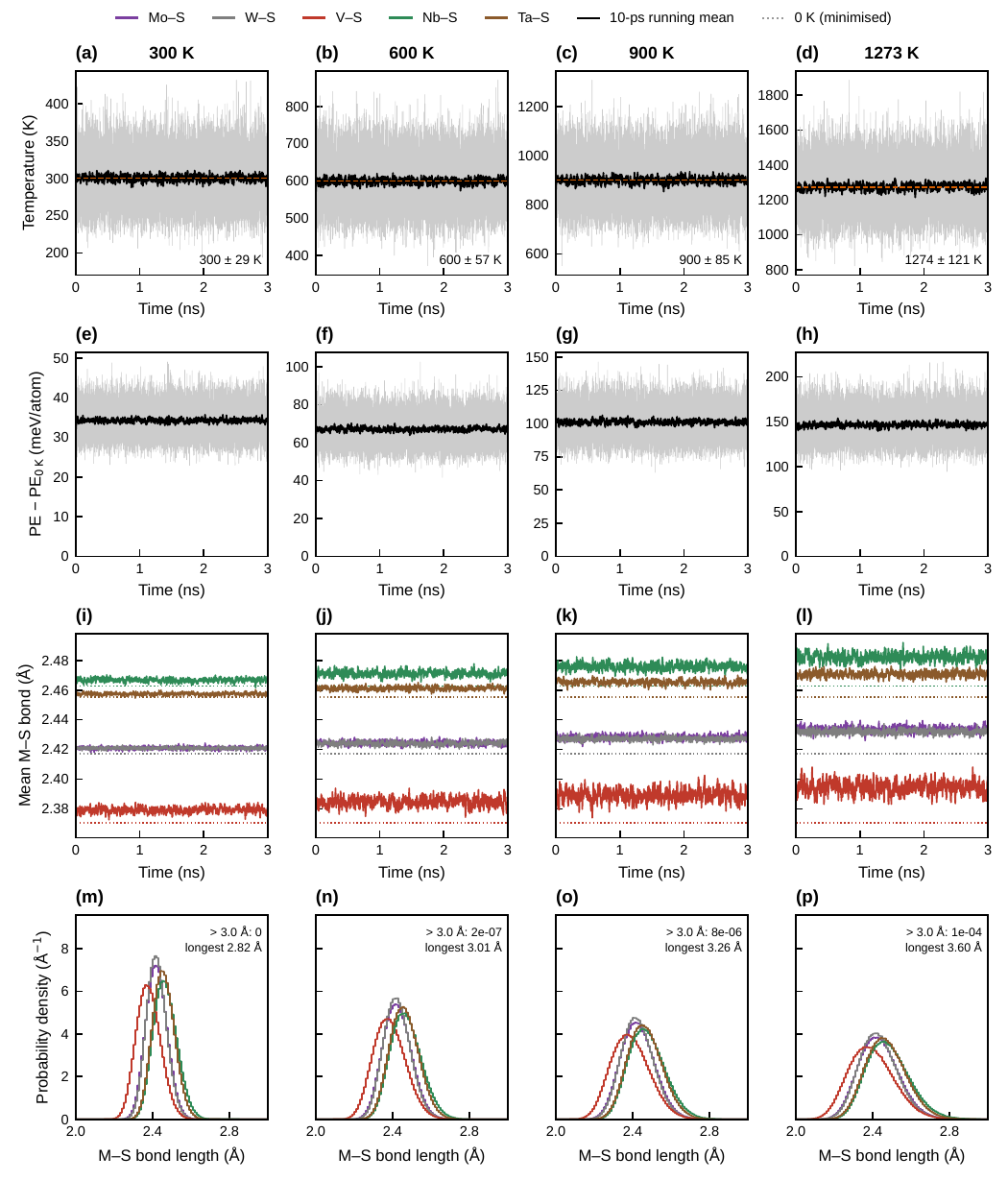}
  \caption{NVT molecular dynamics of configuration 109 with the GAP potential at 300, 600, 900 and
  \qty{1273}{K} (columns; \qty{3}{ns}, time step \qty{1}{fs}).
  (a-d) Instantaneous temperature (grey) with its \qty{10}{ps} running mean (black); the dashed line is the
  set temperature.
  (e-h) Potential energy relative to the \qty{0}{\kelvin} minimised structure.
  (i-l) Mean bond length of each M{--}S pair (\qty{10}{ps} running mean); the dotted lines are the \qty{0}{\kelvin} values.
  (m-p) Distribution of all M{--}S bond lengths over the run; the text gives the fraction of bonds longer than
  \qty{3.0}{\angstrom} and the longest bond observed.}
  \label{fig:SI_md109_overview}
\end{figure}
\begin{figure}[h]
  \centering
  \includegraphics[width=\textwidth]{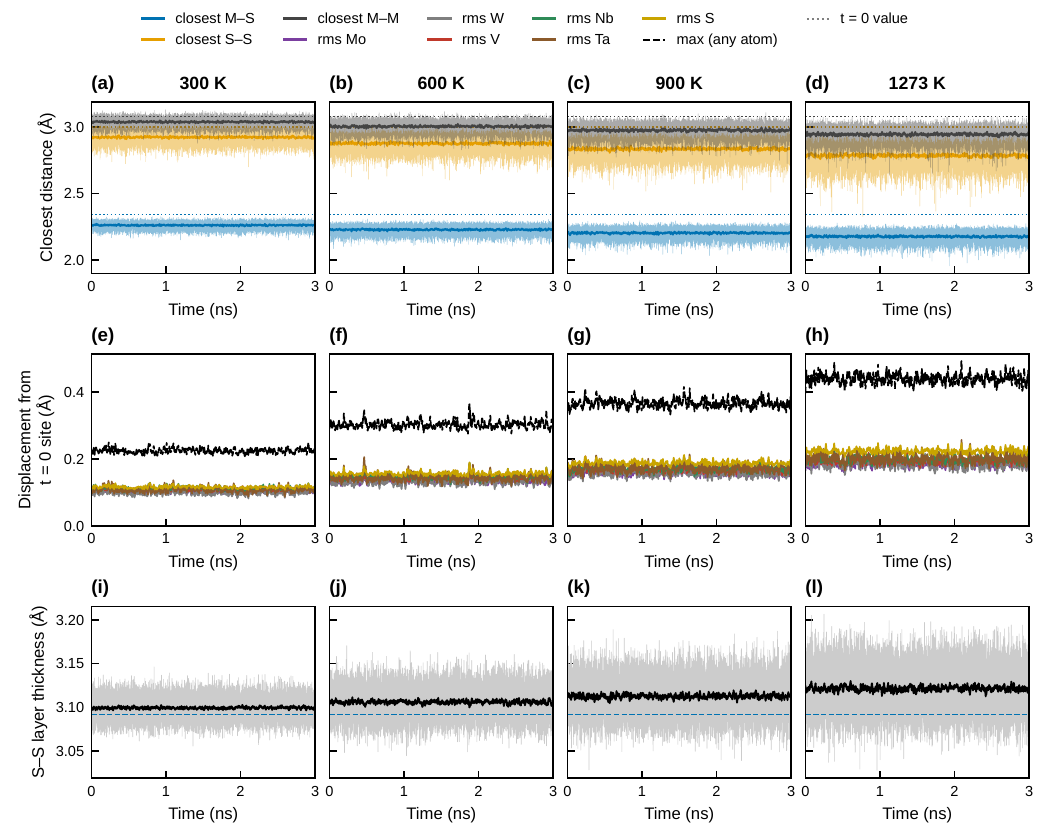}
  \caption{Structural integrity of configuration 109 during the runs of \autoref{fig:SI_md109_overview}.
  (a-d) Closest M{--}S, S{--}S and M{--}M distances in each frame (light) and their \qty{10}{ps} running means
  (dark); the dotted lines are the values at $t=0$.
  (e-h) Root-mean-square displacement of each element from its site at $t=0$ (centre-of-mass motion removed)
  and the largest displacement of any atom (dashed).
  (i-l) Layer thickness, i.e.\ the distance between the top and bottom S planes; the dashed line is the value
  at $t=0$.}
  \label{fig:SI_md109_checks}
\end{figure}
\begin{figure}[h]
  \centering
  \includegraphics[width=\textwidth]{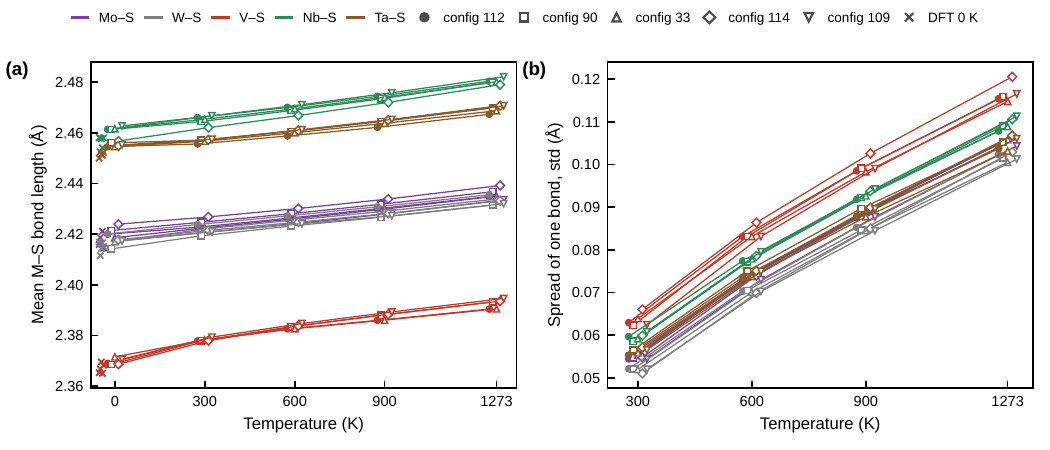}
  \caption{M--S bonds of the five configurations (marker shape; filled symbols: configuration 112) as a
  function of temperature. (a) Mean bond length of each M--S pair; the points at \qty{0}{\kelvin} are the GAP-minimised
  structures and the crosses are the DFT-relaxed structures. (b) Thermal spread (standard deviation) of a single
  M--S bond. Averages exclude the first \qty{10}{ps} of each run.}
  \label{fig:SI_md_bond_vs_T}
\end{figure}
\begin{figure}[h]
  \centering
  \includegraphics[width=\textwidth]{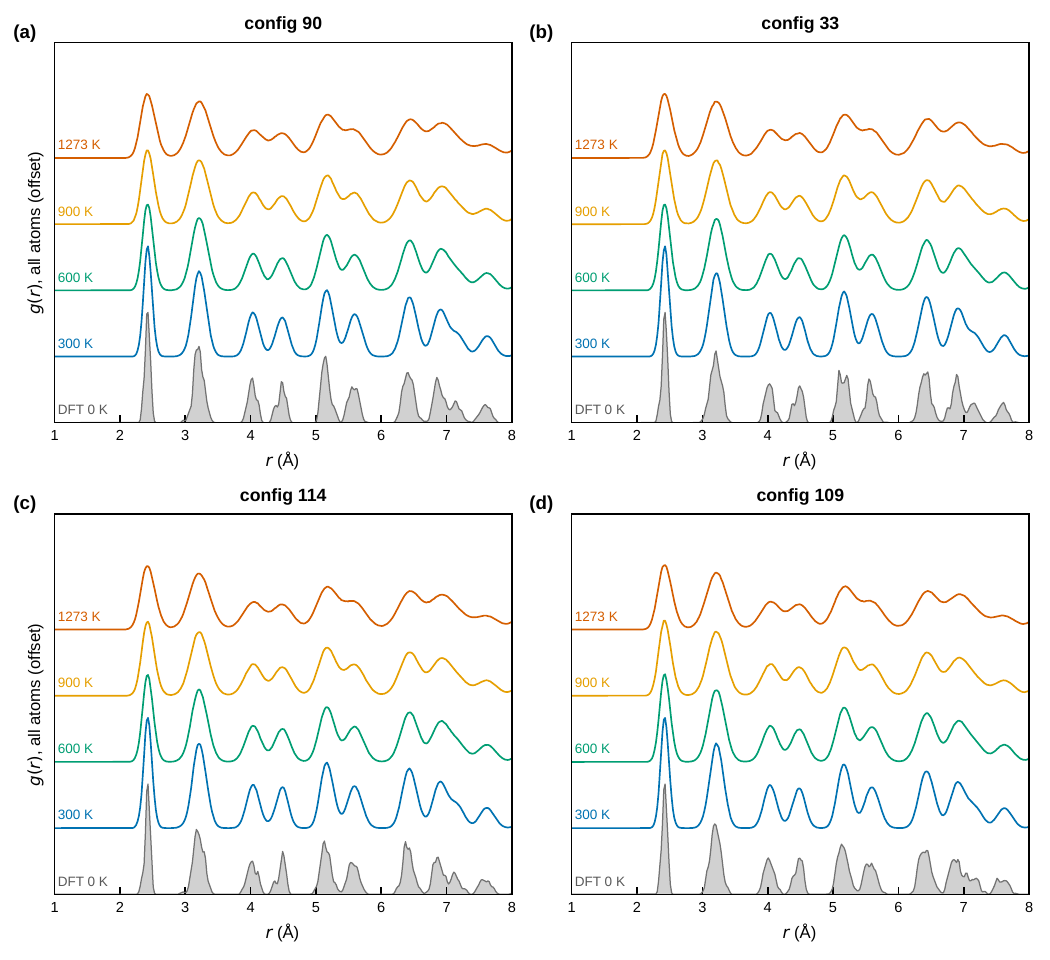}
  \caption{Radial distribution function for configurations (a) 90, (b) 33, (c) 114 and (d) 109. 
  All atom pairs averaged over each \qty{3}{ns} NVT run at 300, 600, 900 and \qty{1273}{K} 
  (curves offset vertically), with the DFT-optimised structure at the bottom (Gaussian
  broadening of \qty{0.02}{\angstrom}, scaled to the height of the \qty{300}{K} curve).} 
  \label{fig:SI_md_rdf_T}
\end{figure}
\begin{figure}[h]
  \centering
  \includegraphics[width=0.75\textwidth]{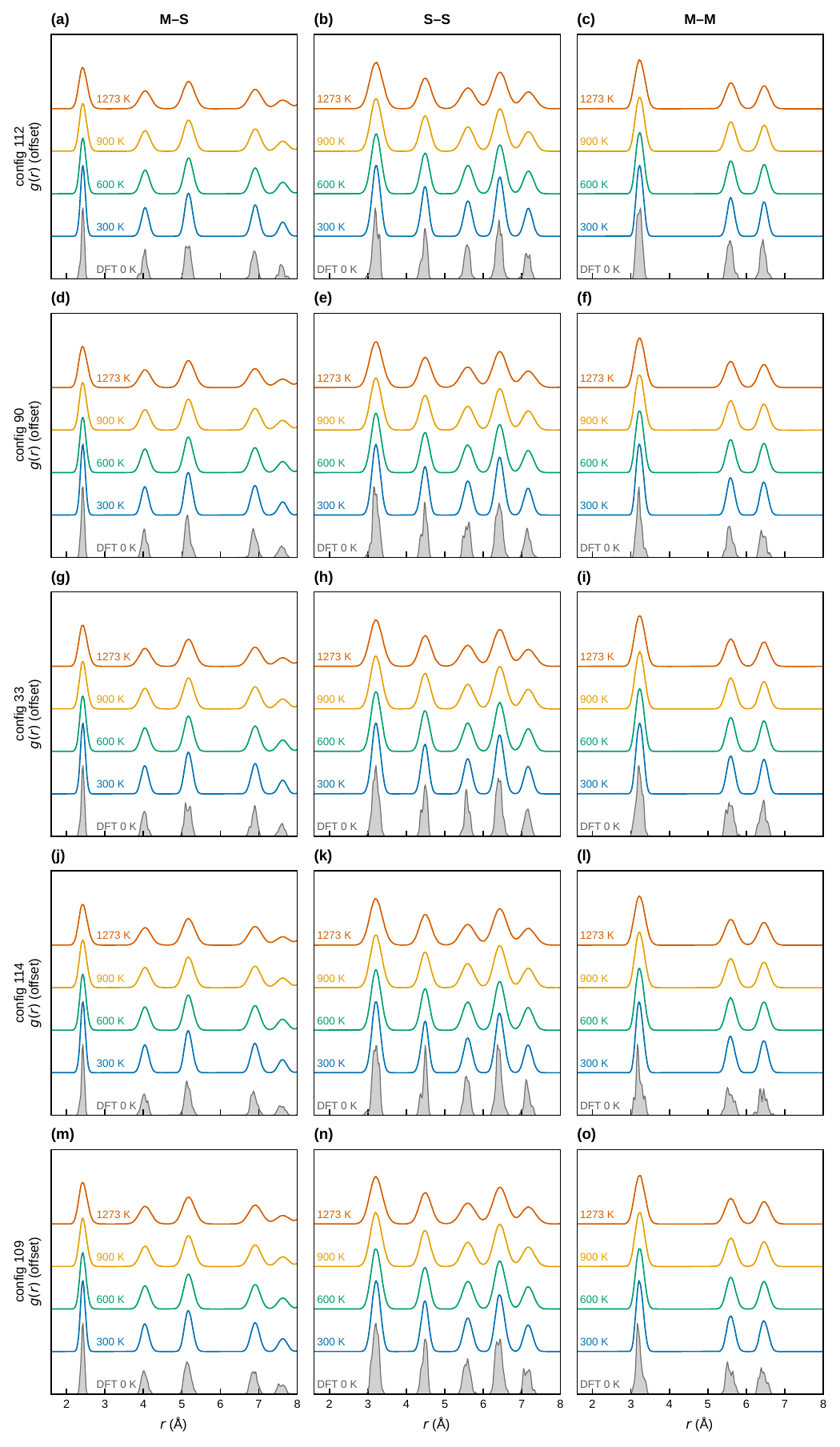}
  \caption{Partial radial distribution functions M{--}S (left), S{--}S (centre) and M{--}M (right) of
  configurations (a-c) 112, (d-f) 90, (g-i) 33, (j-l) 114 and (m-o) 109, averaged over each \qty{3}{ns} NVT run (curves
  offset vertically, temperature increasing upwards). The grey curve at the bottom is the DFT-relaxed
  structure at \qty{0}{\kelvin} (Gaussian broadening of \qty{0.02}{\angstrom}, scaled to the height of the \qty{300}{K}
  curve).}
  \label{fig:SI_md_rdf_partial}
\end{figure}
%
\begin{table}[h]
  \caption{Peaks of the radial distribution function of all atoms, $g(r)$: positions $r_1$ (first M{--}S shell), $r_2$ (M{--}M and S{--}S), $r_3$ (second M{--}S shell) and $r_4$ (S{--}S), full widths at half maximum $w_1$ and $w_2$ of the first two peaks, and number $n_{\text{M--S}}$ of S neighbours of a metal (integral of the partial M{--}S $g(r)$ up to \qty{3.3}{\angstrom}). \qty{0}{\kelvin}: DFT{-}relaxed and GAP-minimised structures, Gaussian broadening of \qty{0.02}{\angstrom}; MD: averages over the NVT runs.}
  \label{tab:SI_rdf}
  \begin{ruledtabular}
  \renewcommand{\arraystretch}{1.0} 
  \begin{tabular}{llccccccc}
    Config. & & $r_1$ & $r_2$ & $r_3$ & $r_4$ & $w_1$ & $w_2$ & $n_{\text{M--S}}$ \\
     & & (\AA) & (\AA) & (\AA) & (\AA) & (\AA) & (\AA) & \\
    \hline
    \multirow{6}{*}{112} & DFT, 0 K & 2.42 & 3.19 & 4.04 & 4.48 & 0.09 & 0.20 & 6.00 \\
     & GAP, 0 K & 2.43 & 3.21 & 4.05 & 4.49 & 0.09 & 0.22 & 6.00 \\
     & GAP, 300 K & 2.43 & 3.22 & 4.04 & 4.48 & 0.16 & 0.25 & 6.00 \\
     & GAP, 600 K & 2.42 & 3.22 & 4.05 & 4.49 & 0.20 & 0.29 & 6.00 \\
     & GAP, 900 K & 2.42 & 3.22 & 4.05 & 4.48 & 0.23 & 0.32 & 6.00 \\
     & GAP, 1273 K & 2.42 & 3.22 & 4.05 & 4.48 & 0.26 & 0.36 & 6.00 \\
    \hline
    \multirow{6}{*}{90} & DFT, 0 K & 2.42 & 3.21 & 4.02 & 4.48 & 0.10 & 0.18 & 6.00 \\
     & GAP, 0 K & 2.43 & 3.17 & 4.02 & 4.49 & 0.10 & 0.19 & 6.00 \\
     & GAP, 300 K & 2.42 & 3.21 & 4.04 & 4.48 & 0.16 & 0.25 & 6.00 \\
     & GAP, 600 K & 2.42 & 3.21 & 4.04 & 4.49 & 0.20 & 0.29 & 6.00 \\
     & GAP, 900 K & 2.42 & 3.21 & 4.04 & 4.48 & 0.23 & 0.33 & 6.00 \\
     & GAP, 1273 K & 2.42 & 3.22 & 4.05 & 4.48 & 0.26 & 0.37 & 6.00 \\
    \hline
    \multirow{6}{*}{33} & DFT, 0 K & 2.43 & 3.21 & 4.03 & 4.49 & 0.09 & 0.20 & 6.00 \\
     & GAP, 0 K & 2.43 & 3.19 & 4.03 & 4.48 & 0.09 & 0.22 & 6.00 \\
     & GAP, 300 K & 2.43 & 3.22 & 4.04 & 4.48 & 0.16 & 0.26 & 6.00 \\
     & GAP, 600 K & 2.42 & 3.22 & 4.04 & 4.48 & 0.20 & 0.30 & 6.00 \\
     & GAP, 900 K & 2.42 & 3.21 & 4.04 & 4.48 & 0.23 & 0.33 & 6.00 \\
     & GAP, 1273 K & 2.42 & 3.22 & 4.05 & 4.49 & 0.26 & 0.37 & 6.00 \\
    \hline
    \multirow{6}{*}{114} & DFT, 0 K & 2.42 & 3.18 & 4.02 & 4.49 & 0.08 & 0.20 & 6.00 \\
     & GAP, 0 K & 2.43 & 3.19 & 4.03 & 4.50 & 0.07 & 0.24 & 6.00 \\
     & GAP, 300 K & 2.43 & 3.21 & 4.04 & 4.49 & 0.16 & 0.25 & 6.00 \\
     & GAP, 600 K & 2.42 & 3.21 & 4.04 & 4.49 & 0.20 & 0.29 & 6.00 \\
     & GAP, 900 K & 2.42 & 3.21 & 4.04 & 4.48 & 0.23 & 0.33 & 6.00 \\
     & GAP, 1273 K & 2.42 & 3.21 & 4.05 & 4.48 & 0.26 & 0.37 & 6.00 \\
    \hline
    \multirow{6}{*}{109} & DFT, 0 K & 2.42 & 3.19 & 4.01 & 4.49 & 0.10 & 0.19 & 6.00 \\
     & GAP, 0 K & 2.43 & 3.22 & 4.02 & 4.50 & 0.10 & 0.19 & 6.00 \\
     & GAP, 300 K & 2.42 & 3.22 & 4.04 & 4.48 & 0.16 & 0.25 & 6.00 \\
     & GAP, 600 K & 2.42 & 3.22 & 4.04 & 4.48 & 0.20 & 0.30 & 6.00 \\
     & GAP, 900 K & 2.42 & 3.21 & 4.05 & 4.48 & 0.23 & 0.33 & 6.00 \\
     & GAP, 1273 K & 2.42 & 3.22 & 4.05 & 4.48 & 0.26 & 0.37 & 6.00 \\
  \end{tabular}
  \end{ruledtabular}
\end{table}
\begin{table}[h]
  \caption{M--S bond lengths (\AA) of the five configurations: \qty{0}{\kelvin} values of the DFT- and GAP-relaxed
  structures, and mean $\pm$ standard deviation over all 30 bonds of each metal and all frames of the NVT
  runs (3 ns, after the first 10 ps). The standard deviation is the thermal spread of a single bond.}
  \label{tab:SI_md_bonds}
  \begin{ruledtabular}
  \renewcommand{\arraystretch}{1.0} 
  \begin{tabular}{llccccc}
    Config. & $T$ & Mo--S & W--S & V--S & Nb--S & Ta--S \\
    \hline
    \multirow{6}{*}{112} & 0 K (DFT) & 2.416 & 2.417 & 2.365 & 2.458 & 2.450 \\
     & 0 K (GAP) & 2.420 & 2.420 & 2.369 & 2.461 & 2.454 \\
     & 300 K & 2.423 $\pm$ 0.054 & 2.424 $\pm$ 0.052 & 2.378 $\pm$ 0.063 & 2.466 $\pm$ 0.060 & 2.456 $\pm$ 0.055 \\
     & 600 K & 2.426 $\pm$ 0.073 & 2.427 $\pm$ 0.070 & 2.383 $\pm$ 0.083 & 2.470 $\pm$ 0.077 & 2.459 $\pm$ 0.074 \\
     & 900 K & 2.430 $\pm$ 0.088 & 2.431 $\pm$ 0.085 & 2.386 $\pm$ 0.099 & 2.474 $\pm$ 0.092 & 2.462 $\pm$ 0.088 \\
     & 1273 K & 2.435 $\pm$ 0.104 & 2.436 $\pm$ 0.102 & 2.390 $\pm$ 0.115 & 2.480 $\pm$ 0.108 & 2.467 $\pm$ 0.104 \\
    \hline
    \multirow{6}{*}{90} & 0 K (DFT) & 2.419 & 2.412 & 2.367 & 2.458 & 2.452 \\
     & 0 K (GAP) & 2.421 & 2.414 & 2.369 & 2.461 & 2.456 \\
     & 300 K & 2.425 $\pm$ 0.055 & 2.419 $\pm$ 0.052 & 2.378 $\pm$ 0.062 & 2.464 $\pm$ 0.059 & 2.457 $\pm$ 0.057 \\
     & 600 K & 2.428 $\pm$ 0.074 & 2.423 $\pm$ 0.070 & 2.383 $\pm$ 0.083 & 2.469 $\pm$ 0.077 & 2.460 $\pm$ 0.075 \\
     & 900 K & 2.432 $\pm$ 0.089 & 2.427 $\pm$ 0.085 & 2.388 $\pm$ 0.099 & 2.473 $\pm$ 0.092 & 2.464 $\pm$ 0.090 \\
     & 1273 K & 2.437 $\pm$ 0.105 & 2.431 $\pm$ 0.102 & 2.393 $\pm$ 0.116 & 2.480 $\pm$ 0.109 & 2.470 $\pm$ 0.105 \\
    \hline
    \multirow{6}{*}{33} & 0 K (DFT) & 2.416 & 2.414 & 2.369 & 2.458 & 2.452 \\
     & 0 K (GAP) & 2.419 & 2.417 & 2.372 & 2.462 & 2.455 \\
     & 300 K & 2.422 $\pm$ 0.055 & 2.422 $\pm$ 0.052 & 2.378 $\pm$ 0.064 & 2.465 $\pm$ 0.059 & 2.457 $\pm$ 0.056 \\
     & 600 K & 2.426 $\pm$ 0.074 & 2.426 $\pm$ 0.070 & 2.383 $\pm$ 0.083 & 2.470 $\pm$ 0.078 & 2.460 $\pm$ 0.074 \\
     & 900 K & 2.430 $\pm$ 0.089 & 2.429 $\pm$ 0.085 & 2.386 $\pm$ 0.098 & 2.474 $\pm$ 0.093 & 2.464 $\pm$ 0.088 \\
     & 1273 K & 2.435 $\pm$ 0.106 & 2.434 $\pm$ 0.100 & 2.390 $\pm$ 0.115 & 2.480 $\pm$ 0.109 & 2.469 $\pm$ 0.103 \\
    \hline
    \multirow{6}{*}{114} & 0 K (DFT) & 2.421 & 2.417 & 2.365 & 2.454 & 2.451 \\
     & 0 K (GAP) & 2.424 & 2.418 & 2.369 & 2.457 & 2.455 \\
     & 300 K & 2.427 $\pm$ 0.055 & 2.421 $\pm$ 0.051 & 2.378 $\pm$ 0.066 & 2.462 $\pm$ 0.060 & 2.457 $\pm$ 0.057 \\
     & 600 K & 2.430 $\pm$ 0.074 & 2.425 $\pm$ 0.070 & 2.384 $\pm$ 0.086 & 2.467 $\pm$ 0.079 & 2.461 $\pm$ 0.075 \\
     & 900 K & 2.434 $\pm$ 0.089 & 2.428 $\pm$ 0.085 & 2.388 $\pm$ 0.103 & 2.472 $\pm$ 0.094 & 2.465 $\pm$ 0.090 \\
     & 1273 K & 2.439 $\pm$ 0.107 & 2.433 $\pm$ 0.103 & 2.394 $\pm$ 0.121 & 2.479 $\pm$ 0.111 & 2.471 $\pm$ 0.107 \\
    \hline
    \multirow{6}{*}{109} & 0 K (DFT) & 2.414 & 2.415 & 2.368 & 2.458 & 2.452 \\
     & 0 K (GAP) & 2.417 & 2.417 & 2.371 & 2.463 & 2.455 \\
     & 300 K & 2.421 $\pm$ 0.054 & 2.421 $\pm$ 0.052 & 2.379 $\pm$ 0.062 & 2.467 $\pm$ 0.061 & 2.457 $\pm$ 0.057 \\
     & 600 K & 2.424 $\pm$ 0.073 & 2.424 $\pm$ 0.070 & 2.385 $\pm$ 0.083 & 2.471 $\pm$ 0.080 & 2.461 $\pm$ 0.075 \\
     & 900 K & 2.428 $\pm$ 0.088 & 2.427 $\pm$ 0.084 & 2.389 $\pm$ 0.099 & 2.476 $\pm$ 0.094 & 2.465 $\pm$ 0.090 \\
     & 1273 K & 2.434 $\pm$ 0.104 & 2.432 $\pm$ 0.101 & 2.395 $\pm$ 0.117 & 2.482 $\pm$ 0.111 & 2.471 $\pm$ 0.106 \\
  \end{tabular}
  \end{ruledtabular}
\end{table}
\begin{table}[h]
  \caption{Stability of the NVT runs (3 ns, averages after the first 10 ps). $T$: mean $\pm$ standard
  deviation of the instantaneous temperature; drift: slope of the potential energy; $d_\mathrm{min}$: closest
  M--S, S--S and M--M distance over the whole run; $u_\mathrm{rms}$: root-mean-square displacement of the
  metal and S atoms from their initial sites; $u_\mathrm{max}$: largest displacement of any atom; $h$: S--S layer
  thickness. In all runs, every metal keeps its six initial S neighbours and no S atom changes side of the layer.}
  \label{tab:SI_md_stability}
  \begin{ruledtabular}
  \renewcommand{\arraystretch}{1.0}
  \begin{tabular}{llcccccc}
    Config. & $T_\mathrm{set}$ & $T$ & drift & $d_\mathrm{min}$ M--S/S--S/M--M & $u_\mathrm{rms}$ M/S & $u_\mathrm{max}$ & $h$ \\
     & (K) & (K) & (meV/atom/ns) & (\AA) & (\AA) & (\AA) & (\AA) \\
    \hline
    \multirow{4}{*}{112} & 300 & 300 $\pm$ 29 & $+$0.009 & 2.11/2.68/2.90 & 0.110/0.116 & 0.40 & 3.099 $\pm$ 0.011 \\
     & 600 & 600 $\pm$ 57 & $+$0.033 & 2.06/2.60/2.83 & 0.143/0.156 & 0.54 & 3.105 $\pm$ 0.015 \\
     & 900 & 899 $\pm$ 85 & $-$0.041 & 2.03/2.43/2.56 & 0.167/0.187 & 0.69 & 3.111 $\pm$ 0.018 \\
     & 1273 & 1273 $\pm$ 121 & $-$0.017 & 1.96/2.41/2.73 & 0.194/0.221 & 0.94 & 3.119 $\pm$ 0.022 \\
    \hline
    \multirow{4}{*}{90} & 300 & 300 $\pm$ 29 & $+$0.014 & 2.14/2.68/2.89 & 0.108/0.116 & 0.41 & 3.102 $\pm$ 0.011 \\
     & 600 & 599 $\pm$ 57 & $-$0.004 & 2.06/2.63/2.83 & 0.142/0.157 & 0.56 & 3.109 $\pm$ 0.015 \\
     & 900 & 900 $\pm$ 85 & $-$0.007 & 2.04/2.55/2.77 & 0.169/0.189 & 0.76 & 3.116 $\pm$ 0.018 \\
     & 1273 & 1274 $\pm$ 121 & $+$0.022 & 1.98/2.38/2.70 & 0.198/0.225 & 0.99 & 3.124 $\pm$ 0.022 \\
    \hline
    \multirow{4}{*}{33} & 300 & 300 $\pm$ 29 & $-$0.040 & 2.14/2.69/2.85 & 0.106/0.116 & 0.38 & 3.102 $\pm$ 0.011 \\
     & 600 & 600 $\pm$ 57 & $-$0.010 & 2.07/2.60/2.75 & 0.140/0.156 & 0.61 & 3.108 $\pm$ 0.015 \\
     & 900 & 900 $\pm$ 85 & $-$0.092 & 2.03/2.50/2.76 & 0.167/0.187 & 0.75 & 3.114 $\pm$ 0.018 \\
     & 1273 & 1273 $\pm$ 121 & $+$0.048 & 1.94/2.43/2.65 & 0.195/0.222 & 0.89 & 3.122 $\pm$ 0.022 \\
    \hline
    \multirow{4}{*}{114} & 300 & 300 $\pm$ 29 & $-$0.028 & 2.14/2.72/2.85 & 0.110/0.116 & 0.39 & 3.102 $\pm$ 0.011 \\
     & 600 & 600 $\pm$ 57 & $-$0.084 & 2.06/2.60/2.75 & 0.145/0.157 & 0.57 & 3.109 $\pm$ 0.015 \\
     & 900 & 900 $\pm$ 85 & $-$0.079 & 2.01/2.52/2.72 & 0.173/0.190 & 0.70 & 3.116 $\pm$ 0.019 \\
     & 1273 & 1272 $\pm$ 121 & $-$0.157 & 1.96/2.39/2.57 & 0.202/0.227 & 0.99 & 3.125 $\pm$ 0.022 \\
    \hline
    \multirow{4}{*}{109} & 300 & 300 $\pm$ 29 & $-$0.001 & 2.15/2.70/2.88 & 0.109/0.117 & 0.40 & 3.099 $\pm$ 0.011 \\
     & 600 & 600 $\pm$ 57 & $+$0.009 & 2.08/2.60/2.83 & 0.143/0.158 & 0.55 & 3.106 $\pm$ 0.015 \\
     & 900 & 900 $\pm$ 85 & $+$0.010 & 2.03/2.51/2.72 & 0.168/0.189 & 0.75 & 3.113 $\pm$ 0.019 \\
     & 1273 & 1274 $\pm$ 121 & $+$0.202 & 1.95/2.33/2.64 & 0.197/0.225 & 0.98 & 3.122 $\pm$ 0.022 \\
  \end{tabular}
  \end{ruledtabular}
\end{table}
%
\section{Raman and IR spectra of the selected configurations}
\label{SI:vib}
\begin{figure}[h]
  \centering
  \includegraphics[width=0.9\textwidth]{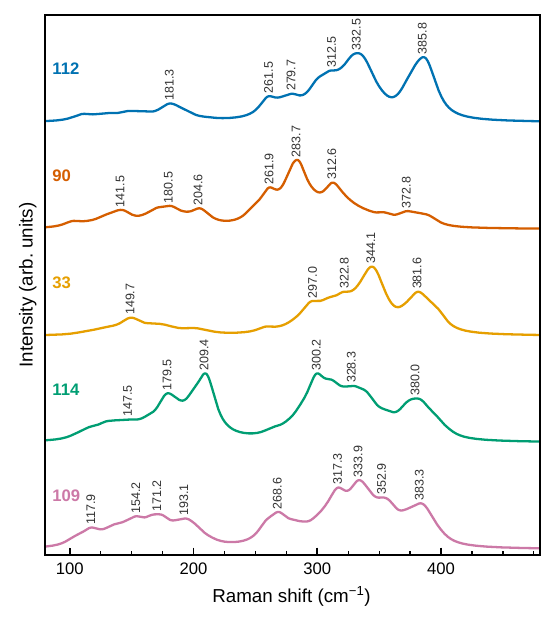}
  \caption{Simulated Raman spectra of the five HEA configurations of the main text, each normalised to
  its own maximum and offset vertically. Numbers give the peak positions in \unit{\per\centi\metre}
  (peaks higher than 25\% of the maximum).}
  \label{fig:SI_raman_hea}
\end{figure}

\begin{figure}[p]
  \centering
  \includegraphics[width=\textwidth]{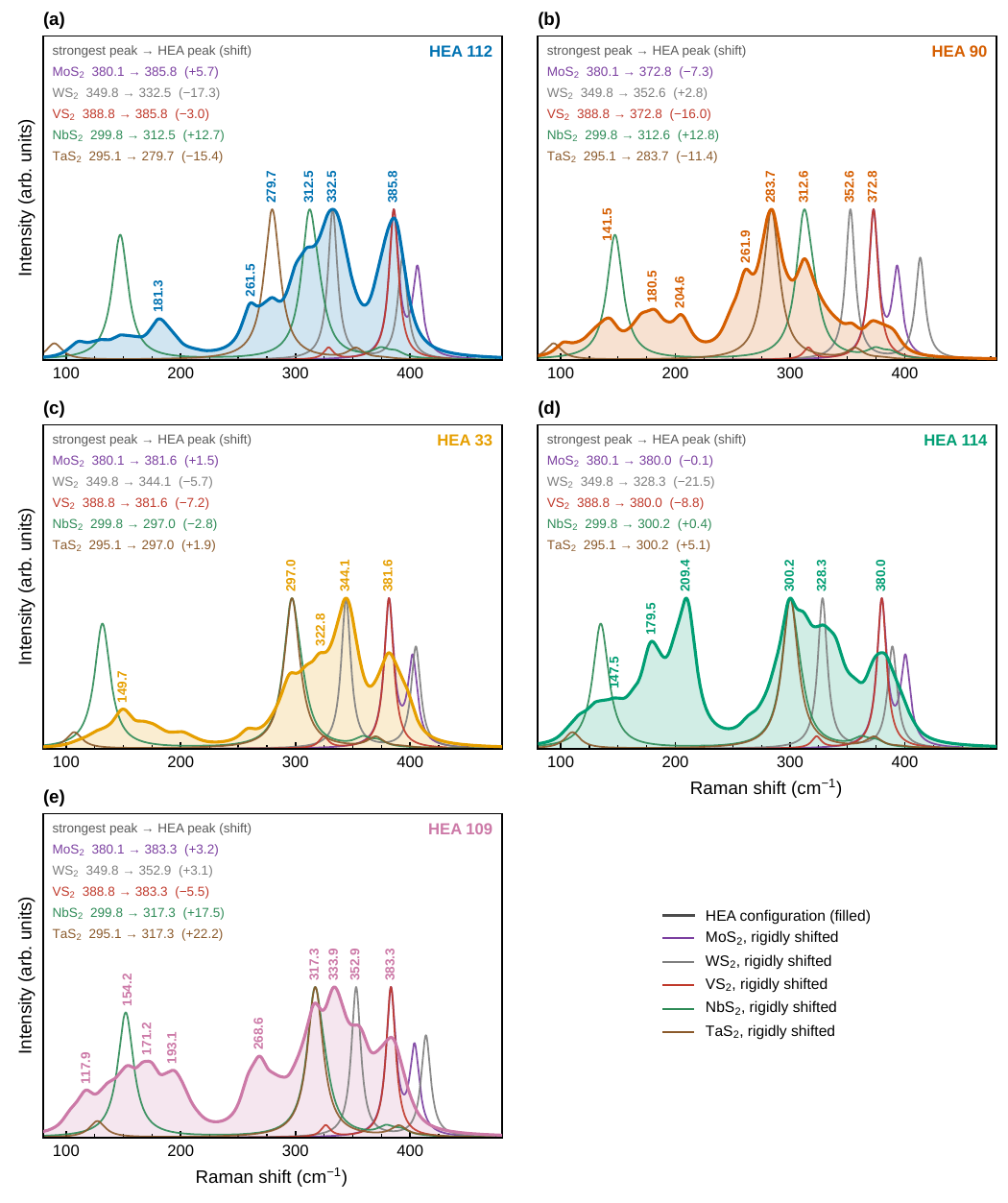}
  \caption{Raman spectra of the five HEA configurations (filled) with the spectra of the pristine
  MS$_2$ monolayers, each rigidly shifted so that its strongest peak coincides with the nearest peak of
  the HEA spectrum. Each panel lists the strongest pristine peak, the HEA peak it is moved onto and the
  shift, all in \unit{\per\centi\metre}. All spectra are normalised to their own maximum.}
  \label{fig:SI_raman_shifted_onto_hea}
\end{figure}

\begin{figure}[h]
  \centering
  \includegraphics[width=0.9\textwidth]{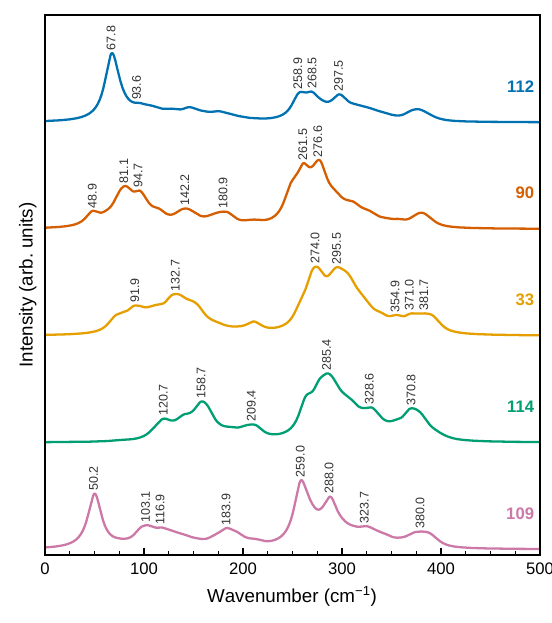}
  \caption{As \autoref{fig:SI_raman_hea}, for the IR spectra.}
  \label{fig:SI_ir_hea}
\end{figure}

\begin{figure}[p]
  \centering
  \includegraphics[width=\textwidth]{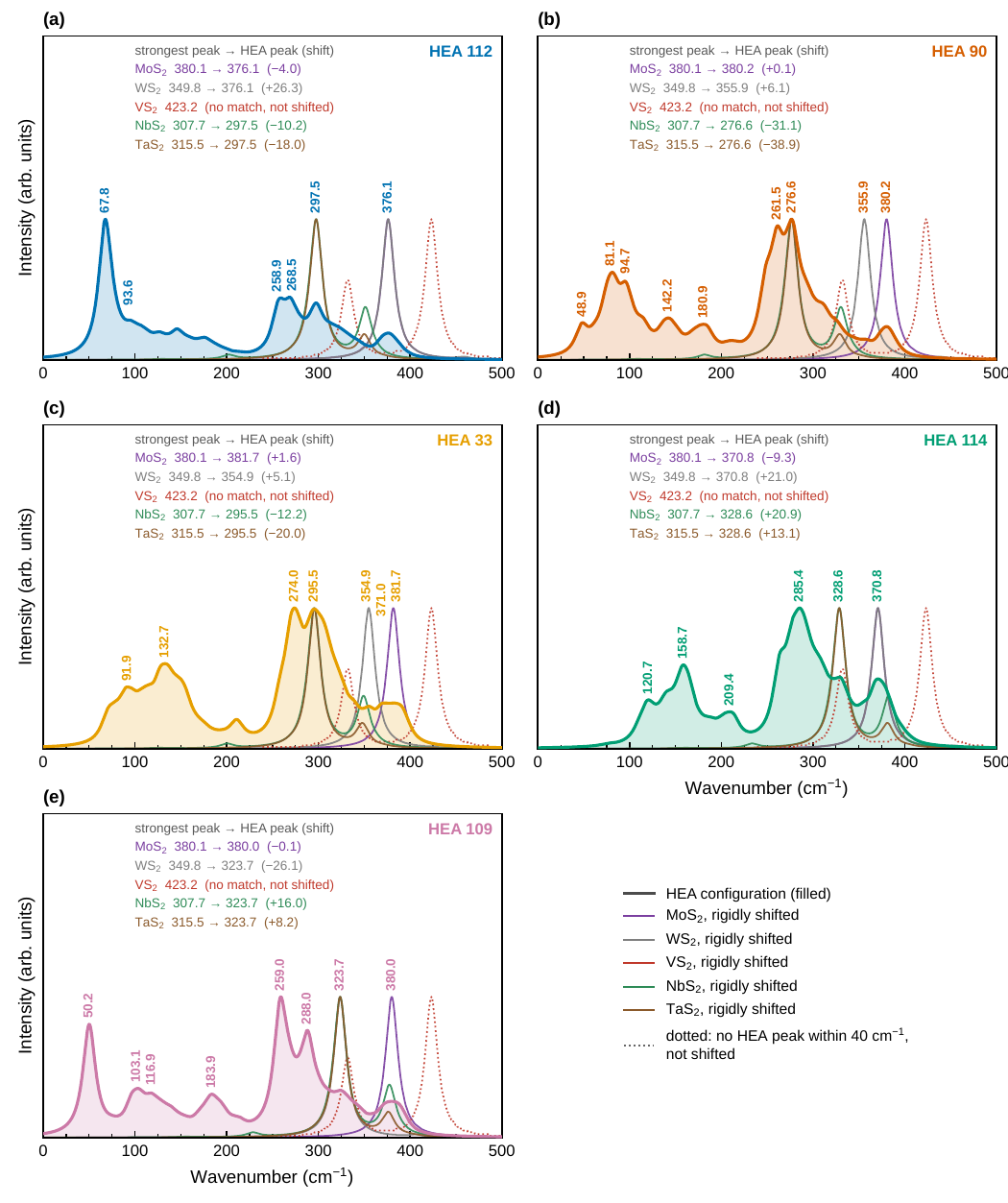}
  \caption{As \autoref{fig:SI_raman_shifted_onto_hea}, for the IR spectra. The VS\textsubscript{2} spectrum (dotted) has
  no HEA peak within \qty[per-mode=reciprocal]{40}{\per\centi\metre} of its strongest peak and is not shifted.}
  \label{fig:SI_ir_shifted_onto_hea}
\end{figure}

\autoref{fig:SI_raman_hea} and \autoref{fig:SI_ir_hea} show the Raman and IR spectra of the five configurations discussed in the main text. 
All spectra are normalised to their own maximum, i.e.\ every intensity of a spectrum is divided by the largest intensity of that spectrum in the plotted range
(\qtyrange[per-mode=reciprocal]{80}{480}{\per\centi\metre} 
for Raman and 
\qtyrange[per-mode=reciprocal]{0}{500}{\per\centi\metre} for IR).
The Raman intensities of the HEA cells were computed for the modes above 
\qty[per-mode=reciprocal]{\approx100}{\per\centi\metre}.
\autoref{fig:SI_raman_shifted_onto_hea} and \autoref{fig:SI_ir_shifted_onto_hea} compare each configuration with the pristine monolayers. 
Each pristine spectrum is shifted rigidly, i.e.\ all its points are moved by the same amount along the wavenumber axis, without fitting. 
The shift moves the strongest peak of the pristine spectrum onto the nearest peak of the HEA spectrum, where peaks are local maxima higher than 10\% of the maximum of the HEA spectrum and shifts are limited to 
$\pm$\qty[per-mode=reciprocal]{40}{\per\centi\metre}.
The shifts are listed in \autoref{tab:SI_raman_shifts} and \autoref{tab:SI_ir_shifts}.
%
%
%
%
\begin{table}[h]
  \caption{Rigid shifts (\unit{\per\centi\metre}) of the pristine Raman spectra onto each HEA
  configuration. The position of the strongest pristine peak is given in brackets.}
  \label{tab:SI_raman_shifts}
  \begin{ruledtabular}
  \begin{tabular}{cccccc}
    $c$ & MoS$_2$ (380.1) & WS$_2$ (349.8) & VS$_2$ (388.8) & NbS$_2$ (299.8) & TaS$_2$ (295.1) \\
    \hline
    112 & $+5.7$ & $-17.3$ & $-3.0$  & $+12.7$ & $-15.4$ \\
    90  & $-7.3$ & $+2.8$  & $-16.0$ & $+12.8$ & $-11.4$ \\
    33  & $+1.5$ & $-5.7$  & $-7.2$  & $-2.8$  & $+1.9$  \\
    114 & $-0.1$ & $-21.5$ & $-8.8$  & $+0.4$  & $+5.1$  \\
    109 & $+3.2$ & $+3.1$  & $-5.5$  & $+17.5$ & $+22.2$ \\
  \end{tabular}
  \end{ruledtabular}
\end{table}
%
\begin{table}[h]
  \caption{Rigid shifts (\unit{\per\centi\metre}) of the pristine IR spectra onto each HEA
  configuration. The position of the strongest pristine peak is given in brackets; n/a: no HEA peak
  within \qty{40}{\per\centi\metre}. $^{a}$The WS$_2$ peak lies almost midway between the HEA peaks at
  \num{328.6} and \qty{370.8}{\per\centi\metre} (shifts of $-21.2$ and $+21.0$).}
  \label{tab:SI_ir_shifts}
  \begin{ruledtabular}
  \begin{tabular}{cccccc}
    $c$ & MoS$_2$ (380.1) & WS$_2$ (349.8) & VS$_2$ (423.2) & NbS$_2$ (307.7) & TaS$_2$ (315.5) \\
    \hline
    112 & $-4.0$ & $+26.3$ & n/a & $-10.2$ & $-18.0$ \\
    90  & $+0.1$ & $+6.1$  & n/a & $-31.1$ & $-38.9$ \\
    33  & $+1.6$ & $+5.1$  & n/a & $-12.2$ & $-20.0$ \\
    114 & $-9.3$ & $+21.0^{a}$ & n/a & $+20.9$ & $+13.1$ \\
    109 & $-0.1$ & $-26.1$ & n/a & $+16.0$ & $+8.2$  \\
  \end{tabular}
  \end{ruledtabular}
\end{table}
%
\section{X-ray Diffraction}
\label{sec:SI_xrd}
\subsection{Simulated patterns}
\label{sec:SI_xrd_sim}
\begin{figure}[t]
  \centering
  \includegraphics[width=0.75\columnwidth]{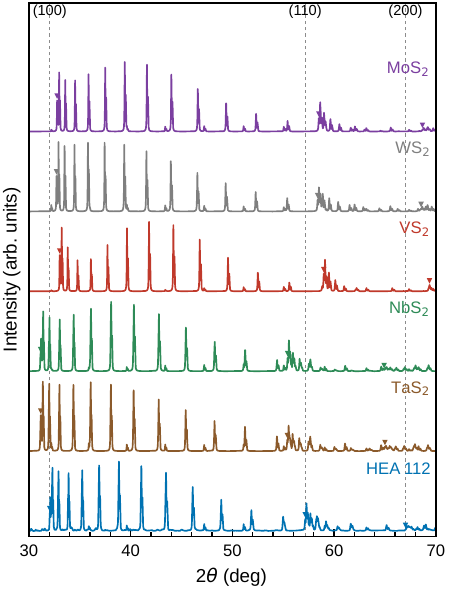}
  \caption{Simulated powder XRD patterns (Cu K$\alpha$) of the pristine MS$_2$ monolayers and of the HEA
  configuration 112, each normalised to its own maximum between \qty{30}{\degree} and \qty{70}{\degree} and offset
  vertically. Triangles mark the in-plane reflections $(100)$, $(110)$ and $(200)$ of each pattern; the dashed
  lines mark those of the HEA. The other sharp peaks are $(hkl)$ reflections with $l \neq 0$ that arise from
  the periodic repetition of the monolayer along $z$ in the simulation cell. Peak positions and lattice
  parameters are listed in \autoref{tab:SI_xrd_pos}.}
  \label{fig:SI_xrd_sim}
\end{figure}

\begin{figure}[h]
  \centering
  \includegraphics[width=0.75\textwidth]{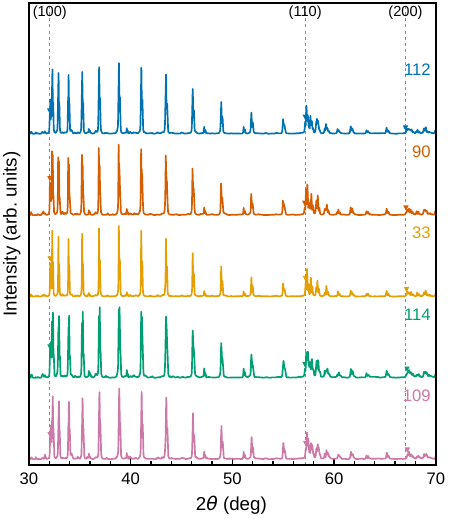}
  \caption{Simulated powder XRD patterns (Cu K$\alpha$) of the five HEA configurations, each normalised to
  its own maximum between 30$^\circ$ and 70$^\circ$ and offset vertically. Triangles mark the in-plane
  reflections $(100)$, $(110)$ and $(200)$ of each pattern; the dashed lines mark those of configuration 112.}
  \label{fig:SI_xrd_hea}
\end{figure}

%
\autoref{fig:SI_xrd_sim} shows the simulated patterns of the pristine monolayers and of configuration 112,
from which the positions marked at the top of \autoref{fig:SI_xrd_sim} are taken, and
\autoref{fig:SI_xrd_hea} those of the five HEA configurations. 
Because the simulation cell repeats the monolayer every \qty{25}{\angstrom} along $z$, each in-plane reflection is followed by a series of sharp $(hkl)$ peaks whose spacing depends only on the cell height; the patterns are therefore shown from 30$^\circ$, and only the in-plane ($l=0$) reflections $(100)$, $(110)$ and $(200)$, which belong to the monolayer, are used. 
Their positions and the lattice parameters obtained from them are listed in \autoref{tab:SI_xrd_pos}.
%
%
%
\begin{table}[h]
  \caption{Positions $2\theta$ of the in-plane reflections (Cu K$\alpha_1$, $\lambda = \qty{1.5406}{\angstrom}$)
  and in-plane lattice parameter $a$ obtained from each of them, for the pristine monolayers and the five HEA
  configurations. The average of the five pristine lattice parameters (Vegard's law, from $(110)$) is
  \qty{3.214}{\angstrom}.}
  \label{tab:SI_xrd_pos}
  \setlength{\tabcolsep}{4pt}
  \begin{ruledtabular}
  \begin{tabular}{lcccccc}
    System & \multicolumn{3}{c}{$2\theta$ (deg)} & \multicolumn{3}{c}{$a$ (\AA)} \\
           & $(100)$ & $(110)$ & $(200)$ & $(100)$ & $(110)$ & $(200)$ \\
    \hline
    MoS$_2$  & 32.77 & 58.50 & 68.69 & 3.153 & 3.153 & 3.153 \\
    WS$_2$   & 32.71 & 58.38 & 68.55 & 3.159 & 3.159 & 3.159 \\
    VS$_2$   & 33.02 & 58.98 & 69.38 & 3.130 & 3.130 & 3.126 \\
    NbS$_2$  & 31.14 & 55.43 & 64.92 & 3.314 & 3.313 & 3.314 \\
    TaS$_2$  & 31.16 & 55.40 & 65.00 & 3.312 & 3.314 & 3.311 \\
    \hline
    HEA, config. 112  & 32.05 & 57.16 & 67.02 & 3.222 & 3.220 & 3.222 \\
    HEA, config. 90   & 32.06 & 57.11 & 67.07 & 3.221 & 3.223 & 3.220 \\
    HEA, config. 33   & 32.09 & 57.20 & 67.12 & 3.218 & 3.218 & 3.218 \\
    HEA, config. 114  & 32.07 & 57.13 & 67.17 & 3.220 & 3.222 & 3.216 \\
    HEA, config. 109  & 32.08 & 57.22 & 67.21 & 3.219 & 3.217 & 3.214 \\
  \end{tabular}
  \end{ruledtabular}
\end{table}
%
\subsection{Fit of the experimental pattern}
\label{sec:SI_xrd_fit}
\paragraph*{Geometric model of the film.}
All calculations in this work are done on the monolayer, whereas the film consists of small crystallites,
each a stack of a few layers. 
To compute its diffraction pattern, we build a purely geometric model of such a crystallite from the calculated monolayer. 
The DFT-optimised monolayer of one configuration (75 atoms, $5\times5$ supercell) is copied $N$ times along $z$, at a distance $d$ between consecutive layers; the atomic positions inside each layer are those of DFT, and only the in-plane lattice is scaled uniformly by a fitted factor. 
Each layer either sits directly above the layer below, with probability $p_0$, or is shifted in-plane by
$+(\mathbf{a}_1+2\mathbf{a}_2)/3$ or $-(\mathbf{a}_1+2\mathbf{a}_2)/3$ (one third of a lattice vector), with
probabilities $p_+$ and $p_-$; the shifts of successive layers are independent (Hendricks--Teller
model \cite{SI_Hendricks1942}). 
Only translations are included, so rotated stackings such as 2H are not part of the model. 
The stack is not relaxed, neither with DFT nor with the GAP, and no interaction between the layers
is computed: $N$, $d$ and the shift probabilities are parameters of the fit. 
Relaxing the stack could shift the peaks slightly, but is not expected to change the results. Since powder diffraction cannot distinguish $p_+$ from $p_-$, only the probability $1-p_0$ that a layer is shifted is meaningful, and it describes the disorder of the stacking rather than a stacking preferred by the material.
\paragraph*{Calculated intensity.}
The intensity is computed along the reciprocal-lattice rods $\mathbf{G}$ of the $5\times5$ supercell,
\begin{equation}
I_\mathbf{G}(Q_z) = |F_\mathrm{L}(\mathbf{G},Q_z)|^2 \Big[ 1 + \frac{2}{N}\sum_{k=1}^{N-1}(N-k)\,
\mathrm{Re}\big(c_\mathbf{G}^{\,k}\, e^{iQ_z d k}\big) \Big],
\qquad c_\mathbf{G} = p_0 + p_+ e^{i\phi_\mathbf{G}} + p_- e^{-i\phi_\mathbf{G}},
\label{eq:SI_rod}
\end{equation}

where
\begin{equation}
F_\mathrm{L}(\mathbf{Q}) = \sum_j f_j(Q)\, e^{i\mathbf{Q}\cdot\mathbf{r}_j}\,
e^{-\frac{1}{2}(Q_\parallel^2 U_{\parallel,j} + Q_z^2 U_{z,j})}
\end{equation}
is the structure factor of the monolayer, with atomic form factors including the anomalous terms for
Cu K$\alpha$ and the in-plane and out-of-plane mean-square displacements $U_\parallel$, $U_z$ of each element
from the MD at \qty{300}{\kelvin} (Debye--Waller factor), and $\phi_\mathbf{G} = \mathbf{G}\cdot(\mathbf{a}_1+2\mathbf{a}_2)/3$ is the phase of the shift for rod $\mathbf{G}$. For $h-k = 3n$ (indices of the primitive cell) $\phi_\mathbf{G}$ is a multiple of $2\pi$ and $c_\mathbf{G} = 1$, so the $(00l)$ and $(11l)$ reflections are not affected by the shifts and stay sharp. 
For the other rods, the shifts spread the intensity along $Q_z$, and the $(10l)$ reflections merge into a continuous band, which starts sharply at $(100)$ and decays slowly towards higher angles. 
Rods that exist only because of the $5\times5$ supercell (superstructure of the metal arrangement) carry no correlation between layers and are given an in-plane coherence of \qty{2}{\nano\metre}, so that they contribute as weak diffuse scattering. 
A non-integer $N$ is treated as a mixture of the two neighbouring integers.
The rods are broadened in plane by the finite size of the layers (disks with a lognormal distribution of
diameters) and averaged over all orientations of the crystallites,
\begin{equation}
P(Q) = \frac{1}{Q}\sum_\mathbf{G}\int_0^Q I_\mathbf{G}(Q_z)\, T\, \frac{s(q_\parallel - G)}{2\pi q_\parallel}\, \mathrm{d}Q_z,
\qquad q_\parallel = \sqrt{Q^2 - Q_z^2},
\end{equation}
where $s$ is the in-plane line profile of the layers and $T$ a March--Dollase distribution (parameter $r$) of
the $c$ axis about the film normal, evaluated for the crystallites that diffract at the tilt
$\psi = \theta - \omega$ of the grazing-incidence geometry ($r > 1$: the layers tend to stand up from the
substrate). 
This average yields the asymmetric line shape of two-dimensional crystallites \cite{SI_warren1941}
without further assumptions. 
The pattern is multiplied by the polarisation factor $(1+\cos^2 2\theta)/2$, by an
additional isotropic displacement factor $e^{-2Bs^2}$ ($s = \sin\theta/\lambda$) for static disorder, and by the
absorption factor of a film at fixed incidence angle $\omega = 2.5^\circ$, $[1-e^{-\mu t k}]/k$ with
$k = 1/\sin\omega + 1/\sin(2\theta-\omega)$ (thickness $t = \qty{300}{\nano\metre}$ assumed); the Cu K$\alpha_2$
component is included. 
The background, i.e.\ the smooth intensity underneath the peaks, which comes from disordered parts of the film, the substrate and air scattering, is described by a Chebyshev polynomial (6 terms) plus a $1/2\theta$ term and is not interpreted.

\begin{figure}[h]
  \centering
  \includegraphics[width=\textwidth]{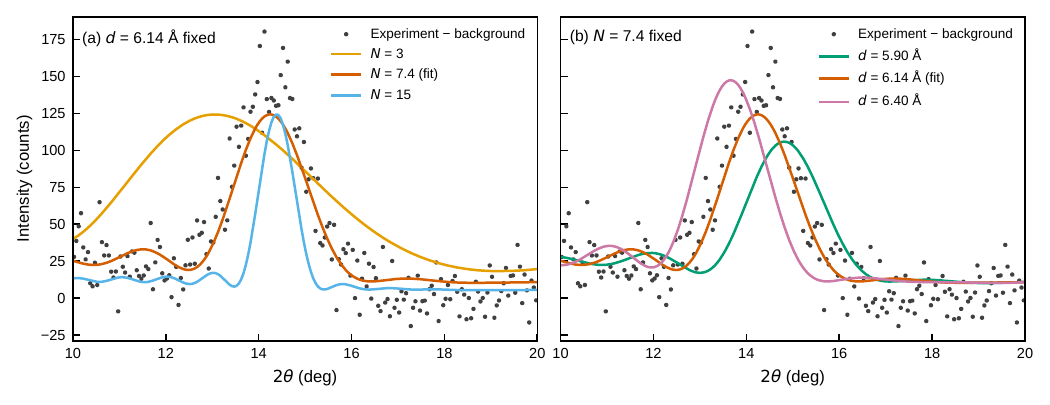}
  \caption{The $(001)$ reflection of the film (dots, background subtracted) and the model of
  configuration 112 with all parameters at their fitted values except (a) the number of layers $N$ and (b) the
  interlayer distance $d$. The width of the peak fixes $N$ and its position fixes $d$.}
  \label{fig:SI_xrd_001}
\end{figure}

\begin{figure}[h]
  \centering
  \includegraphics[width=\textwidth]{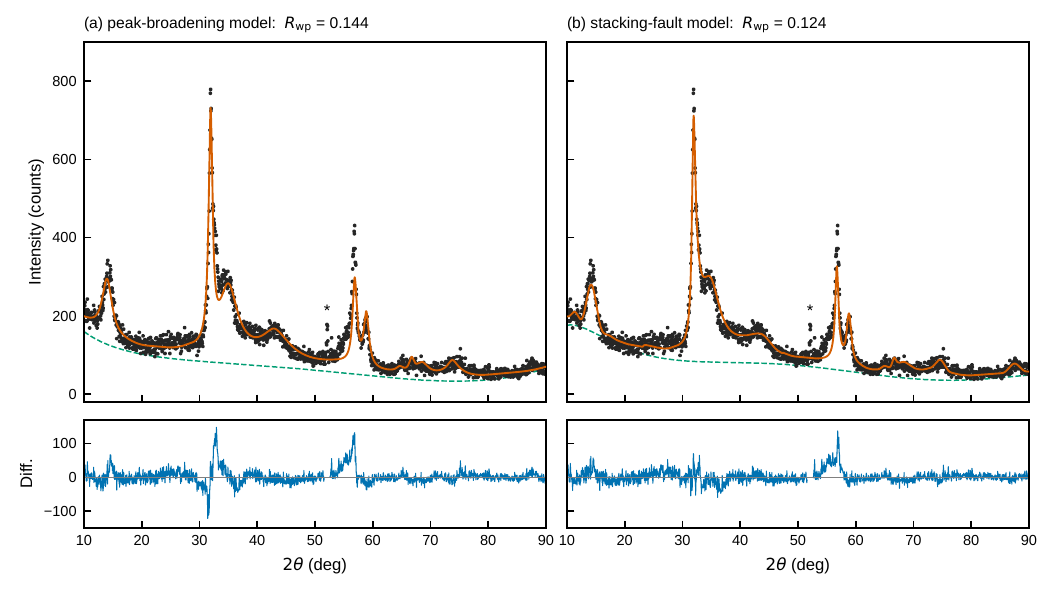}
  \caption{Fit of the experimental GIXRD pattern with configuration 112: (a) every reflection of a one-layer
  stacking broadened by an ellipsoidal crystallite shape; (b) stacking-fault model used in the main text.
  Dots: experiment; line: fit; dashed line: background; lower panels: difference.}
  \label{fig:SI_xrd_models}
\end{figure}

\noindent Note: The stacking-fault intensity follows the Hendricks–Teller treatment of randomly stacked layers \cite{SI_Hendricks1942, Warren1969book, DritsTchoubar1990}, and the powder average of the rods that of two-dimensional crystallites \cite{SI_warren1941, Yang1996}; preferred orientation is described with the March–Dollase function \cite{March1932,Dollase1986} and the atomic form factors are taken from Refs. \cite{Waasmaier1995, Chantler2000}.
\paragraph*{Fit.}
The calculated pattern is fitted to the measured one by weighted least squares (weights $1/I_\mathrm{obs}$)
over $2\theta = 10^\circ$--$90^\circ$, excluding the sharp reflection at 52.1$^\circ$, which does not belong to the film. 
Unlike the simulation cell of \autoref{sec:SI_xrd_sim}, in which the monolayers are \qty{25}{\angstrom} apart,
the model of the film stacks the layers at their real distance, so its reflections below 30$^\circ$, in particular $(001)$, are physical and the whole measured range is used. 
The scale and the background are obtained by linear least squares at every step. 
The refined parameters are the in-plane lattice parameter $a$, the interlayer distance $d$, the number of layers $N$, the mean diameter and width of the size distribution, $p_0$ and $p_+$ ($p_- = 1 - p_0 - p_+$), $r$ and $B$. 
The quality of the fit is given by the weighted profile factor
$R_\mathrm{wp} = [\sum w (I_\mathrm{obs} - I_\mathrm{calc})^2 / \sum w I_\mathrm{obs}^2]^{1/2}$, 
to be compared with $R_\mathrm{exp} = 0.087$ expected from counting statistics. 
The errors given below are statistical only.
\paragraph*{Interlayer distance and number of layers.}
Both are fixed mainly by the $(001)$ reflection at $2\theta = 14.25^\circ$, which arises from the spacing of the
layers (\autoref{fig:SI_xrd_001}). 
Its position sets $d$: a smaller $d$ moves it to higher angle, a larger $d$ to lower angle. 
Its width sets $N$: a thick stack gives a sharp peak and a thin one a broad peak. Simple estimates
from the peak alone agree with the fit. 
Bragg's law, $d = \lambda/(2\sin\theta)$ with $\lambda = \qty{1.5406}{\angstrom}$ and $\theta = 7.13^\circ$, gives \qty{6.21}{\angstrom}, close to the fitted $d = \qty{6.14}{\angstrom}$; the difference arises because the $(001)$ peak of a thin stack is broad and its maximum is shifted slightly to lower angle, which the model takes into account. 
The Scherrer formula, $L = K\lambda/(\beta\cos\theta)$ with $K = 0.9$ and the measured width 
$\beta = 1.75^\circ = \qty{0.0305}{rad}$, gives $L = \qty{4.6}{\nano\metre}$, i.e.\ about
7.5 layers of \qty{0.61}{\nano\metre}, close to the fitted $N = 7.4$. The $(111)$ reflection, whose position depends on both $a$ and $d$, provides an independent check: with the fitted values it is calculated at 58.80$^\circ$ and measured at 58.83$^\circ$.
%
\paragraph*{Results.}
With configuration 112 the model gives $R_\mathrm{wp} = 0.124$, against \num{0.144} for a simpler model in which every
reflection of a one-layer stacking is only broadened by an ellipsoidal crystallite shape
(\autoref{fig:SI_xrd_models}); the improvement comes from the $(10l)$ band, which reproduces the tail of $(100)$
and the broad maxima at about 35$^\circ$ and 43$^\circ$. 
The refined parameters are
$a_\mathrm{exp} = \qty{3.246}{\angstrom}$, $d = \qty{6.14}{\angstrom}$, $N = 7.4$ ($L_c = N d = \qty{4.5}{\nano\metre}$), $1 - p_0 = 0.51$, $r = 1.78$ and $B = \qty{1.5}{\angstrom\squared}$. 
The lattice parameter of the film is
larger than that given by the position of the $(100)$ maximum alone, \qty{3.236}{\angstrom}, because the shifts
move the maximum of the $(10l)$ band to higher angle.
%

The only systematic misfit left is a broad shoulder on the low-angle side of $(110)$, between 54$^\circ$ and
56.7$^\circ$. 
Adding an empirical peak for it (centre 56.3$^\circ$, FWHM 2.7$^\circ$, corresponding to
$a \approx \qty{3.27}{\angstrom}$) lowers $R_\mathrm{wp}$ to \num{0.101} and changes $a$ by only \qty{0.002}{\angstrom}, with $d$, $N$, $1-p_0$ and $r$ unchanged, but raises $B$ to \qty{3.0}{\angstrom\squared}; the value of $B$ is therefore only indicative. 
The shoulder lies between the $(110)$ reflections of the HEA and of NbS$_2$/TaS$_2$
and may come from regions of the film with a larger lattice parameter, for example richer in Nb and Ta. 
The $(111)$ reflection of the film at 58.83$^\circ$ lies close to the $(110)$ reflections of the simulated MoS$_2$, WS$_2$ and VS$_2$ patterns (58.38$^\circ${--}58.98$^\circ$).

\begin{figure}[h]
  \centering
  \includegraphics[width=\textwidth]{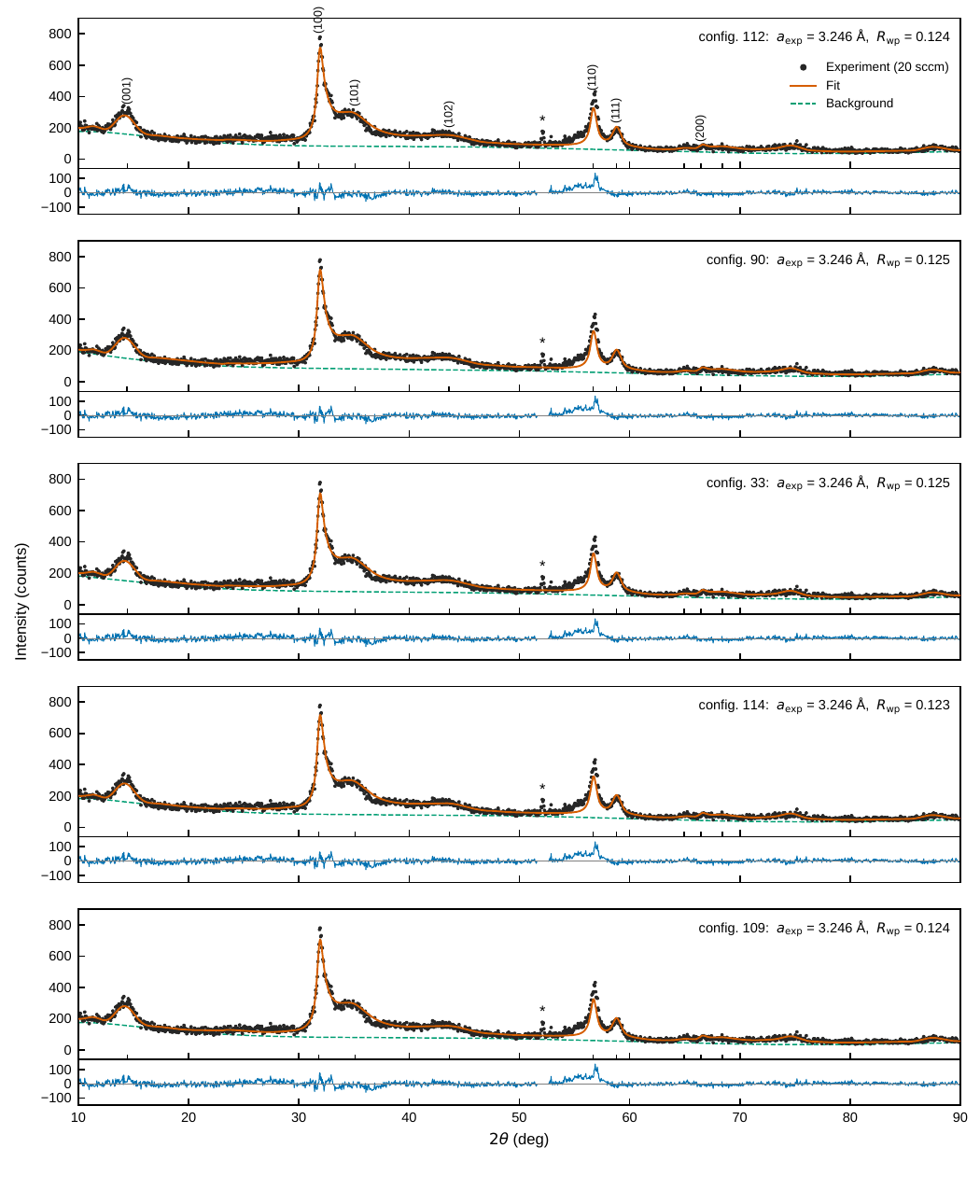}
  \caption{Fit of the experimental GIXRD pattern of the 20 sccm film with the stacking-fault model built from
  each of the five configurations. Dots: experiment; line: fit; dashed line: background; lower panels:
  difference; *: reflection not belonging to the film, excluded from the fit.}
  \label{fig:SI_xrd_fit_all_configs}
\end{figure}

All five configurations give the same fit, with $a_\mathrm{exp} = \qty{3.246}{\angstrom}$ and $R_\mathrm{wp} = $ 0.123{--}0.125 (\autoref{fig:SI_xrd_fit_all_configs}, Table \ref{tab:SI_xrd_fit}).
%
%
\begin{table}[h]
  \caption{Fit of the experimental GIXRD pattern (20 sccm film) with the stacking-fault model built from each HEA
  configuration: in-plane lattice parameter of the DFT- and GAP-optimised structures (taken from Table II of main text)
  and of the film, $a_\mathrm{exp}$, with its relative difference from the calculated values; interlayer distance $d$; mean number of layers per crystallite $N$; area-weighted mean in-plane diameter $D_A$ (approximate, see text); probability $1-p_0$ that a layer is shifted with respect to the layer below; March--Dollase parameter $r$; additional (static) displacement parameter $B$; weighted profile $R$ factor ($R_\mathrm{exp} = 0.087$ from counting statistics). The stack is a geometric model and is not relaxed.}
  \label{tab:SI_xrd_fit}
  \begin{ruledtabular}
  \begin{tabular}{lcccccccccccc}
    Config. & \multicolumn{3}{c}{$a$ (\AA)} & \multicolumn{2}{c}{$a_\mathrm{exp}/a - 1$ (\%)} & $d$ & $N$ & $D_A$ & $1-p_0$ & $r$ & $B$ & $R_\mathrm{wp}$ \\
     & DFT & GAP & exp. & DFT & GAP & (\AA) & & (nm) & & & (\AA$^2$) & \\
    \hline
    112 & 3.219 & 3.233 & 3.246 & $+0.84$ & $+0.40$ & 6.14 & 7.4 & 18 & 0.51 & 1.78 & 1.5 & 0.124 \\
    90 & 3.219 & 3.230 & 3.246 & $+0.85$ & $+0.49$ & 6.14 & 7.2 & 17 & 0.51 & 1.76 & 1.5 & 0.125 \\
    33 & 3.218 & 3.231 & 3.246 & $+0.86$ & $+0.46$ & 6.14 & 7.3 & 18 & 0.51 & 1.77 & 1.5 & 0.125 \\
    114 & 3.217 & 3.231 & 3.246 & $+0.91$ & $+0.46$ & 6.14 & 7.3 & 18 & 0.51 & 1.78 & 1.4 & 0.123 \\
    109 & 3.216 & 3.233 & 3.246 & $+0.95$ & $+0.41$ & 6.14 & 7.3 & 18 & 0.51 & 1.76 & 1.4 & 0.124 \\
  \end{tabular}
  \end{ruledtabular}
\end{table}
%
\section{Work function}
\label{SI:wf}
\begin{figure}[h]
  \centering
  \includegraphics[width=\textwidth]{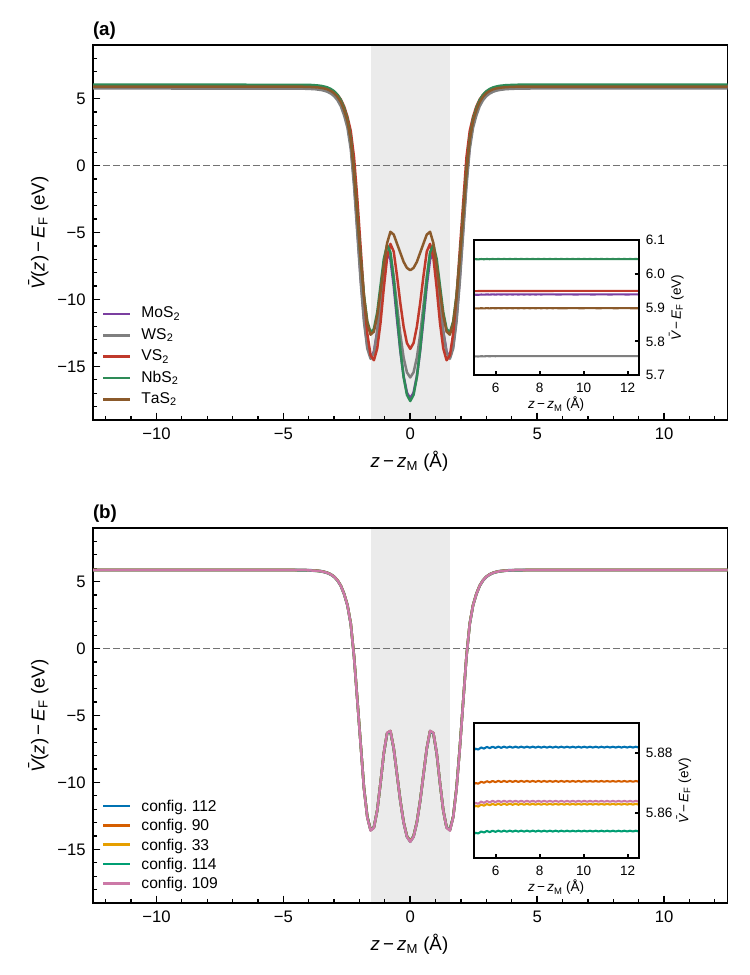}
  \caption{Planar-averaged electrostatic potential $\bar{V}(z)$ relative to the Fermi energy
  $E_\mathrm{F}$ of (a)~the pristine MS$_2$ monolayers and (b)~the five HEA configurations;
  $z_\mathrm{M}$ is the position of the metal plane and the shaded area marks the layer between the
  two sulfur planes. The plateau far from the layer is the work function $\Phi$
  (\autoref{tab:SI_wf}). The insets enlarge the vacuum region.}
  \label{fig:SI_wf_planar}
\end{figure}

\autoref{tab:SI_wf} lists the vacuum level $E_\mathrm{vac}$, the Fermi energy $E_\mathrm{F}$ and the
work function $\Phi = E_\mathrm{vac} - E_\mathrm{F}$ of all systems.
$E_\mathrm{vac}$ and $E_\mathrm{F}$ are given on the energy scale of the DFT calculation, whose zero
is the average electrostatic potential of the cell; only their difference, $\Phi$, can be compared
between systems.
$E_\mathrm{vac}$ is the average of the planar-averaged electrostatic potential $\bar{V}(z)$ over the
flat region in the middle of the vacuum, i.e.\ more than \qty{7}{\angstrom} from the metal plane
(cell height \qty{25}{\angstrom}).
\autoref{fig:SI_wf_planar} shows $\bar{V}(z)$, relative to $E_\mathrm{F}$, for all systems.
In every system the potential is flat on both sides of the layer, and the vacuum levels of the two
sides agree to within \qty{0.01}{\milli\electronvolt}, so that no dipole correction is needed.
The pristine monolayers differ mainly inside the layer, where the depth of the potential reflects the
metal, whereas the profiles of the five HEA configurations coincide on the scale of the figure; their
vacuum plateaus, enlarged in the insets, differ by at most \qty{28}{\milli\electronvolt}.

%
\begin{table}[h]
  \caption{Vacuum level $E_\mathrm{vac}$, Fermi energy $E_\mathrm{F}$ and work function
  $\Phi = E_\mathrm{vac} - E_\mathrm{F}$ (all in \unit{\electronvolt}). $\Phi$ was computed from the
  unrounded values, so the last digit may differ from the difference of the rounded columns.}
  \label{tab:SI_wf}
  \begin{ruledtabular}
  \begin{tabular}{lccc}
    System & $E_\mathrm{vac}$ & $E_\mathrm{F}$ & $\Phi$ \\
    \hline
    MoS$_2$ & 3.166 & $-2.773$ & 5.939 \\
    WS$_2$  & 3.081 & $-2.675$ & 5.756 \\
    VS$_2$  & 2.895 & $-3.054$ & 5.949 \\
    NbS$_2$ & 3.011 & $-3.032$ & 6.044 \\
    TaS$_2$ & 2.581 & $-3.317$ & 5.898 \\
    \hline
    HEA, config 112 & 2.929 & $-2.953$ & 5.882 \\
    HEA, config 90  & 2.929 & $-2.941$ & 5.871 \\
    HEA, config 33  & 2.930 & $-2.933$ & 5.863 \\
    HEA, config 114 & 2.933 & $-2.921$ & 5.854 \\
    HEA, config 109 & 2.935 & $-2.929$ & 5.864 \\
  \end{tabular}
  \end{ruledtabular}
\end{table}

\section{Structural Information of Stable Configurations}
\label{SI:relaxed_geo}
The fully relaxed atomic structures of the most stable equimolar (MoNbTaVW)S\textsubscript{2} high-entropy alloy supercells are provided in the following subsections in POSCAR format. 
Each supercell contains 75 atoms (Mo$_5$Nb$_5$Ta$_5$V$_5$W$_5$S$_{50}$), i.e., 25 metal atoms randomly distributed over the transition-metal sublattice and 50 sulfur atoms. 
\subsection{Configuration 112}
\label{SI:config112}
\begin{verbatim}
Mo5 Nb5 S50 Ta5 V5 W5                    
1.0000000000000000
16.1017138077697624   0.01668116758716650   0.00000000000000000
-8.0364137289266502   13.9241238386128199   0.00000000000000000
0.0000000000000000    0.00000000000000000   25.0000000000000000
Mo Nb S Ta V W
5  5  50 5 5 5
Direct
     0.9977541074063936    0.1978707444513432    0.5000000030693172
     0.8013689643020214    0.4016719679816987    0.4999999940277959
     0.1999222512353356    0.6028467086389213    0.4999999965772943
     0.5964665178076558    0.5945045352318630    0.5000000000980691
     0.0024619268517370    0.7973091784315045    0.4999999961399126
     0.8043517987877138    0.0084080607921564    0.5000000040249120
     0.0031382132421385    0.0052328657478427    0.4999999994297823
     0.5963833880989853    0.3988867697290530    0.4999999975801188
     0.7989227949894797    0.5947079430355657    0.4999999947432201
     0.7999915101075601    0.7987571756372205    0.4999999986888426
     0.1376020383975671    0.0712242069746409    0.5629587947005774
     0.3367250871635505    0.0705364776203936    0.5631659486997194
     0.5376287568776164    0.0692975029504268    0.5601464147926905
     0.7293858678190617    0.0705666230192890    0.5609190973813156
     0.9350450950933435    0.0698205744344449    0.5643207459557615
     0.1334225187346547    0.2692786936203291    0.5610797359149231
     0.3331287655411401    0.2674381997152366    0.5629320283340462
     0.5331285547471296    0.2616959286399023    0.5610430787874634
     0.7340669395296189    0.2660890946635703    0.5592152260795018
     0.9297047557218576    0.2641420014760953    0.5612928482800825
     0.1386817709040784    0.4670415628921489    0.5613837826467432
     0.3282580284948793    0.4603029039217787    0.5619141071799376
     0.5296831781708207    0.4651760235361451    0.5640147266825508
     0.7321011164525473    0.4619139939991825    0.5634365890780728
     0.9339722952460150    0.4681778163891011    0.5628147488340858
     0.1322923766244699    0.6654990482798979    0.5630177409191717
     0.3330341651687760    0.6690892481561050    0.5610187609278015
     0.5326752156459236    0.6657254152704155    0.5622864955754352
     0.7290683618193867    0.6628263015511777    0.5621038156163749
     0.9349171943218266    0.6658854813735431    0.5626270944918303
     0.1341686823493831    0.8645853596072470    0.5627051119201951
     0.3316461000203930    0.8626988999927292    0.5621558713184953
     0.5281552394974938    0.8701007773848772    0.5593214634175112
     0.7336911343681649    0.8701321680228773    0.5628551344145976
     0.9377467021672174    0.8660237784820808    0.5617151686287798
     0.1376020358620593    0.0712242078892729    0.4370412065933849
     0.3367250857939226    0.0705364742060297    0.4368340482033872
     0.5376287619997484    0.0692975075883169    0.4398535907641659
     0.7293858693797617    0.0705666254674355    0.4390809080789424
     0.9350450971743953    0.0698205777146512    0.4356792592246848
     0.1334225168812040    0.2692786950670580    0.4389202695842158
     0.3331287626242982    0.2674382008604894    0.4370679716486734
     0.5331285558629943    0.2616959220415183    0.4389569195559773
     0.7340669341641828    0.2660890860272573    0.4407847759119932
     0.9297047579660145    0.2641420023407903    0.4387071566166227
     0.1386817719350120    0.4670415614025529    0.4386162200350441
     0.3282580264437058    0.4603029037547728    0.4380858943878323
     0.5296831737053580    0.4651760226149406    0.4359852724520934
     0.7321011170732348    0.4619139944534680    0.4365634026817062
     0.9339723011447845    0.4681778152664080    0.4371852457450700
     0.1322923819253603    0.6654990494498492    0.4369822586787251
     0.3330341674226105    0.6690892502057153    0.4389812428750517
     0.5326752162333221    0.6657254190397026    0.4377135062122903
     0.7290683538095308    0.6628263005090460    0.4378961810882416
     0.9349171958225995    0.6658854861278343    0.4373728998030089
     0.1341686823691930    0.8645853585954271    0.4372948875444251
     0.3316461010878271    0.8626989005855705    0.4378441288599291
     0.5281552412147050    0.8701007774725724    0.4406785397352948
     0.7336911376223085    0.8701321751361419    0.4371448695193934
     0.9377467027888713    0.8660237827231825    0.4382848325982778
     0.2035896411533358    0.0007398509818017    0.4999999974406961
     0.3994534969540169    0.9981774170781463    0.4999999995649134
     0.0032906366115313    0.4025740648888672    0.4999999993068158
     0.3965800712805105    0.5940596373440030    0.5000000004087539
     0.6004582351952479    0.8027015297490934    0.5000000028504914
     0.6032332416387689    0.0067200190965977    0.5000000014283046
     0.6036489815471455    0.2000783636737609    0.5000000042554333
     0.8023105807502944    0.2000525590329758    0.5000000016394210
     0.1988639526246985    0.3959717180923263    0.5000000023161491
     0.3925894813200301    0.7988824775450141    0.5000000005612151
     0.2018938217286461    0.2037903979486586    0.5000000033388758
     0.3993172166906689    0.2026613229621788    0.4999999972213800
     0.4009541429472350    0.3987609407680708    0.5000000023489690
     0.0016141662417226    0.6007321295672909    0.4999999901127126
     0.1982509713031888    0.7966954430804165    0.4999999938505226
\end{verbatim}
\subsection{Configuration 90}
\label{SI:config90}
\begin{verbatim}
Mo5 Nb5 S50 Ta5 V5 W5                    
1.0000000000000000
16.1154318784585939    0.0019449356398522    0.0000000000000000
-8.0560388498952555   13.9166161486336417    0.0000000000000000
 0.0000000000000000    0.0000000000000000   25.0000000000000000
Mo Nb S Ta V W
5  5  50 5 5 5
Direct
     0.3996581870348948    0.0031318393991100    0.5000000018205668
     0.0006705770652005    0.0024331046861974    0.5000000038306780
     0.5946224909551034    0.3968859324595684    0.4999999931804808
     0.1990515446526686    0.5965869169192075    0.4999999989576978
     0.8033871227204799    0.8019714160632034    0.4999999964351005
     0.8088720515773318    0.0026638542828792    0.5000000044110493
     0.5958096410071789    0.2033983614794721    0.5000000013064110
     0.7971364915067767    0.4029089513561956    0.5000000018121682
     0.0063568682530644    0.4018435480520921    0.5000000050336197
     0.3994931788201299    0.7989276558287202    0.5000000028027856
     0.1367195184172587    0.0693735495745855    0.5595263956538616
     0.3303397181131871    0.0685257637649457    0.5605706351400599
     0.5309048769455984    0.0679233098973882    0.5632523234419153
     0.7355322713176093    0.0700525610533589    0.5608708979523129
     0.9402495096796031    0.0711158071085123    0.5638857403867874
     0.1367527450484332    0.2668856911204358    0.5613855035570978
     0.3303309830777366    0.2659459750394504    0.5594865603303703
     0.5307812397420548    0.2685218205602879    0.5648612554893546
     0.7364137763826051    0.2645969366030556    0.5607046151292039
     0.9324448989564971    0.2637003654956089    0.5611620711540937
     0.1371433606410832    0.4662433169368114    0.5642927714012913
     0.3328721532534047    0.4619435469389472    0.5608798588081356
     0.5273106560969509    0.4614412549404037    0.5619850361584042
     0.7294007833694423    0.4704839696369731    0.5610966007896607
     0.9349044813452707    0.4679066792350071    0.5624145317640128
     0.1330129057127536    0.6632803408256436    0.5624967343800051
     0.3304245764325202    0.6619083182246115    0.5629531587261596
     0.5277476072819667    0.6607348363386265    0.5632922897832202
     0.7365927740219789    0.6665851947855614    0.5613723051139274
     0.9307303240439211    0.6629323438769390    0.5615500856186486
     0.1336680155957097    0.8723419455161689    0.5615603022689939
     0.3329120192393304    0.8689896603751635    0.5624176526237292
     0.5351080850324905    0.8658061139672881    0.5626559134189577
     0.7358691045017807    0.8664082555789108    0.5627264601704525
     0.9338381538779217    0.8704871570700021    0.5635298339727797
     0.1367195057294168    0.0693735384779117    0.4404736052916576
     0.3303397210812662    0.0685257612564740    0.4394293614758809
     0.5309048846547455    0.0679233089455380    0.4367476848377976
     0.7355322653093381    0.0700525567626045    0.4391291167401334
     0.9402495076145883    0.0711158085314665    0.4361142695098366
     0.1367527376158648    0.2668856888422866    0.4386144963829466
     0.3303309755459836    0.2659459663908947    0.4405134305103715
     0.5307812434283932    0.2685218112108316    0.4351387384650494
     0.7364137807831099    0.2645969333876650    0.4392953915467004
     0.9324448949928836    0.2637003628175307    0.4388379415526842
     0.1371433480730744    0.4662433116243889    0.4357072292955522
     0.3328721491607709    0.4619435498535864    0.4391201329899488
     0.5273106539235428    0.4614412516491910    0.4380149456235571
     0.7294007944064020    0.4704839691741043    0.4389033917953284
     0.9349044824945025    0.4679066742960032    0.4375854735123781
     0.1330129078036038    0.6632803443732058    0.4375032673448231
     0.3304245772299907    0.6619083244957480    0.4370468397544514
     0.5277476026412202    0.6607348439096086    0.4367077013102971
     0.7365927837941584    0.6665851999271197    0.4386276809125604
     0.9307303329798557    0.6629323509161793    0.4384499130919011
     0.1336680234843186    0.8723419480747054    0.4384397029304026
     0.3329120158522585    0.8689896581875018    0.4375823552734133
     0.5351080826333927    0.8658061214394053    0.4373440918935003
     0.7358691075499380    0.8664082660496604    0.4372735374926862
     0.9338381651969954    0.8704871661344740    0.4364701667040833
     0.5965546042465490    0.0013589062089565    0.5000000071431998
     0.6003207028926596    0.5976867883444044    0.4999999898372849
     0.2000134840417545    0.8017144195135941    0.5000000034898804
     0.6006760877877255    0.7940593426675306    0.4999999980480041
     0.9945179717643953    0.7996888568084870    0.5000000010339249
     0.2001433417309926    0.0014744063255065    0.5000000011349783
     0.1984112753341876    0.1993227763201233    0.4999999972747204
     0.8031721209482896    0.1995254953184187    0.5000000063826633
     0.4006896654847125    0.3975021202399844    0.4999999934082879
     0.8003783172274908    0.6032396026859601    0.4999999958634547
     0.4032880568052377    0.2025304860574693    0.4999999958427281
     0.0037302808759421    0.2014555991410063    0.5000000053299587
     0.1995207969167367    0.3998732663064359    0.4999999984503916
     0.3940266440819755    0.5953809752260497    0.4999999976573335
     0.0021594141617811    0.5994959471167220    0.5000000000412754
\end{verbatim}
\subsection{Configuration 33}
\label{SI:config33}
\begin{verbatim}
Mo5 Nb5 S50 Ta5 V5 W5                    
1.0000000000000000
16.0925785114523805   -0.0003074512969118   0.00000000000000000
-8.0465609463908532   13.9358879529260644   0.00000000000000000
0.0000000000000000    0.00000000000000000   25.0000000000000000
Mo Nb S Ta V W
5  5  50 5 5 5
Direct
     0.3973610930979135    0.0008188594136982    0.4999999982221084
     0.3980704726593913    0.1981532317413439    0.5000000005551115
     0.4029634297275990    0.4007011572074290    0.5000000051510227
     0.8022180162650940    0.3960210513813536    0.4999999932360808
     0.0038064683345951    0.8042591234168484    0.4999999985869863
     0.2006378986566233    0.6032424218433690    0.4999999984696117
     0.4064762679563501    0.6015961675967958    0.5000000022667592
     0.8013691192913228    0.5988871243159863    0.4999999952133649
     0.4044066560503552    0.8017797704550702    0.5000000010061001
     0.7978222827371582    0.7961779319344302    0.4999999996021174
     0.1315718741726499    0.0631968974503039    0.5619257119145900
     0.3324676548077434    0.0673560427869617    0.5626382146154967
     0.5332838270106990    0.0667887832232594    0.5596604200404727
     0.7299959858999969    0.0698168784865842    0.5614418974160174
     0.9392092813373694    0.0670657512186139    0.5601421541768303
     0.1261299766234245    0.2601007815419081    0.5615022501078428
     0.3364788083503782    0.2679351188997430    0.5633555669978350
     0.5324428419933369    0.2597647789816406    0.5613582988286936
     0.7298883653476480    0.2643954807995641    0.5614848807469812
     0.9341110755146360    0.2655306873845689    0.5608015449317421
     0.1262456640026528    0.4635945590903390    0.5604176959993907
     0.3330644930696351    0.4643898262698798    0.5623142945267148
     0.5342773730039738    0.4685730636971925    0.5622417535994373
     0.7342236543805143    0.4628533814488023    0.5624794853206723
     0.9345265251148973    0.4647248708304446    0.5615557107025921
     0.1348110640191962    0.6723777085793046    0.5628532545005669
     0.3339668102356583    0.6660701231955812    0.5632200318115537
     0.5368778392178086    0.6693899928925120    0.5647562784169224
     0.7295302170667100    0.6626689652419628    0.5634262348408825
     0.9344804344949793    0.6714112770133411    0.5637835414993617
     0.1344481223341987    0.8695419850693327    0.5631722599628191
     0.3328310154280842    0.8694268714350670    0.5620770087346685
     0.5402080493919783    0.8725527259746855    0.5611022429119075
     0.7330970249364483    0.8684309708892002    0.5622666743819380
     0.9361324906412776    0.8701154317291540    0.5613682253190362
     0.1315718736711489    0.0631968969572938    0.4380742887162796
     0.3324676517814140    0.0673560433602134    0.4373617843072779
     0.5332838303364511    0.0667887841895265    0.4403395756615751
     0.7299959914244383    0.0698168741191907    0.4385581060333124
     0.9392092806702834    0.0670657516160205    0.4398578557613675
     0.1261299755387952    0.2601007830943232    0.4384977529215206
     0.3364788123276767    0.2679351258210829    0.4366444366939177
     0.5324428404501589    0.2597647799221576    0.4386417008079064
     0.7298883640481080    0.2643954728083244    0.4385151109029977
     0.9341110776836672    0.2655306823649894    0.4391984560863946
     0.1262456660744959    0.4635945582267738    0.4395823006322104
     0.3330644960714579    0.4643898255578236    0.4376857161606225
     0.5342773646761856    0.4685730608801535    0.4377582514844534
     0.7342236458226168    0.4628533767158842    0.4375205046991795
     0.9345265300987435    0.4647248726398487    0.4384442813035534
     0.1348110668479237    0.6723777117764485    0.4371467432718745
     0.3339668151561241    0.6660701252889965    0.4367799694955536
     0.5368778341731044    0.6693899948239377    0.4352437230138975
     0.7295302136373039    0.6626689686927705    0.4365737609989964
     0.9344804371309579    0.6714112795161924    0.4362164536544952
     0.1344481258008727    0.8695419880957331    0.4368277409467467
     0.3328310124642400    0.8694268690174454    0.4379229917944016
     0.5402080488836560    0.8725527196743030    0.4388977584531801
     0.7330970259796317    0.8684309783798483    0.4377333329475093
     0.9361324929888823    0.8701154370866888    0.4386317737216814
     0.8017668112436098    0.0056755544388309    0.5000000057583662
     0.2049525127834144    0.1986180786581215    0.4999999997169979
     0.1966378085803555    0.3951656914078612    0.5000000021277131
     0.5973326245230339    0.3948659928141112    0.4999999997300435
     0.9984193983261846    0.6038865887775344    0.4999999970118907
     0.6013385436640775    0.0044169573525252    0.4999999982651886
     0.0032302521178167    0.9962755094513938    0.5000000017805490
     0.5990545942506387    0.1959588680613180    0.4999999976888034
     0.9922966836059572    0.1972035487513111    0.5000000043070756
     0.9901037046635466    0.3951479635907234    0.4999999976754452
     0.2011399395758247    0.9996804085316100    0.5000000014533228
     0.8005395142786327    0.2036952202790445    0.4999999986559160
     0.5968393857584786    0.6020915453288040    0.4999999986735375
     0.1999275208030866    0.8054716837993183    0.4999999999368967
     0.5993580589146745    0.7973936346952470    0.5000000021331488
\end{verbatim}
\subsection{Configuration 114}
\label{SI:config114}
\begin{verbatim}
Mo5 Nb5 S50 Ta5 V5 W5                    
1.0000000000000000
16.0798766933676660   -0.0271443922558780   0.00000000000000000
-8.0634450269860984   13.9459446219424468   0.00000000000000000
0.0000000000000000    0.00000000000000000   25.0000000000000000
Mo Nb S Ta V W
5  5  50 5 5 5
Direct
     0.2009104631433516    0.9999778914531490    0.5000000015811779
     0.7994112757582172    0.4012715157178306    0.4999999992157669
     0.9971770632012777    0.4016502135323847    0.5000000006136176
     0.4044732293286089    0.6019261771836923    0.4999999955432131
     0.9989394866092454    0.7982972914611608    0.5000000008640981
     0.5974889660833256    0.9983719618092906    0.5000000019153745
     0.8006537272103458    0.2047982049407082    0.4999999999002682
     0.9973994834947320    0.2042342072143413    0.4999999986143493
     0.1979458289786251    0.4028840897826385    0.4999999978856309
     0.5975902224444454    0.7910572956560102    0.5000000029355505
     0.1329689926615885    0.0636278893904176    0.5634901034766813
     0.3342713013527359    0.0655374007963729    0.5619301835559938
     0.5315998290117250    0.0717139588756623    0.5595537664176078
     0.7341466637771887    0.0650587372530893    0.5614799473502998
     0.9292128711230347    0.0639291955906529    0.5604458148666254
     0.1349431728020249    0.2652194524180373    0.5622925314189686
     0.3380781486751729    0.2673331036167426    0.5626649387644278
     0.5345606038610171    0.2675898952380891    0.5620848945479651
     0.7271493637327424    0.2674041878816809    0.5613518409911222
     0.9324250087097568    0.2710042871256775    0.5641590977054562
     0.1292197719229975    0.4659489399745738    0.5639283182388297
     0.3353884577655393    0.4684792706204277    0.5624210541798078
     0.5405229520742836    0.4692614110046165    0.5634225331019849
     0.7345108182526729    0.4677297751104064    0.5626805516169640
     0.9309024107759782    0.4668953721830889    0.5627191665046780
     0.1308505422558852    0.6676597754452018    0.5623020873806581
     0.3368071242678781    0.6693654914239175    0.5606294560465841
     0.5342535284934868    0.6618875642144459    0.5648034300030744
     0.7344048238153889    0.6671424929590444    0.5621438502507843
     0.9322282453880320    0.6654793667366192    0.5623328900103886
     0.1310951148939026    0.8689540792665795    0.5637925406275954
     0.3396000504876611    0.8698072382768132    0.5586659983203575
     0.5311260794251282    0.8630876635757758    0.5619748168810190
     0.7389676110357328    0.8679990052267073    0.5593229051564634
     0.9264579973055120    0.8589542529213006    0.5615268934715729
     0.1329689893079546    0.0636278819621410    0.4365098957600466
     0.3342713029268793    0.0655373985712515    0.4380698219957537
     0.5315998282215801    0.0717139586687097    0.4404462390938022
     0.7341466610161120    0.0650587381888741    0.4385200548936083
     0.9292128673550479    0.0639291880246375    0.4395541842219970
     0.1349431711479880    0.2652194507046346    0.4377074637296658
     0.3380781514004312    0.2673331031351935    0.4373350567257859
     0.5345606000809866    0.2675898857151893    0.4379151042650165
     0.7271493606495767    0.2674041856462992    0.4386481590051403
     0.9324250115104604    0.2710042920062818    0.4358409004767125
     0.1292197714262002    0.4659489446317693    0.4360716842778913
     0.3353884519219861    0.4684792650632374    0.4375789369599232
     0.5405229522231139    0.4692614155990142    0.4365774619758795
     0.7345108218099483    0.4677297750247220    0.4373194422265669
     0.9309024137147475    0.4668953747163798    0.4372808330048414
     0.1308505487473894    0.6676597821414489    0.4376979168629163
     0.3368071218201933    0.6693654955020065    0.4393705406241679
     0.5342535413632277    0.6618875771042170    0.4351965684610377
     0.7344048253634625    0.6671424946774849    0.4378561482673788
     0.9322282453232305    0.6654793667914802    0.4376671101613141
     0.1310951158698614    0.8689540767863022    0.4362074626224199
     0.3396000565135608    0.8698072458853758    0.4413340086179574
     0.5311260786794350    0.8630876640860023    0.4380251868860370
     0.7389676106819039    0.8679990037960721    0.4406770979078800
     0.9264579989340191    0.8589542532068464    0.4384731088092693
     0.9989656892601815    0.9929104678847196    0.5000000011862227
     0.2033787639414513    0.1963932761810909    0.4999999973646538
     0.4085863176490108    0.4054990560996146    0.4999999973085636
     0.1961343327177616    0.5971619669874002    0.5000000013209558
     0.1981638412855844    0.8056772117267528    0.5000000031095695
     0.4004142097083019    0.0061224640757587    0.5000000048603042
     0.7985963184244440    0.9992262690976119    0.4999999996411120
     0.5959287555309558    0.2054820192345446    0.5000000013365238
     0.4042185244742100    0.7978870580602404    0.5000000006933405
     0.7956603932143764    0.7918208753040048    0.5000000012192842
     0.4008636878977541    0.1970142990103625    0.5000000000614762
     0.6032466530371750    0.4000984811084365    0.4999999957609376
     0.6025158253402267    0.6017434018185099    0.4999999993628137
     0.8000075354677705    0.6001297569312456    0.4999999991386233
     0.9966164239222124    0.5975549229670136    0.4999999998476383
\end{verbatim}
\subsection{Configuration 109}
\label{SI:config109}
\begin{verbatim}
Mo5 Nb5 S50 Ta5 V5 W5                    
1.0000000000000000     
16.0832649752914527   -0.0184467113120496  0.00000000000000000
-8.0576010454363143   13.9298899880714391  0.00000000000000000
0.0000000000000000    0.0000000000000000   25.0000000000000000
Mo Nb S Ta V W
5  5  50 5 5 5
Direct
     0.8016801252012726    0.0016226791714971    0.4999999953118390
     0.1972084746714629    0.2027904145859836    0.4999999968190281
     0.5996190792569536    0.3999911696462490    0.5000000041956838
     0.2006927379624486    0.5981771866541763    0.4999999989564614
     0.3942728629595322    0.5985773929964324    0.4999999982345216
     0.1931272385095397    0.0002359981432889    0.5000000043921773
     0.3976658758004277    0.3974203378288195    0.4999999970740845
     0.5996906610627732    0.6002925314363168    0.4999999992760351
     0.8014782815719954    0.6048913939558176    0.5000000092020400
     0.1991898096360813    0.7936747860207944    0.4999999998352322
     0.1262372326055328    0.0690073426998907    0.5614309490887308
     0.3330403714094317    0.0650332205856472    0.5615340045359574
     0.5378222385149556    0.0669040829239123    0.5629002384516681
     0.7361563947362839    0.0673132638660618    0.5634172536397060
     0.9361805937744094    0.0701047496651910    0.5600571379361767
     0.1281444225290826    0.2657247053091822    0.5613564589090316
     0.3277291419086765    0.2647659131718783    0.5643595995887054
     0.5351845149762582    0.2670648000962004    0.5623870599908329
     0.7409805483874123    0.2693127346633020    0.5627067085398352
     0.9350298936459325    0.2643121906756676    0.5618520909411302
     0.1320710649292280    0.4647739233459234    0.5625995258066752
     0.3296884949739010    0.4669531448364239    0.5626210190768717
     0.5343073205919850    0.4662873587140126    0.5629238192241530
     0.7360424891054151    0.4630159851002418    0.5600951748059657
     0.9297946602608889    0.4602606240881641    0.5620234734448175
     0.1334352282936493    0.6617642516698936    0.5640263188594759
     0.3336396680595897    0.6689626661995334    0.5612334023426726
     0.5287399419183814    0.6662755652093963    0.5623194717736197
     0.7362876815533355    0.6734323985802675    0.5630770426525231
     0.9375794464096074    0.6702620809663102    0.5614628716033039
     0.1273477208717821    0.8620364671347289    0.5608490628748015
     0.3350669459379745    0.8627079274275218    0.5625003832093185
     0.5303066799485450    0.8679612577470621    0.5614018288817846
     0.7373623640173292    0.8699682426499535    0.5627920766371517
     0.9379166526440130    0.8713442950242580    0.5612545406484628
     0.1262372311015980    0.0690073318377316    0.4385690532254358
     0.3330403607037411    0.0650332179936228    0.4384659983901642
     0.5378222413924618    0.0669040946157509    0.4370997581251856
     0.7361563976271484    0.0673132678257176    0.4365827393343196
     0.9361805975136974    0.0701047472230840    0.4399428540084713
     0.1281444265920939    0.2657247054132483    0.4386435288723831
     0.3277291433099023    0.2647659039396970    0.4356403976148613
     0.5351845168686395    0.2670648020320598    0.4376129464663734
     0.7409805481652540    0.2693127405314044    0.4372932930653732
     0.9350298950658384    0.2643121986384855    0.4381479016043031
     0.1320710596713468    0.4647739269815219    0.4374004694032934
     0.3296885009770776    0.4669531486143441    0.4373789733266591
     0.5343073322223333    0.4662873661774753    0.4370761803403411
     0.7360424873347213    0.4630159855313423    0.4399048331132036
     0.9297946619351976    0.4602606336757020    0.4379765305736569
     0.1334352222084974    0.6617642500688063    0.4359736845022368
     0.3336396696139445    0.6689626674864115    0.4387665967852996
     0.5287399436755180    0.6662755473232167    0.4376805271342405
     0.7362876835586789    0.6734323947157534    0.4369229678611006
     0.9375794405938151    0.6702620789098717    0.4385371427057763
     0.1273477168955637    0.8620364690466218    0.4391509428457852
     0.3350669423269323    0.8627079230418460    0.4374996147394583
     0.5303066761393751    0.8679612569647759    0.4385981642820482
     0.7373623690923026    0.8699682371690329    0.4372079211934121
     0.9379166508863079    0.8713442869032519    0.4387454664132733
     0.4026447774525437    0.9962169453933214    0.4999999962659274
     0.3992364449223871    0.2053890845180888    0.5000000035297845
     0.6085050622052677    0.2039095044657770    0.5000000026756979
     0.0009350968781590    0.5943089424555623    0.5000000066769559
     0.6012434541758225    0.8040915033903531    0.4999999978576426
     0.9942197441432015    0.9988501982077693    0.4999999985717878
     0.9983317173456925    0.2012739146539602    0.4999999912783650
     0.7991456726963904    0.3961654940187742    0.5000000028638425
     0.3968184939581718    0.7975975220128575    0.4999999977901410
     0.9958109763096843    0.8024075159802564    0.5000000098830881
     0.6062897951862851    0.0030920704345192    0.4999999945099489
     0.8018541027871251    0.1969267322350277    0.4999999978694376
     0.1981361765273988    0.3996703225737477    0.4999999936670392
     0.0007691680245543    0.3981102050775860    0.4999999993547988
     0.8059207432792022    0.8065477791316482    0.5000000045184336
\end{verbatim}
%
\makeatletter
\def\@bibsetup#1{\setlength{\topsep}{\z@}\NATx@bibsetnum{\ref{LastBibItemSI}}}%
\renewenvironment{thebibliography}[1]{%
 \NAT@thebibliography{#1}%
 \let\@TBN@opr\present@bibnote
 \@FMN@list
}{%
 \auto@bib@innerbib
 \edef\@currentlabel{\arabic{NAT@ctr}}%
 \label{LastBibItemSI}%
 \endNAT@thebibliography
 \aftergroup\auto@bib@empty
}%
\makeatother


\begin{thebibliography}{10}

\bibitem{Yeh2004}
J.-W. Yeh, S.-K. Chen, S.-J. Lin, J.-Y. Gan, T.-S. Chin, T.-T. Shun, C.-H.
  Tsau, and S.-Y. Chang.
\newblock Nanostructured high-entropy alloys with multiple principal elements:
  novel alloy design concepts and outcomes.
\newblock {\em Advanced engineering materials}, 6:299, 2004.

\bibitem{Cantor2004}
B.~Cantor, I.~T. Chang, P.~Knight, and A.~Vincent.
\newblock Microstructural development in equiatomic multicomponent alloys.
\newblock {\em Materials Science and Engineering: A}, 375:213, 2004.

\bibitem{Rost2015}
C.~M. Rost, E.~Sachet, T.~Borman, A.~Moballegh, E.~C. Dickey, D.~Hou, J.~L.
  Jones, S.~Curtarolo, and J.-P. Maria.
\newblock Entropy-stabilized oxides.
\newblock {\em Nat. Commun.}, 6(1):8485, 2015.

\bibitem{Castle2018}
E.~Castle, T.~Csanádi, S.~Grasso, J.~Dusza, and M.~Reece.
\newblock Processing and properties of high-entropy ultra-high temperature
  carbides.
\newblock {\em Sci. Rep.}, 8(1):8609, 2018.

\bibitem{Zaid2020a}
H.~Zaid, A.~Aleman, and S.~Kodambaka.
\newblock Size-dependent yielding and strain-hardening of
  compositionally-enriched body-centered cubic {VNbTaMoW} alloy.
\newblock {\em Scr. Mater.}, 178:518--521, 2020.

\bibitem{Gild2016}
J.~Gild, Y.~Zhang, T.~Harrington, S.~Jiang, T.~Hu, M.~C. Quinn, W.~M. Mellor,
  N.~Zhou, K.~Vecchio, and J.~Luo.
\newblock High-entropy metal diborides: A new class of high-entropy materials
  and a new type of ultrahigh temperature ceramics.
\newblock {\em Sci. Rep.}, 6(1):37946, 2016.

\bibitem{Zaid2021a}
H.~Zaid, K.~Tanaka, M.~Liao, M.~S. Goorsky, S.~Kodambaka, and H.~Kindlund.
\newblock Self-organized growth of 111-oriented {(VNbTaMoW)N} nanorods on
  {MgO}(001).
\newblock {\em Nano Lett.}, 21(1):577--582, 2021.

\bibitem{Zaid2021b}
H.~Zaid, K.~Tanaka, C.~V. Ciobanu, J.-M. Yang, S.~Kodambaka, and H.~Kindlund.
\newblock Growth of elastically-stiff, nanostructured, high-entropy alloy
  nitride, {(VNbTaMoW)N}/{Al$_2$O$_3$}(0001) thin film.
\newblock {\em Scripta Materialia}, 197:113813, 2021.

\bibitem{Nemani2023}
S.~K. Nemani, M.~Torkamanzadeh, B.~C. Wyatt, V.~Presser, and B.~Anasori.
\newblock Functional two-dimensional high-entropy materials.
\newblock {\em Communications Materials}, 4:16, 2023.

\bibitem{Divilov2024}
S.~Divilov, H.~Eckert, D.~Hicks, C.~Oses, C.~Toher, R.~Friedrich, M.~Esters,
  M.~J. Mehl, A.~C. Zettel, Y.~Lederer, et~al.
\newblock Disordered enthalpy–entropy descriptor for high-entropy ceramics
  discovery.
\newblock {\em Nature}, 625:66, 2024.

\bibitem{Ying2021}
T.~Ying, T.~Yu, Y.-S. Shiah, C.~Li, J.~Li, Y.~Qi, and H.~Hosono.
\newblock High-entropy van der waals materials formed from mixed metal
  dichalcogenides, halides, and phosphorus trisulfides.
\newblock {\em Journal of the American Chemical Society}, 143:7042, 2021.

\bibitem{Chiu2014}
M.~H. Chiu, M.~Y. Li, W.~J. Zhang, W.~T. Hsu, W.~H. Chang, M.~Terrones,
  H.~Terrones, and L.~J. Li.
\newblock Spectroscopic signatures for interlayer coupling in
  {MoS$_2$}–{WSe$_2$} van der waals stacking.
\newblock {\em ACS Nano}, 8(9):9649--9656, 2014.

\bibitem{Lee2012}
H.~S. Lee, S.~W. Min, Y.~G. Chang, M.~K. Park, T.~Nam, H.~Kim, J.~H. Kim,
  S.~Ryu, and S.~Im.
\newblock {MoS$_2$} nanosheet phototransistors with thickness-modulated optical
  energy gap.
\newblock {\em Nano Lett.}, 12(7):3695--3700, 2012.

\bibitem{Ramasubramaniam2012}
A.~Ramasubramaniam.
\newblock Large excitonic effects in monolayers of molybdenum and tungsten
  dichalcogenides.
\newblock {\em Phys. Rev. B}, 86(11):115409, 2012.

\bibitem{Zeng2012}
H.~L. Zeng, J.~F. Dai, W.~Yao, D.~Xiao, and X.~D. Cui.
\newblock Valley polarization in {MoS$_2$} monolayers by optical pumping.
\newblock {\em Nat. Nanotechnol.}, 7(8):490--493, 2012.

\bibitem{Zhang2012}
Y.~Zhang, J.~Ye, Y.~Matsuhashi, and Y.~Iwasa.
\newblock Ambipolar {MoS$_2$} thin flake transistors.
\newblock {\em Nano Lett.}, 12(3):1136--1140, 2012.

\bibitem{Butler2013}
S.~Z. Butler, S.~M. Hollen, L.~Y. Cao, Y.~Cui, J.~A. Gupta, H.~R. Gutierrez,
  T.~F. Heinz, S.~S. Hong, J.~X. Huang, A.~F. Ismach, et~al.
\newblock Progress, challenges, and opportunities in two-dimensional materials
  beyond graphene.
\newblock {\em ACS Nano}, 7(4):2898--2926, 2013.

\bibitem{Cazalilla2014}
M.~A. Cazalilla, H.~Ochoa, and F.~Guinea.
\newblock Quantum spin hall effect in two-dimensional crystals of
  transition-metal dichalcogenides.
\newblock {\em Phys. Rev. Lett.}, 113(7):077201, 2014.

\bibitem{Mak2014}
K.~F. Mak, K.~L. McGill, J.~Park, and P.~L. McEuen.
\newblock The valley hall effect in {MoS$_2$} transistors.
\newblock {\em Science}, 344(6191):1489--1492, 2014.

\bibitem{Li2014}
L.~K. Li, Y.~J. Yu, G.~J. Ye, Q.~Q. Ge, X.~D. Ou, H.~Wu, D.~L. Feng, X.~H.
  Chen, and Y.~B. Zhang.
\newblock Black phosphorus field-effect transistors.
\newblock {\em Nat. Nanotechnol.}, 9(5):372--377, 2014.

\bibitem{Piacentini2023}
Agata Piacentini, Alwin Daus, Zhenxing Wang, Max~C. Lemme, and Daniel Neumaier.
\newblock Potential of transition metal dichalcogenide transistors for flexible
  electronics applications.
\newblock {\em Adv. Electron. Mater.}, 9:2300181, 2023.

\bibitem{Daus2021}
A.~Daus, S.~Vaziri, V.~Chen, et~al.
\newblock High-performance flexible nanoscale transistors based on transition
  metal dichalcogenides.
\newblock {\em Nat Electron}, 4:495--501, 2021.

\bibitem{Pu2012}
J.~Pu, Y.~Yomogida, K.~K. Liu, L.~J. Li, Y.~Iwasa, and T.~Takenobu.
\newblock Highly flexible {MoS$_2$} thin-film transistors with ion gel
  dielectrics.
\newblock {\em Nano Lett.}, 12(8):4013--4017, 2012.

\bibitem{Henckel2018}
D.~A. Henckel, O.~M. Lenz, K.~M. Krishnan, and B.~M. Cossairt.
\newblock Improved {HER} catalysis through facile, aqueous electrochemical
  activation of nanoscale {WSe$_2$}.
\newblock {\em Nano Lett.}, 18(4):2329--2335, 2018.

\bibitem{Wu2021}
Longfei Wu and Jan~P. Hofmann.
\newblock Comparing the intrinsic her activity of transition metal
  dichalcogenides: Pitfalls and suggestions.
\newblock {\em ACS Energy Lett.}, 6(7):2619--2625, 2021.

\bibitem{Wang2022}
R.~Wang, J.~Huang, X.~Zhang, J.~Han, Z.~Zhang, T.~Gao, L.~Xu, S.~Liu, P.~Xu,
  and B.~Song.
\newblock Two-dimensional highentropy metal phosphorus trichalcogenides for
  enhanced hydrogen evolution reaction.
\newblock {\em ACS nano}, 16:3593, 2022.

\bibitem{Akhound2025}
M.~A. Akhound, K.~W. Jacobsen, and K.~S. Thygesen.
\newblock Activating the basal plane of {2D} transition metal dichalcogenides
  via high-entropy alloying.
\newblock {\em Journal of the American Chemical Society}, 147:5743, 2025.

\bibitem{Cavin2021}
J.~Cavin, A.~Ahmadiparidari, L.~Majidi, A.~S. Thind, S.~N. Misal, A.~Prajapati,
  Z.~Hemmat, S.~Rastegar, A.~Beukelman, M.~R. Singh, K.~A. Unocic,
  A.~Salehi-Khojin, and R.~Mishra.
\newblock {2D} high-entropy transition metal dichalcogenides for carbon dioxide
  electrocatalysis.
\newblock {\em Adv. Mater.}, 33(2100347), 2021.

\bibitem{Pedersen2023}
Pernille~D. Pedersen, Tejs Vegge, Thomas Bligaard, and Heine~A. Hansen.
\newblock Trends in {CO$_2$} reduction on transition metal dichalcogenide
  edges.
\newblock {\em ACS Catal.}, 13(4):2341--2350, 2023.

\bibitem{Deshpande2022}
Aditya Deshpande, Christian Ratsch, Cristian~V. Ciobanu, and Suneel Kodambaka.
\newblock Entropy stabilization of two-dimensional transition metal
  dichalcogenide alloys: A density functional theory study.
\newblock {\em J. Appl. Phys.}, 131(234302), 2022.

\bibitem{Chin2025}
Ching-Wei Chin, Chong-Chi Chi, Kai-Yuan Hsiao, Yi-Hsin Fang, Dun-Jie Jhan,
  Chun-Hui Lin, Ming-Pei Lu, and Ming-Yen Lu.
\newblock Entropy-stabilized multi-element {2D TMDs}: Tunable electronics and
  enhanced thermal stability.
\newblock {\em Chemical Engineering Journal}, 524(169215), 2025.

\bibitem{SWang2025}
Shuxi Wang, Sai~Varun Sunkara, Sukriti Manna, Alireza Ahmadiparidari, Khagesh
  Kumar, Taimin Yang, Shahriar Namvar, Pardis Seraji, Zhehao Huang, Jordi
  Cabana, Subramanian K. R.~S. Sankaranarayanan, Yuzi Liu, Anirudha~V. Sumant,
  and Amin Salehi-Khojin.
\newblock Self-lubricating tribo-catalytic activity of 2d high entropy alloy
  nanoflakes.
\newblock {\em Small}, 21(16):2500322, 2025.

\bibitem{Behler2007}
J.~Behler and M.~Parrinello.
\newblock Generalized neural-network representation of high-dimensional
  potential-energy surfaces.
\newblock {\em Physical Review Letters}, 98:146401, 2007.

\bibitem{Shapeev2016}
A.~V. Shapeev.
\newblock Moment tensor potentials: A class of systematically improvable
  interatomic potentials.
\newblock {\em Multiscale Modeling \& Simulation}, 14:1153, 2016.

\bibitem{Mishin2021}
Y.~Mishin.
\newblock Machine-learning interatomic potentials for materials science.
\newblock {\em Acta Materialia}, 214:116980, 2021.

\bibitem{Bartok2010}
A.~P. Bartók, M.~C. Payne, R.~Kondor, and G.~Csányi.
\newblock Gaussian approximation potentials: The accuracy of quantum mechanics,
  without the electrons.
\newblock {\em Phys. Rev. Lett.}, 104:136403, 2010.

\bibitem{Bartok2013}
A.~P. Bartók, R.~Kondor, and G.~Csányi.
\newblock On representing chemical environments.
\newblock {\em Phys. Rev. B}, 87:184115, 2013.

\bibitem{Bartok2015}
A.~P. Bartók and G.~Csányi.
\newblock Gaussian approximation potentials: A brief tutorial introduction.
\newblock {\em Int. J. Quantum Chem.}, 115:1051, 2015.

\bibitem{Byggmastar2019}
J.~Byggm{\"a}star, A.~Hamedani, K.~Nordlund, and F.~Djurabekova.
\newblock Machine-learning interatomic potential for radiation damage and
  defects in tungsten.
\newblock {\em Phys. Rev. B}, 100:144105, 2019.

\bibitem{Byggmastar2020a}
J.~Byggm{\"a}star, K.~Nordlund, and F.~Djurabekova.
\newblock Gaussian approximation potentials for body-centered-cubic transition
  metals.
\newblock {\em Phys. Rev. Materials}, 4:093802, 2020.

\bibitem{Nikoulis2021}
G.~Nikoulis, J.~Byggm{\"a}star, J.~Kioseoglou, K.~Nordlund, and F.~Djurabekova.
\newblock Machine-learning interatomic potential for {W-Mo} alloys.
\newblock {\em J. Phys.: Condens. Matter}, 33:315403, 2021.

\bibitem{Byggmastar2021}
J.~Byggm{\"a}star, K.~Nordlund, and F.~Djurabekova.
\newblock Modeling refractory high-entropy alloys with efficient
  machine-learned interatomic potentials: Defects and segregation.
\newblock {\em Phys. Rev. B}, 104:104101, 2021.

\bibitem{Byggmastar2022a}
J.~Byggm{\"a}star, K.~Nordlund, and F.~Djurabekova.
\newblock Simple machine-learned interatomic potentials for complex alloys.
\newblock {\em Phys. Rev. Materials}, 6:083801, 2022.

\bibitem{Fellman2026}
A~Fellman, J.~Byggm{\"a}star, F~Granberg, F.~Djurabekova, and K.~Nordlund.
\newblock Radiation damage and phase stability of alxcrcufeniy alloys using a
  machine-learned interatomic potential.
\newblock {\em Phys. Rev. Materials}, 10:023801, 2026.

\bibitem{Byggmastar2022b}
J.~Byggm{\"a}star, G~Nikoulis, A~Fellman, F~Granberg, and K.~Nordlund.
\newblock Multiscale machine-learning interatomic potentials for ferromagnetic
  and liquid iron.
\newblock {\em J. Phys.: Condens. Matter}, 34:305402, 2022.

\bibitem{Byggmastar2026}
Jesper Byggmästar.
\newblock Machine-learned interatomic potential for titanium carbide mxenes:
  application to ion irradiation simulations.
\newblock 18(36):19696--19707, 2026.

\bibitem{Zhang2020}
J.~Zhang, X.~Liu, S.~Bi, J.~Yin, G.~Zhang, and M.~Eisenbach.
\newblock Robust data-driven approach for predicting the configurational energy
  of high entropy alloys.
\newblock {\em Materials \& Design}, 185:108247, 2020.

\bibitem{Liu2021}
X.~Liu, J.~Zhang, J.~Yin, S.~Bi, M.~Eisenbach, and Y.~Wang.
\newblock Monte carlo simulation of order-disorder transition in refractory
  high entropy alloys: A data-driven approach.
\newblock {\em Computational Materials Science}, 187:110135, 2021.

\bibitem{Zhang2020b}
Y.~Zhang, C.~Wen, C.~Wang, S.~Antonov, D.~Xue, Y.~Bai, and Y.~Su.
\newblock Phase prediction in high entropy alloys with a rational selection of
  materials descriptors and machine learning models.
\newblock {\em Acta Materialia}, 185:528, 2020.

\bibitem{Dai2020}
D.~Dai, T.~Xu, X.~Wei, G.~Ding, Y.~Xu, J.~Zhang, and H.~Zhang.
\newblock Using machine learning and feature engineering to characterize
  limited material datasets of high-entropy alloys.
\newblock {\em Computational Materials Science}, 175:109618, 2020.

\bibitem{Pei2020}
Z.~Pei, J.~Yin, J.~A. Hawk, D.~E. Alman, and M.~C. Gao.
\newblock Machine-learning informed prediction of high-entropy solid solution
  formation: Beyond the hume-rothery rules.
\newblock {\em npj Computational Materials}, 6:50, 2020.

\bibitem{Deng2023}
B.~Deng, P.~Zhong, K.~Jun, J.~Riebesell, K.~Han, C.~J. Bartel, and G.~Ceder.
\newblock Chgnet as a pretrained universal neural network potential for
  charge-informed atomistic modelling.
\newblock {\em Nature Machine Intelligence}, pages 1031--1041, 2023.

\bibitem{Batatia2022}
I.~Batatia, D.~P. Kovacs, G.~Simm, C.~Ortner, and G.~Csányi.
\newblock Mace: Higher order equivariant message passing neural networks for
  fast and accurate force fields.
\newblock {\em Advances in neural information processing systems}, 35:11423,
  2022.

\bibitem{Yang2024}
H.~Yang, C.~Hu, Y.~Zhou, X.~Liu, Y.~Shi, J.~Li, G.~Li, Z.~Chen, S.~Chen,
  C.~Zeni, et~al.
\newblock Mattersim: A deep learning atomistic model across elements,
  temperatures and pressures.
\newblock {\em arXiv preprint arXiv:2405.04967}, 2024.

\bibitem{MartinBarrios2024}
R.~Martin-Barrios, E.~Navas-Conyedo, X.~Zhang, Y.~Chen, and
  J.~Gulín-González.
\newblock An overview about neural networks potentials in molecular dynamics
  simulation.
\newblock {\em Int. J. Quantum Chem.}, 124(11):e27389, 2024.

\bibitem{Kocer2021}
Emir Kocer, Tsz~Wai Ko, and Jörg Behler.
\newblock Neural network potentials: A concise overview of methods.
\newblock {\em arXiv:2107.03727}, 2021.

\bibitem{Blochel1994}
P.~E. Blöchl.
\newblock Projector augmented-wave method.
\newblock {\em Phys. Rev. B}, 50:17953, 1994.

\bibitem{Kresse1999}
G.~Kresse and D.~Joubert.
\newblock From ultrasoft pseudopotentials to the projector augmented-wave
  method.
\newblock {\em Phys. Rev. B}, 59:1758, 1999.

\bibitem{Kresse1993}
G.~Kresse and J.~Hafner.
\newblock Ab initio molecular dynamics for liquid metals.
\newblock {\em Phys. Rev. B}, 47:558(R), 1993.

\bibitem{Kresse1994}
G.~Kresse and J.~Hafner.
\newblock Ab initio molecular-dynamics simulation of the
  liquid-metal–amorphous-semiconductor transition in germanium.
\newblock {\em Phys. Rev. B}, 49:14251, 1994.

\bibitem{Kresse1996a}
G.~Kresse and J.~Furthmüller.
\newblock Efficiency of ab-initio total energy calculations for metals and
  semiconductors using a plane-wave basis set.
\newblock {\em Comput. Mater. Sci.}, 6:15, 1996.

\bibitem{Kresse1996b}
G.~Kresse and J.~Furthmüller.
\newblock Efficient iterative schemes for ab initio total-energy calculations
  using a plane-wave basis set.
\newblock {\em Phys. Rev. B}, 54:11169, 1996.

\bibitem{Perdew1996}
J.~P. Perdew, K.~Burke, and M.~Ernzerhof.
\newblock Generalized gradient approximation made simple.
\newblock {\em Phys. Rev. Lett.}, 77:3865, 1996.

\bibitem{Thompson2022}
A.~P. Thompson, H.~M. Aktulga, R.~Berger, D.~S. Bolintineanu, W.~M. Brown,
  P.~S. Crozier, P.~J. in~’t Veld, A.~Kohlmeyer, S.~G. Moore, T.~D. Nguyen,
  R.~Shan, M.~J. Stevens, J.~Tranchida, C.~Trott, and S.~J. Plimpton.
\newblock Lammps - a flexible simulation tool for particle-based materials
  modeling at the atomic, meso, and continuum scales.
\newblock {\em Computer Physics Communications}, 271:108171, 2022.

\bibitem{Togo1}
A.~Togo.
\newblock First-principles phonon calculations with phonopy and phono3py.
\newblock {\em J. Phys. Soc. Jpn.}, 92:012001, 2023.

\bibitem{Togo2}
T.~Tadano A.~Togo, L.~Chaput and I.~Tanaka.
\newblock Implementation strategies in phonopy and phono3py.
\newblock {\em J. Phys.:Condens. Matter}, 35:353001, 2023.

\bibitem{Zunger1990}
A.~Zunger, S.-H. Wei, L.~G. Ferreira, and J.~E. Bernard.
\newblock Special quasirandom structures.
\newblock {\em Phys. Rev. Lett.}, 65:353--356, 1990.

\bibitem{vandeWalle2002}
A.~van~de Walle, M.~Asta, and G.~Ceder.
\newblock The alloy theoretic automated toolkit: A user guide.
\newblock {\em Calphad}, 26:539--553, 2002.

\bibitem{vesta}
K.~Momma and F.~Izumi.
\newblock Vesta 3 for three-dimensional visualization of crystal, volumetric
  and morphology data.
\newblock {\em J. Appl. Crystallogr.}, 44:1272, 2011.

\bibitem{ovito}
Alexander Stukowski.
\newblock Visualization and analysis of atomistic simulation data with
  ovito–the open visualization tool.
\newblock {\em Modelling Simul. Mater. Sci. Eng.}, 18:015012, 2010.

\bibitem{QUIP}
Quip - quantum mechanics and interatomic potentials.
\newblock \url{https://github.com/libAtoms/QUIP}.
\newblock www.libatoms.org.

\bibitem{Hajinazar2021}
Samad Hajinazar, Aidan Thorn, Ernesto~D. Sandoval, Saba Kharabadze, and
  Aleksey~N. Kolmogorov.
\newblock Maise: Construction of neural network interatomic models and
  evolutionary structure optimization.
\newblock {\em Computer Physics Communications}, 259:107679, 2021.

\bibitem{Suneel2023}
Gokay Adabasi, Aditya Deshpande, Koichi Tanaka, Joshua Ancheta, Emmanuel
  Maldonado, Mehmet Özdoğan, Suneel Kodambaka, and Mehmet~Z. Baykara.
\newblock Nanoscale friction of high entropy alloy sulfide thin films in
  comparison with molybdenum disulfide.
\newblock {\em Appl. Phys. Lett.}, 123(261603), 2023.

\bibitem{Suneel2024}
Koichi Tanaka, Hicham Zaid, Toshihiro Aoki, Aditya Deshpande, Koki Hojo,
  Cristian~V. Ciobanu, and Suneel Kodambaka.
\newblock Growth of highly oriented {(VNbMoTaW){S$_2$}} layers.
\newblock {\em Nano Lett.}, 24(1):493--500, 2024.

\bibitem{Zahra2020}
Zahra Hemmat, John Cavin, Alireza Ahmadiparidari, Alexander Ruckel, Sina
  Rastegar, Saurabh~N. Misal, Leily Majidi, Khagesh Kumar, Shuxi Wang, Jinglong
  Guo, Radwa Dawood, Francisco Lagunas, Prakash Parajuli, Anh~Tuan Ngo,
  Larry~A. Curtiss, Sung~Beom Cho, Jordi Cabana, Robert~F. Klie, Rohan Mishra,
  and Amin Salehi-Khojin.
\newblock Quasi-binary transition metal dichalcogenide alloys: Thermodynamic
  stability prediction, scalable synthesis, and application.
\newblock {\em Advanced Materials}, 32(26):1907041, 2020.

\bibitem{Togo2015}
Atsushi Togo and Isao Tanaka.
\newblock First principles phonon calculations in materials science.
\newblock {\em Scripta Materialia}, 108:1--5, 2015.

\bibitem{cada2011}
Martin Čada, Zdeněk Hubička, Petr Adámek, Jan Klusoň, and Lubomír
  Jastrabík.
\newblock Time-resolved plasma parameters in the hipims discharge with ti
  target in {Ar/{O$_2$}} atmosphere.
\newblock {\em Surface and Coatings Technology}, 205:S317--S321, 2011.

\bibitem{Skelton2017}
J.~M. Skelton, L.~A. Burton, A.~J. Jackson, F.~Oba, S.~C. Parker, and A.~Walsh.
\newblock Lattice dynamics of the tin sulphides {Sn{S$_2$}}, {SnS} and
  {Sn{S$_3$}}: vibrational spectra and thermal transport.
\newblock {\em Physical Chemistry Chemical Physics}, 19(12452), 2017.

\bibitem{phonopy-spectroscopy}
Phonopy-spectroscopy.
\newblock \url{https://github.com/skelton-group/Phonopy-Spectroscopy}.

\bibitem{Lee2010}
Changgu Lee, Hugen Yan, Louis~E. Brus, Tony~F. Heinz, James Hone, and Sunmin
  Ryu.
\newblock Anomalous lattice vibrations of single$-$ and few$-$layer
  {Mo{S$_2$}}.
\newblock {\em ACS Nano}, 4(5):2695--2700, 2010.

\bibitem{Berkdemir2013}
Ayse Berkdemir, Humberto~R. Gutiérrez, Andrés~R. Botello-Méndez, Néstor
  Perea-López, Ana~Laura Elías, Chen-Ing Chia, Bei Wang, Vincent~H. Crespi,
  Florentino López-Urías, Jean-Christophe Charlier, Humberto Terrones, and
  Mauricio Terrones.
\newblock Identification of individual and few layers of {W{S$_2$}} using raman
  spectroscopy.
\newblock {\em Sci Rep}, 3(1755), 2013.

\bibitem{Su2020}
Jianwei Su, Mingshan Wang, Yuan Li, Fakun Wang, Qiao Chen, Peng Luo, Junbo Han,
  Shun Wang, Huiqiao Li, and Tianyou Zhai.
\newblock Sub-millimeter-scale monolayer p-type {H}-phase {VS$_2$}.
\newblock {\em Advanced Functional Materials}, 30(17):2000240, 2020.

\bibitem{McMullan1983}
W.~G. McMullan and J.~C. Irwin.
\newblock Raman scattering from {2H} and {3R--NbS$_2$}.
\newblock {\em Solid State Communications}, 45(7):557--560, 1983.

\bibitem{ValenciaIbanez2022}
Santiago Valencia-Ib{\'a}{\~n}ez, Diego Jim{\'e}nez-Guerrero, Juan~D.
  Salcedo-Pimienta, Karen Vega-Bustos, Gabriel C{\'a}rdenas-Chirivi, Leonardo
  L{\'o}pez-Manrique, Jos{\'e}~A. Galvis, Yenny Hern{\'a}ndez, and Paula
  Giraldo-Gallo.
\newblock Raman spectroscopy of few-layers {TaS$_2$} and {Mo}-doped {TaS$_2$}
  with enhanced superconductivity.
\newblock {\em Advanced Electronic Materials}, 8(10):2200457, 2022.

\bibitem{warren1941}
B.~E. Warren.
\newblock X-ray diffraction in random layer lattices.
\newblock {\em Phys. Rev.}, 59:693--698, 1941.

\bibitem{Hendricks1942}
Sterling Hendricks and Edward Teller.
\newblock X‐ray interference in partially ordered layer lattices.
\newblock {\em The Journal of Chemical Physics}, 10(3):147--167, 1942.

\end{thebibliography}

\begin{thebibliography}{1}

\bibitem{SI_Hendricks1942}
Sterling Hendricks and Edward Teller.
\newblock X‐ray interference in partially ordered layer lattices.
\newblock {\em The Journal of Chemical Physics}, 10(3):147--167, 1942.

\bibitem{SI_warren1941}
B.~E. Warren.
\newblock X-ray diffraction in random layer lattices.
\newblock {\em Phys. Rev.}, 59:693--698, 1941.

\bibitem{Warren1969book}
Bertram~Eugene Warren.
\newblock {\em X-Ray Diffraction}.
\newblock Addison-Wesley Publishing Company, Reading, MA, 1969.

\bibitem{DritsTchoubar1990}
I.G. Wood, V.A. Drits, and C.~Tchoubar.
\newblock X-ray diffraction by disordered lamellar structures. theory and
  applications to microdivided silicates and carbons.
\newblock {\em Clay Minerals}, 27(2):265–267, 1990.

\bibitem{Yang1996}
D.~Yang and R.F. Frindt.
\newblock Powder x-ray diffraction of turbostratically stacked layer systems.
\newblock {\em Journal of Materials Research}, 11:1733–1738, 1996.

\bibitem{March1932}
Arthur March.
\newblock Mathematische theorie der regelung nach der korngestalt bei affiner
  deformation.
\newblock {\em Zeitschrift für Kristallographie - Crystalline Materials},
  81(1--6):285--297, 1932.

\bibitem{Dollase1986}
W.~A. Dollase.
\newblock Correction of intensities for preferred orientation in powder
  diffractometry: application of the march model.
\newblock {\em Journal of Applied Crystallography}, 19(4):267--272, 1986.

\bibitem{Waasmaier1995}
D.~Waasmaier and A.~Kirfel.
\newblock New analytical scattering-factor functions for free atoms and ions.
\newblock {\em Acta Crystallographica Section A}, 51(3):416--431, 1995.

\bibitem{Chantler2000}
C.~T. Chantler.
\newblock Detailed tabulation of atomic form factors, photoelectric absorption
  and scattering cross section, and mass attenuation coefficients in the
  vicinity of absorption edges in the soft x-ray (z = 30–36, z = 60–89, e =
  0.1–10 kev) – addressing convergence issues of earlier work.
\newblock {\em Journal of Synchrotron Radiation}, 8(4):1124--1124, 2001.

\end{thebibliography}
\end{document}